\documentclass{PoS}

\usepackage{amsmath}
\usepackage{amsfonts}
\usepackage{amssymb}
\usepackage{dsfont}
\usepackage[ansinew]{inputenc}

\newtheorem{definition}{Definition}[section]
\newtheorem{theorem}{Theorem}[section]
\newtheorem{lemma}{Lemma}[section]
\newtheorem{proposition}{Proposition}[section]
\newtheorem{corollary}{Corollary}

\title{Introduction to Loop Quantum Gravity}

\ShortTitle{Introduction to Loop Quantum Gravity}

\author{\speaker{Simone Mercuri}
         \\
        Institute for Gravitation and the Cosmos,\\The Pennsylvania State University,\\Physics Department, 104 Davey Lab, PMB 67,\\University Park, PA 16802, USA\\
        E-mail: \email{mercuri@gravity.psu.edu}}

\abstract{The questions I have been asked during the 5th International School on Field Theory and Gravitation, have compelled me to give an account of the premises that I consider important for a beginner's approach to Loop Quantum Gravity. After a description of some general arguments and an introduction to the canonical theory of gravity, I review the background independent approach to quantum gravity, giving only a brief survey of Loop Quantum Gravity.}

\FullConference{5th International School on Field Theory and Gravitation\\
		 April 20-24, 2009\\
		 Cuiabá city, Brazil}

\begin{document}

\section{Preface}
Loop Quantum Gravity (LQG) is a background independent and mathematically rigorous canonical quantization of the gravitational field. The Organizers of the 5th International School on Field Theory and Gravitation have asked me to give an understandable account of the main techniques and results of this theory and I have been pleased to fulfill their requests. 
I immediately realized that the ideas of LQG were capturing the interest of many students, who asked me a lot of clarifications about some mathematical tools and physical aspects of background independent theories. I honestly think to have answered their questions and clarified many details, but, at the same time, I had the impression that many of them were getting confused by the canonical formulation of gravity and by the Dirac quantization procedure.

Generally speaking, my impression has been that the study of canonical quantization of gauge theories and, especially, gravity presupposes the knowledge of some arguments which are often not treated in details in the basic courses. I had the same experience when I started working on quantum gravity during my PhD. Usually, the approach to General Relativity (GR) proposed in the basic courses is based on the Lagrangian mechanics, while the canonical formulation of the theory is only marginally described, without deepening into the general pictures of theories with constraints. On the one hand, this is understandable from the perspective of academic and practical purposes, but, on the other hand, the Hamiltonian, or better, the canonical formulation of gauge theories and gravity remain an obscure argument among students. This is a common problem, which cannot be neglected when one attempts to describe the problem of quantum gravity and, more specifically, LQG.

Therefore, motivated by the questions asked me during coffee breaks and lunches, I have decided to slightly shift the focus of this proceeding, giving to the readers the opportunity to begin with quantum gravity from what I consider its natural starting point, namely the physical formulation of the problem.

So, Section \ref{sec2} is entirely devoted to describe some simple motivations which compel one to formulate a quantum theory of gravity, digressing on the physical implications of such a theory on the existing concepts of space and time. Section \ref{sec3} contains some preliminary arguments, which I consider as fundamental to understand the following discussion. They are quite simple and well known arguments, nevertheless they have to be necessarily clear before going on to face the canonical theory of gravity and the Ashtekar-Barbero formulation of GR. So, I collected in Section \ref{sec3} an extremely brief description of the causal structure of space-time, group theory, Dirac canonical formulation of gauge theories, and the initial value formulation of theories with gauge freedom, using a simple language and neglecting many complicated details. This should also serve as an easily accessible account of the main definitions and concepts used throughout the paper. In Section \ref{sec4} I describe the canonical formulation of GR starting from the 3+1 splitting of space-time. The mathematical procedure that allows to write the Einstein equations in the Hamiltonian form is described in details; the Section concludes with a description of the initial value problem in gravity. Section \ref{sec5} is dedicated to the connections formulation of canonical GR, better known as Ashtekar--Barbero (AB) formulation of gravity. This argument is particularly interesting in view of quantization, because by using the so-called AB connections, the constraints of GR can be rewritten in a more suitable form for quantization. Interestingly enough, the use of AB connections introduces a quantum ambiguity known as Barbero--Immirzi (BI) parameter, which affects the eigenvalues of geometrical quantum operators. The physical interpretation of the BI parameter and its correlation with the topological sector of the theory is at present an argument of active discussion. Recently, the idea that this parameter is in fact a field has attracted the interest of many researchers (me included!). I think that this could represent an interesting bridge with particles physics and could have consequences in Cosmology that deserve to be studied. For these reasons I invite the readers to refer to the original papers to get more information. Finally, in Section \ref{sec6}, I face the problem of quantization, starting from a brief description of the Dirac procedure and the Wheeler-DeWitt (WDW) equation. The last part of this Section is dedicated to the description of the main ideas of LQG, without entering in the complicated details of the theory. As I have explained before, in this paper I prefer to focus on introductory arguments to LQG, more than on the theory itself, which is beautifully described in many books and reviews, written by the major experts in this field \cite{AshLew2004,Rov2004,Thi2001}. For this reason I refer the readers for more details to the standard Literature on LQG, with the hope that this paper could help them to face the argument more confidently. 

At the end, I also added two appendices, one on differential forms, which are commonly used in Literature, but sometimes little known among students; while the other is about the topological sector of gauge theories, and aims to clarify some concepts which apply in canonical quantum gravity as well.

Throughout the paper, I will use an extremely simple approach, sometimes neglecting some interesting but slightly involved details. This obviously will affect the completeness and the rigor of the discussion, but, I am sure, will be appreciated by beginners and the students of the International School of Field Theory and Gravitation, who can find in this paper a simple description of many arguments. My hope is to give them the possibility to get a ``first order understanding'' of the main concepts of canonical quantum gravity, without being discouraged by the rigorous mathematical formulation of the problem. I strongly suggest interested readers to delve into the ``higher order descriptions'' of the more complete and rigorous books and reviews cited above.

\section{What is Quantum Gravity?}\label{sec2}
\noindent The present knowledge in Physics is the result of the new and revolutionary ideas born in the last century, which later led to the formulation of the two major physical theories describing the four interactions: Quantum Mechanics (QM) and General Relativity (GR). They have, on the one hand, opened the way to a great number of scientific discoveries and technical developments, but, on the other hand, they destroyed the coherence of prerelativistic classical physics \cite{Rov2004}, since the basic assumptions of each one of the two theories are contradicted by the other. QM is formulated using a Newtonian absolute (fixed, non-dynamical) space-time. On the contrary GR describes the dynamics of space-time itself, which is no more an external set of clocks and rods, but a physical interacting field, namely the gravitational field. The basic physical lesson of GR is contained in the following simple sentence: \emph{Geometry tells matter how to move; matter tells geometry how to curve}, which expresses in a very suggestive way the fact that the theory describes both the dynamics of space-time (or gravitational field) and the motion of the bodies subjected to the gravitational field. But it also contains the seeds of an issue, namely the separation of the physical world in matter and geometry. 

This dichotomy in Physics, together with the fact that on the one side we have the present description of what is in general intended as matter, namely the electromagnetic, weak and strong interactions, unified in the language of Quantum Field Theory (QFT) and on the other side gravity (or geometry) described by the pure classical theory of General Relativity, creates a sort of ``scientific discomfort'' \cite{Ame-Cam2006}. This is not only a philosophical problem, but assumes the distinguishing features of a real scientific problem as soon as one considers measurements in which both quantum and gravitational effects cannot be neglected. In fact, QM and GR are hugely successful in their own range of applicability, but they seem to be subjected to a sort of ``reciprocal exclusion principle''. In particular, QM describes microscopic phenomena involving fundamental particles, ignoring completely gravity, while GR describes macroscopic systems, whose quantum properties are in general (safely) neglected. So far, no experimental evidences are available on systems in which neither gravity nor quantum effects can be neglected, but we already know that our current theories would not be able to describe such phenomena. This situation is usual in Physics and it is, in general, the prelude to the formalization of a well stated scientific problem \cite{Ame-Cam2006}. Specifically, the goal of obtaining quantitative predictions about the outcomes of certain measurements on extremely energetic gravitational systems is often referred as the \emph{Quantum Gravity Problem}.

It is clear from what said above that new elements could be necessary in order to make our current theories able to face a certain class of physical phenomena. Then, it is natural to wonder whether these new elements affect low energy processes. In other words, should we expect, even at low energies, small QG corrections to the predictions of our current theories? It is worth stressing that even a little deviation from the predictions of standard physics ascribable to any QG effects, found in the experimental data of current and future experiments, would have an enormous impact on the research. We recall as an example the Lamb shift, which motivated and stimulated the studies about QED. From this perspective, it could be important to answer to the following question: \emph{How far we are from an experimental evidence of a QG effect.} We cannot give a completely satisfying answer to it, nevertheless we can use a dimensional argument to state that the new effects should modify the usual predictions with additional terms proportional to the factor $\left(E/E_{Pl}\right)^n$, where $E$ is the typical energy scale of the experiments, $E_{Pl}$ is the Planck energy ($E_{Pl}\equiv \left(G_N\right)^{-1/2}\approx 10^{28} eV$), while $n$ is a positive integer number. At this point one may be surprised by the compelling necessity to quantize gravity felt by physicists, since at the LHC, the most powerful accelerator ever projected, we can reach a very small energy if compared to the Planck scale. Specifically, the ratio between the Planck and the reachable energy is of the order of $E_{LHC}/E_{Pl}\approx 10^{-15}$, that means we are fifteen orders of magnitude below the scale at which we expect to see the quantum effects of the gravitational field. Even though this fact is true, it is absolutely false that this is a good reason for abandoning the program of constructing a consistent theory of QG. The motivations are connected with the fact that there exist in Nature particles of energies much larger than those we can produce with the accelerators, moreover during its evolution our Universe experimented regimes in which the energy available was (most likely) even larger of the Planck scale. Furthermore, even though the factor $E/E_{Pl}$ is extremely small at the present available energies and, consequently, the QG effects cannot be directly experimented, it exists the concrete possibility that some astrophysical phenomena can behave as magnifying glasses, being able to make them visible in near future \cite{Ame-CamEllMav1998,Pir2004}. In other words, the fact that QG effects are expected to be very tiny does not mean they are absolutely untestable, clearly as one should expect the opportunities to make such tests are rather rare. We also emphasize that it could be not necessary to reach the Planck energy to see some QG effects.  In this respect, we recall that a class of extremely energetic phenomena called \emph{Gamma Ray Bursts} (GRB) could represent a really important laboratory to test QG predictions, in fact they seem to be the natural candidates to verify whether the fundamental hypothesis about a discrete structure of space-time will be confirmed by experiments. The peculiar features which make GRB relevant for QG is the extremely wide range of the emitted energies and the cosmological distance of the explosive events.

Concluding, as usual in Physics, from the pure empirical point of view, the new framework possibly introduced by a consistent and complete QG theory could represent a tiny deviation from what we already know, being only a further small modification of those empirical laws which give us a pictorial description of how Nature works. But from the theoretical point of view, it could represent not only the completion of that revolution QM and GR introduced in the last century, but also open the way to the discovery of complete new aspects of Nature, exactly what QFT and GR have been doing during the last one-hundred years. However, in spite of their great empirical success, QM and GR have left us with a fragmented understanding of the physical world, this requires a new synthesis, which is a major challenge in today's fundamental Physics \cite{Rov2004}.

In this sense, Quantum Gravity can solve the dichotomy present in our current understanding of physical phenomena and, moreover, it could give us predictions on those regimes in which the quantum and gravity effects merge.

\subsection{Why we need a Quantum Theory of Gravity?}\label{WDWNQTG?}

Above we introduced the so called QG problem, which gains the status of a true scientific problem as soon as one considers physical systems in which both the gravitational and the quantum mechanical effects play an important role. Moreover, the failure of the existing theories as soon as we push them near their extreme margins of applicability, suggests to quantize gravity. We also digressed on the empirical content of such a problem, affirming that obtaining an experimental evidence of a QG effect is extremely complicated, but not impossible; even though we expect a very tiny modification of the existing laws describing the physical systems. Therefore, as often occurs, one may object to the above pleaded motivations, saying that they are mainly suggested by philosophical reasons. Namely, the hope of finding the way to conciliate the basic assumptions of two very different and complementary theories is not really required by a scientific problem. Put differently, this attempt has very little to share with Physics, simply because the physical effects are so tiny to be actually undetectable. It could be really so! 

For this reason, in this Section three well stated problems, which regards respectively QFT, GR and the merging point of QFT and GR, are discussed in order to emphasize that QG is a true physical problem. It, in fact, can provide information about the behavior of fundamental gravitating quantum systems as, for example, a system of gravitating fermions, or extremely energetic scattering processes, or the Universe itself in its initial expansion.

The questions we present below are generally connected with the fundamental structures of the theories and mainly concern the problem of singularities. In fact, it is worth recalling that the theory describing the gravitational interaction fails in giving a fully satisfactory description of the observed Universe \cite{Haw1979}. GR, indeed, leads inevitably to space-time singularities as a number of theorems mainly due to Hawking and Penrose  demonstrate. The singularities occur both at the beginning of the expansion of our Universe and in the collapse of gravitating objects to form Black Holes \cite{HawPen1970}. Classical GR breaks completely down at these singularities, or rather it results to be an incomplete theory, because it neither gives a description of the singularities themselves, nor provides the boundary conditions for fields in the singular points. The appearance of singularities in extremal situations reflects both on QFT and on GR itself, generating a more subtle question when we trivially try to merge the two theories as we are going to argue.

This failure of current theories represents a good reason to pose a scientific problem and its solution is widely believed to be in the formulation of a consistent and complete theory of QG.

It remains to treat in a more systematic way the problem regarding the empirical content of a QG theory. This argument is postponed at the end of the Section, where we will trace the way to get information about QG from the astrophysical phenomena named Gamma Ray Bursts, motivating the belief that the necessary experimental evidences which could support the QG research are not so far.

\subsubsection{Planck scale collisions}

The question is:\footnote{Some of the ideas presented below are extracted from \cite{Ame-Cam2006} and \cite{Rov2004}.} What would happen if we managed to collide an electron-positron pair of energy \emph{per particle} of $10^{28} eV$? We are unable to give an answer to this question, because the energy in the center of mass is greater than the Planck energy. So, what would happen in the center of mass during such a collision is completely out of our understanding. But, according to our present physical theories describing the collisions between fundamental particles, there should be nothing peculiar in the setup of such an experiment. However, the same hugely successful theories are not able to provide us with a consistent prediction for the outcome of this experiment. The reason of this failure is related to the fact that, in such an experiment, we cannot neglect the gravitational properties of the involved particles at the moment of the collision. But, we do not have any scientific information on how taking into account such an effect in the framework of QFT. In other words, when the gravitational field is so intense that space-time geometry evolves on a very short time scale, QFT cannot be consistently applied any longer. Or, from another perspective, we can say that when the gravitational effects are so strong to produce the emergence of space-time singularities, field theory falls into troubles. Summarizing, we are able to extract numbers (predictions) from QFT when the curved space-time is static (or slowly varying) and non-singular, but we are not able to handle situations in which the gravitational field is so intense to give rise to a fast varying and singular space-time \cite{Ame-Cam2006}.

The incompatibility between QM and GR in treating the proposed scattering problem can be further analyzed. In this respect, we have to remember that GR governs consistently the space-time and particles dynamics. In particular, given the Lagrangian for the matter, once the Einstein equations have been solved, we can predict the trajectories of the particles. But, in the framework of QFT, particles are asymptotic states of quantum operators and during the collision they do not follow any classical trajectory. The whole dynamics of the collision is contained in the S-matrix, which gives the evolution from the initial ($\left|in\right>$) to the final ($\left|out\right>$) state. During the interaction, the intermediate state is a pure quantum superposition of all the possible states compatible with the quantum numbers of the initial state; namely, we can associate to particles a semi-classical (fuzzy) trajectory only asymptotically, in other words much before or much later the collision. One could try to apply the formalism of GR to the formally classical trajectories contained in the path-integral, but the problem remains ill-defined and, generally speaking, it will be affected by divergences if the energy of the particle are sufficiently high to generate a significant geometrodynamics. This fact is suitable for a pictorial description. The path-integral formulation of QM consists in summing on all the possible trajectories connecting the initial and final quantum state, with a weight proportional to the exponential of the classical action. The major role in the sum is played by those trajectories near the classical one, because the contribution of those ``far away'' from the classical one is suppressed by the weight factor. If the gravitational field is weak, we can safely approximate the space-time near the classical trajectory with the Minkowski flat space-time, and assume that the curvature does not affect the trajectories which enter in the sum. But, if the gravitational field is intense, then also those trajectories close enough to the classical one are affected by the curvature of the space-time and this effect should be taken into account in the path-integral sum. The case of fast varying or singular gravitational field is even worst, because in that case the trajectories could fall into a singularity or oscillate very fast. This breaks down completely the formalism, by introducing remarkable and uncontrollable effects into the sum.

\subsubsection{Singularities}

The study of singularities in GR is an absolutely fascinating argument. Here we give an extremely brief account of this huge argument, which represents one of the crucial point suggesting that a quantum theory of gravity is, in fact, necessary. Near singularities classical GR becomes inconsistent and incomplete, as we already stressed before and, differently from Newtonian gravity, they represent an inevitable feature of the theory. Indeed, in Newtonian gravity the $r=0$ singularity, appearing in the complete collapse of a spherical non-rotating shell of dust (namely, when all the matter reach simultaneously the origin), can be easily avoided slightly perturbing the spherical symmetry of the collapsing shell, for example giving it a little rotation. On the contrary, in GR the singular behavior of space-time cannot be avoided. All the solutions we have of Einstein equations show a singular behavior. But, since all of them are characterized by symmetries, one may think that, as it happens in Newtonian gravity, the relaxation of symmetries could allow to avoid singular points. But the Hawking, Penrose \emph{et al.} theorems demonstrate that this is not the case. Wald says \cite{Wal1984}:
\begin{quotation}
Although the singularity theorems do not prove that the singularities of classical General Relativity must involve unbounded large curvature, they strong suggest the occurrence in Cosmology and gravitational collapse of conditions in which quantum or other effects which invalidate classical General Relativity will play a dominant role. 
\end{quotation}
So the singularity theorems do not use the natural and, in a certain sense, more physical notion of unbounded density to characterize the space-time singularities, but the characterization of singularities is based on the notion of incompleteness of geodesics, which, however, contains some unwanted features.

The use of such a notion to characterize space-time singularities is due to the necessity of a diffeomorphisms invariant criterion. In this respect, it is worth noting that the singular points in $r=0$ of Schwarzschild or Robertson-Walker metrics, rigorously speaking, are not point of those space-times, otherwise the metrics would not be well defined everywhere on the manifold $M$. Moreover, the criterion based on the bad behavior of higher order scalars constructed with the curvature tensor does not work in some cases, so it is clear that a more satisfactory definition has to be introduced.

So far, the criterion of geodesics incompleteness seems to be the most appropriate. Its physical meaning is suggested directly by its definition. Indeed, a geodesic is said to be incomplete when it is inextendible in at least one direction; namely, it has a finite range for the affine parameter. As a consequence, a particle falling along a time-like or null inextendible geodesic will end its existence within a finite proper time (or it began its existence a finite proper time ago). So, even though a completely satisfying notion of singularity lacks \cite{Wal1984}, we call ``physically singular'' all those space-times having at least one incomplete geodesic.

At this point, the following famous theorem by Hawking and Penrose can be enunciated:\footnote{See § \ref{csost} for definitions.}
\begin{theorem}
\emph{Singularity theorem of Hawking and Penrose (1970)}. Let us suppose that the space-time $(M,g_{\mu\nu})$ satisfies the following four hypothesis:
\begin{enumerate}
	\item $R_{\mu\nu}u^{\mu}u^{\nu}\geq 0$ for all time-like or null $u^{\mu}$;
	\item it exists at least one point for every time-like or null geodesics at which $R_{\mu\nu}u^{\mu}u^{\nu}\neq 0$;
	\item no closed time-like curve exist;
	\item at lest one of the following three conditions holds:
	 \\i) $(M,g_{\mu\nu})$ possesses a compact achronal set without edge, i.e. $(M,g_{\mu\nu})$ is a closed Universe,
	 \\ii) $(M,g_{\mu\nu})$ possesses a trapped surface,
	 \\iii) there exists a point $p\in M$ such that the expansion of the future (or past) directed null geodesics emanating from $p$ becomes negative along each geodesic in this congruence.
\end{enumerate}
Then $(M,g_{\mu\nu})$ must contain at least one incomplete time-like or null geodesic.
\end{theorem}
The first three conditions of the above theorem are believed to be satisfied in our Universe. The first one in particular can be simply showed to be shared by all those space-times satisfying the Einstein equations and the \emph{strong energy condition},\footnote{Namely for every time-like $u^{\mu}$ we have $T_{\mu\nu}u^{\mu}u^{\nu}\geq -T/2$, where $T_{\mu\nu}$ is the energy-momentum tensor of the matter.} which seems to be plausible for ordinary matter. 

Finally, we can conclude that strong evidences suggest that our Universe is singular; of course we cannot know by which kind of singularity it is characterized, because the above theorem does not give us any insight on this question.\footnote{The particular kind of singularity we obtain from the Einstein equations depends in general on the particular symmetries of the model.} Nevertheless, it suggests that an extension toward QG is necessary. It, in fact, demonstrates that the universally accepted theory of gravity cannot definitely give us a complete and consistent description of the evolution of our Universe.

\subsubsection{Merging General Relativity and Quantum Field Theory}\label{MGRAQFT}

After having described the problem of singularities in GR, we want to address a number of difficulties coming out when one attempts to quantize gravity using the usual formalism of QFT. The main source of problems in extending the formalism of QFT to gravity is represented by the double role played by the metric tensor. It, in fact, represents both the dynamical field describing gravity and the tensor describing the causal structure of the background. The nature of these difficulties is not only conceptual, as for example the consideration that a quantum theory of gravity would imply a quantum, namely discrete, structure of space-time itself. But, more practically, they are correlated to the profound difference existing between GR and other classical field theories: while in the latter the background is always considered as an external and fixed structure, the former is the theory describing the dynamics of the background itself. Therefore, we cannot assume a given structure of space-time \emph{ab initio}, but we have to invent a formalism that allows to quantize a classical theory in a background independent way. This fact makes GR so peculiar that, so far, all the attempts to quantize gravity have encountered fundamental difficulties; only in the last years many of these obstacles have been overcome, leading to a consistent QG \cite{AshLew2004,Rov2004,Thi2001}.

As is well known, the construction of a QFT on Minkowski space-time of a free or perturbatively interacting field is the only procedure we can control. In particular, given a small number of axioms, the Wightman axioms, we can construct a consistent QFT. Let us begin the description of the issues one would find in applying the usual formalism of QFT to gravity, by describing a very simple and well known example, which best illustrates the dichotomy existing in the metric tensor.

One of the Wightman axioms contains the notion of micro-causality. In order to introduce this concept let me consider a scalar field represented by the smeared operator-valued distribution
\begin{equation}
\Phi\left(f\right)=\int_{\mathbb{R}^{n+1}}d^{n+1}\!\!x\ \Phi\left(x\right)f\left(x\right)\,,
\end{equation}
where $f$ is a test function of rapid decrease. Suppose now that the supports of the test functions $f$ and $f^{\prime}$ are space-like separated, then the micro-causality assumption is equivalent to require that:
\begin{equation}
\left[\Phi\left(f\right),\Phi\left(f^{\prime}\right)\right]=0\,. 
\end{equation}
Physically, the above condition assures that a measurement of the field $\Phi$ in the region of space-time contained in the support of the function $f$ cannot be influenced by the measure of the same field in the region contained in the support of the function $f^{\prime}$.

The gravitational interaction is described by a self-interacting spin-$2$ field, so, it is natural to expect, in analogy with the previous example, that the following commutation relation holds
\begin{equation}\label{metric commutator}
\left[g\left(f\right),g\left(f^{\prime}\right)\right]=0\,,
\end{equation}
where $f$ and $f^{\prime}$ are two tensorial test functions with compact supports separated by a space-like distance. But, strictly speaking, the above relation makes no sense. The reason being that we cannot give a consistent meaning to the requirement that the supports of the test functions be space-like separated, unless we already know the state of the gravitational field, namely the metric tensor. But, the commutation relation in line (\ref{metric commutator}) must hold independently of the state of the gravitational field. 

It is clear that the micro-causality condition is the quantum translation of the pure classical concept of causality. Even though it contains information about measurements on quantum field, its basic structure is, however, founded upon the existence of a Minkowski space-time, with its fixed light cone. In a quantum theory of gravity, the causality condition and, as a consequence, the micro-causality axiom, is inextricably bound with the quantum dynamics of the gravitational field, so it has to be, at least, modified or completely replaced by a new requirement. 

In general, the other Wightman axioms are violated in gravity as well. The reason being that the fundamental objects one has to postulate in order to write down the axioms, namely i) a differentiable manifold $\mathbb{R}^{n+1}$, on which a non-dynamical Minkowski metric $\eta$ is defined, together with its fixed future and past causal light cones $J^{+}\cup J^{-}$, ii) a symmetry group, which, in the Minkowski space-time, is the ten parameters Poincar\'e group $\mathcal{P}$, together with its associated infinite dimensional representation acting on the quantum states $U\left(\mathcal{P}\right)$ and iii) an invariant vacuum state $\left|\Omega\right\rangle$, cannot be properly postulated. For a general space-time, in fact, we do not have a symmetry group and a unique invariant vacuum state. Consequently, it neither exists any obvious generalization of the Wightman axioms, nor one can rigorously define any Fock-Hilbert space for the quantum states of the theory. Therefore, the whole formalism falls into troubles.

A possible solution to these issues is the splitting of the metric tensor as would be suggested by a perturbative approach:
\begin{equation}\label{metric tensor separation}
g_{\mu\nu}=\eta_{\mu\nu}+\gamma_{\mu\nu}\,.
\end{equation}
The basic assumption here is that $\gamma_{\mu\nu}$ represents the dynamical variable describing a self-interacting spin-$2$ field, while $\eta_{\mu\nu}$ describes the background metric, which, in general, could be any solution of the classical Einstein equations. This method is of course mathematically correct. It provides some interesting insights on the quantization of a spin-2 field on a fixed (in general curved) space-time and could be useful to describe the interaction between gravitons and matter or the gravitational waves. Nonetheless, it cannot be considered as a good starting point for a complete quantum theory of gravity, because the metric separation (\ref{metric tensor separation}) destroys the full general covariance of the classical theory, namely its main constructing principle. More practically, the infinite perturbative series becomes meaningless if the fluctuations become large. In other words, GR is, in general, a non-renormalizable theory. One can hope that the perturbation theory turns out to be finite as a consequence of possible (magic) cancellations of the divergences, but this hope is unjustified \cite{GorSag1986}. This means that the resulting theory cannot predict any physical result. We still can advance the hypothesis that the super-symmetric extension of this theory has a chance to be a finite theory, because, as is well known, the super-symmetric extension of a classical field theory features, in general, an higher degree of ultraviolet convergence due to fermionic cancellations \cite{Nie1981}. But again the resulting theory is non-renormalizable, even worst neither the eleven dimensional super-gravity theory shows any hoped cancellation property \cite{Des1999} (see also \cite{BerCarDix09} and references therein). Then, although we do not have a complete proof of the failure of the QFT formalism in the case of gravity, it is widely believed that the perturbative approach does not provide a completely consistent answer to the problem of QG .

All these issues have been sometimes pushed to the extreme consequences by some authors, who argued that classical GR is correct at the fundamental level. This position is, however, untenable for at least two reasons: The inevitability of singularities in classical GR; and the interaction of gravity with quantum matter systems, which is a source of troubles as we are going to show.

The question we want to answer is: what is the curvature of space-time associated to a given quantum state of the matter? Let us suppose that the classical Einstein equation holds at the fundamental level, then the most natural candidate to get an answer to this question is:
\begin{equation}\label{semiclassical einstein equations} 
R_{\mu\nu}-\frac{1}{2}\,g_{\mu\nu}R=\left\langle\hat{T}_{\mu\nu}\right\rangle\,.
\end{equation}
Where in the right hand side we have put the expectation value of the energy-momentum tensor of the matter in a given quantum state. Now suppose that the quantum state of the matter is  such that we have probability $1/2$ for the localization of all the matter in a region of space-time denoted as $U_1$ and the same probability for the localization in another region $U_2$, disjoint from the region $U_1$. In other words, we are in the following situation:
\begin{equation}
\left|\text{matter}\right\rangle=
\frac{1}{\sqrt{2}}\left|\text{all matter in $U_1$}\right\rangle+\frac{e^{i\theta}}{\sqrt{2}}\left|\text{all matter in $U_2$}\right\rangle,
\end{equation}
where $U_1$ and $U_2$ are disjoint regions of space-time. 

In this physical situation, the gravitational field, according to equation (\ref{semiclassical einstein equations}), would behave like half of the matter were in $U_1$ and the other half in $U_2$. Now, if we resolve the quantum state by measuring the position of matter, we will find all the matter either in $U_1$ or in $U_2$. Then, the gravitational field should modify in a discontinuous acausal manner, leading to serious difficulties \cite{Wal1984}. The idea that this problem can find its consistent description in the framework of a quantum theory of gravity is widely accepted. It could, hopefully, provide an answer to the following question: how does a quantum particle modify space-time? This question is, to a large extent, equivalent to the previous one and contains the subtle problem regarding the interaction of the quanta of the gravitational field with matter. 

So, QG seems to be necessary as soon as we consider the interaction of quantum matter with the gravitational field, but, to quantize gravity as a usual QFT, we have to face a large number of conceptual and technical issues. This is a long standing problem, nevertheless, it does not contain any indication about a fundamental incompatibility between QM and GR principles. In fact, as stressed by Rovelli \cite{Rov2004}, it is important to distinguish between QM, which is a general mechanical theory and QFT which can be considered as a particular application of the laws of QM to a system with an infinite number of degrees of freedom. As we said above, GR is incompatible with the formalism of QFT for the non-existence of a fixed background structure, but this does not means that it is incompatible with QM at all \cite{Rov1991-1}. So the right question one should pose is the following one: is it possible to construct a quantum theory of a system with an infinite number of degrees of freedom, without assuming a fixed background causal structure? The answer is: yes! It is possible, as the modern background independent theories of quantum General Relativity demonstrate. However, more insights are necessary in order to better face many issues, in particular those connected with measurements and the consistent introduction of measuring devices in the framework of such theories.

\subsection{Space-time, background independence and relationalism in Physics}\label{STBIRP}

The question we want to discuss here represents another important ``open issue'' of the research in QG and regards the fate of classical space-time.

A very interesting feature of GR and QFT is that both are compatible with a classical description of space-time, although they do not share the same construction of such a classical space-time. According to QFT, the strategy to sharply localize a point in space-time requires a limiting procedure on the mass of the devices localizing points; in GR, instead, the localization of a point is a background independent procedure, based on the crossing of two geodesics. Below, I will describe in more details these contrasting features.

The concept of classical space-time is appropriate in Physics as long as the proposed theories allows to localize space-time points sharply. A fundamental requirement for a consistent physical theory is the agreement between the limits enforced on possible measurements by the formalism and the limits imposed, in principle, by the physical measurements procedure. An example to illustrate this important point can be easily constructed by considering the measurement of the angular momentum vector. In classical physics, the angular momentum of a particle can be sharply measured. In other words, we can, in principle, measure each one of the components of the angular momentum vector of a particle simultaneously and with an infinite precision. But, modern physics has radically changed our perspective by imposing some limitations to the measurements allowed by the experimental procedures, namely, we cannot measure simultaneously all the components of the angular momentum with an infinite precision. As a consequence, in the description of such a system, the angular momentum has to be described by a non-classical formalism, which incorporates the experimental limitations found.

This aspect deserves to be further discussed and clarified, because it is correlated with the fate of the classical concept of space-time as described below. Fortunately, a good example exists in the history of physics, which involved authoritative physicists as Einstein, Bohr, Landau and Rosenfeld. A lively and fruitful debate, in fact, animated the scientific community during the period immediately after the birth of quantum electrodynamics. The matter of the discussion regarded the measurement of the electromagnetic field in the framework of quantum electrodynamics, leading to the formulation of the so-called \emph{Bohr-Rosenfeld criteria} for a consistent theory. Before discussing this matter, we want to focus the attention of the reader on the importance of these kind of arguments for theoretical physics. A simple example can clarify this point: In this respect, we want to stress that the study of the synchronization of distant clocks, once an absolute maximum velocity for signals has been assumed, that led Einstein to special relativity.

\subsubsection{Bronstein objection and the fate of space-time in QG}\label{BRCBOFM}

Here, we want to briefly describe the main points of the debate come out in 30s, when a group of really distinguished physicists argued that there was a conceptual disagreement between the uncertainty limits predicted by quantum electrodynamics on a certain class of observations, and the mathematical formalism that the same theory adopts for describing these measurements results. Specifically, the key point of the debate was a physical consideration due to Landau and Peierls. They argued that, according to quantum electrodynamics, the electric field in a generic point $P$ can be measured sharply, namely with zero uncertainty, consequently a measurement procedure allowing to measure it sharply must exist. Obviously, if the situation were different then the theory would be inconsistent. Eventually, the conclusion of Landau and Peierls was that quantum electrodynamics must be rejected as physical theory, because such a zero uncertainty measurement procedure is not possible in Nature, so quantum electrodynamics is inconsistent.

The conclusion of Landau and Peierls was motivated by the following consideration: The measurement procedure of the electric field in a generic point can be ideally performed by using an electrically charged probe, which undergoes the effect of the electric field and accelerates, the outcome consists in the experimental measurement of this acceleration. In principle one can measure the mean value of the electric field in a small region of space, approaching the point asymptotically. Classically, this procedure is perfectly consistent, but, quantum mechanically, it is affected by the Heisenberg principle. In fact, the simultaneous measurements of the localization of the probe, say $\Delta X$ and of the variation of its momentum due to the effect of the electric field, say $\Delta P$ must satisfy the following uncertainty relation $\Delta X\Delta P\geq\hbar$. Moreover, the acceleration of the probe introduces another problem, related to the fact that an accelerated charge emits energy. So, the measurement affects itself by modifying the momentum of the probe, namely the outcome of the experiment.

Such an analysis, in fact, leads to the conclusion that a sharp measurement of the electric field is possible only in very special conditions. Indeed, in order to avoid the two effects described above, the characteristics of the probe has to be adjusted in such a way that, by a limiting procedure, it is possible to reduce the ratio between its electric charge density and its mass density to zero.

But, as noted by Landau and Peierls, the above requirement cannot be fulfilled, because, even though in Nature it exists a great variety of particles with different charge/mass ratios, no one can constitute the ideal probe, namely with a charge/mass ratio equal to zero. As a consequence, quantum electrodynamics must be rejected and a consistent alternative should be sought.

Bohr and Rosenfeld opposed to this viewpoint, claiming the consistency of quantum electrodynamics. The point is that the generations of particles existing in Nature are not a prediction of the theory, rather they are an outside input; in other words, given the particle content, the theory predicts their mutual interactions. In this sense, the failure pointed out by Landau and Peierls cannot be considered as an inconsistency of the theory, because it can be attributed to an external fact. Therefore, it does not affect the logical structure of the physical theory. If we found particles with a vanishing charge/mass ratio, then we would be able to measure sharply the electric field and quantum electrodynamics would not absolutely be in contrast with such a discovery.

The same argument applied to the gravitational field has a remarkable consequence related to the fate of the common accepted notion of space-time. In fact, as Bronstein pointed out, the requirement that the charge/mass ratio vanishes for the probe used to measure the field cannot be applied to the gravitational field. The reason being that the gravitational counterpart of the electric charge is the gravitational mass. So, according to the equivalence principle, the ratio between the ``gravitational charge'' and the inertial mass of the probe cannot be freely adjusted, being equal to one for any particle existing in Nature. Hence, the equivalence principle seems to put a serious restriction on the possibility to sharply measure the gravitational field.

This fact suggests that an unavoidable fundamental limit on the measurements accuracy exists, affecting, as a consequence, the fundamental structure of space-time. Furthermore, this could have important implications in the construction of QG, because, as is well known, QM imposes limitations on the simultaneous measurement of conjugate pairs of fields, but, following Bohr and Rosenfeld, no limitation on the accuracy of a measurement of one single field occurs. The Bronstein's argument suggests to consider the possibility that ordinary quantum mechanics could be inadequate to describe the quantum theory of a geometric field. In this sense, either a modification of the uncertainty relation comes out naturally from the theory, encoding this intrinsic limitation on the measurements of the gravitational field, or, more speculatively, a different quantum mechanics, which reduces to the ordinary one in the appropriate limit, should be sought to face the problem of QG.

\subsubsection{Space-time in classical and quantum mechanics}

In accordance to what said above, it is worth stressing once more that QG could radically change the present concepts of space and time. It could be possible, in fact, that new and unpredictable non-trivial outcomes of the theory oblige us to abandon any intuitive representation of space and time. In this respect, it is important to understand how the notion of space and time changes according to the postulates of different physical theories. This is the motivation of the following discussion, where the different notions of space and time are briefly described.

In classical mechanics space-time is an external fixed and flat structure. In particular, space is an Euclidean three-dimensional stage for the physical phenomena, $\mathds{R}^3$, while time is represented by the oriented one dimensional real axis, $\mathds{R}$. Space and time can be measured with an infinite precision in classical mechanics. The common idea of space-time acquires a proper operative meaning as soon as we imagine the existence of a dense array of ideal synchronized clocks beating the flow of time, while perfect rods sharply measure the distances among clocks, giving in this way an empirical meaning to the points of space-time.

Quantum mechanics is characterized by a novelty with respect to its classical counterpart; we are referring to the uncertainty principle. This principle establishes the impossibility to simultaneously measure with an infinite precision a pairs of conjugate variables. Spatial coordinates are the conjugate variables to the momenta along the same axis and can be measured with an infinite precision at the price of losing any information on the velocities. Therefore, a subtle question about the evolution of the reference system arises. Indeed, as soon as we have realized that the role of time in QM is identical to the one it plays in classical mechanics,\footnote{With the difference that in some extrapolations of the theory, an uncertainty relation between time and energy comes out. However, since time is not an observable in quantum mechanics but only an evolution parameter, the time-energy uncertainty relation must be interpreted differently with respect the coordinates-momenta one, which, instead, is a consequence of the fact that coordinates and momenta are conjugate observables in QM.} we can imagine to construct the same array of dense ideal clocks to give an operative meaning to the space-time points. But, if the clocks had a finite mass, one should worry about their evolution, in fact, if we measure their positions we lose information on their evolution. This fact suggests the possibility that the space and time of quantum mechanics acquire an operative meaning only in the limit of infinite mass clocks. This does not create any embarrassment, because quantum mechanics, as stressed at the very beginning of this section, completely ignores gravity, then its logical consistency can safely rely on the idealization of a physical reference frame constituted by infinitely heavy particles.  

The study of the space-time of quantum field theory introduces even more interesting features, which deserve to be deepened. As clarified above, the background of quantum mechanics is a classical space-time; namely, we can, with a limiting procedure, measure sharply the position of a particle in a space-time, pictorially represented by a dense array of infinite mass clocks. In particular, in order to localize a finite mass particle, we consider an interaction between a probe and the particle. The accuracy of the measure is proportional to the inverse of the energy carried by the probe, namely, if the probe carries an energy, say, $1/\Delta X$, we can localize a particle interacting with the probe with an accuracy of $\Delta X$. The reason being that the probe results to be confined to a region of space-time of size $\Delta X$. This means that the sharp localization implies the injection into the physical system of a greater and greater (in principle infinite) energy. Even though this procedure does not create any problem in classical quantum mechanics, as soon as we consider a ``relativistic quantum mechanics'' a subtle issue crops up. In fact, even though the 4-dimensional picture does not modify the fundamental structure of space-time, still pictorially described by a dense array of extremely massive clocks, it introduces the equivalence between mass and energy, which creates a shortcoming in the above described measuring procedure. Indeed, as soon as the energy carried by the probe becomes higher than the rest mass of the particle being measured, many copies of the original particle are produced as a direct effect of the position measurement (injection of energy into the system).\footnote{From another perspective this is the reason why in place of relativistic quantum mechanics, a quantum field theory is required to describe such phenomena, we need, in fact, a theoretical framework which does not require to fix the number of particles of the system.} 

In order to avoid any misunderstanding, we want to stress that the sharp localization of an infinite mass particle is still possible, because in order to produce copies of the same measured particle an infinite energy is necessary, so, as in ordinary quantum mechanics, no problem exists in the localization of space-time points.

Concluding, we can say that even though the space-time background of quantum field theory is the same ideal dense array of clocks that we have already introduced in ordinary quantum mechanics, in quantum field theory the accuracy of a measurement of the position of a particle is however limited, the limit being imposed by the measuring procedure itself. In other words, the coexistence of the Heisenberg uncertainty principle with Special Relativity prevents from obtaining a sharp measurements of the position of a finite mass particle, even though the concept of classical space-time is preserved. However, this appear to be no longer possible when gravity is present. On one side, in fact, quantum mechanics obliges to consider an infinite mass point particle in order to give sense to a sharp localization procedure, but, on the other side, this is incompatible with GR for obvious reasons and the entire sharp localization procedure falls into troubles.

\subsubsection{Space-time in General Relativity}\label{GRST}

General Relativity is the theory describing the dynamics of space-time. In this sense, even the shortest discussion about GR requires to deepen into the basic principles and the main ideas which led Einstein to the formulation of his geometrical theory of the gravitational field. We are not referring to the construction of the Einstein dynamical equations, but to the more subtle and complicate ``struggle with the meaning of the coordinates.''

Generally speaking, in constructing the field equations of GR one has a number of hints, as, for example, the fact that the static limit of the field equations must be the Newton law. More suggestive is the fact that the mass of a particle is the source of the Newtonian gravitational field: In fact, considering that the mass is a form of energy, as Einstein himself clarified, it is quite reasonable that the energy-momentum tensor turns out to be the source of the relativistic field equations. Finally, the concept that no privileged reference systems exist suggests that the equations must be covariant under a class of general coordinates transformations. Einstein, some years before the publication of his most famous paper, appeared in 1915, learned that the only possible combination of second order partial derivatives of the gravitational field (space-time metric), transforming covariantly under general coordinates transformations, is the Riemann tensor. It became soon clear to his mind how the equations of motion of GR had to look like. But the subtlest and most striking aspect of GR, which Einstein dedicated himself to for a long period, regarded the philosophical content of the theory: The real novelty introduced by GR is that the coordinates have no physical meaning, independently from the value of the physical field and from the trajectories of the physical particles.

The Equations of GR are, in fact, generally covariant, or, in other words, if $e^{\phantom1a}_{\mu}(x)$ is a solution of the field equations, then, given the general coordinates transformation $y=y\left(x\right)$, also $e^{\prime\phantom1a}_{\nu}\left(y\right)$, 
\begin{equation}\label{covariant transformation}
e^{\prime\phantom1a}_{\nu}\left(y\left(x\right)\right)\frac{\partial y^{\nu}\left(x\right)}{\partial x^{\mu}}=e^{\phantom1a}_{\mu}(x)\,,
\end{equation}
is a solution of the field equations. Essentially, this means that the physical laws are the same in all the reference frames, namely in all coordinates systems.

Now, in order to understand the meaning of general covariance, which will be useful for clarifying the structure of space-time of GR, we consider a region of space-time, say $\mathcal{U}$, containing the event $P$, and, assigned in $\mathcal{U}$ the system of coordinates $\mathcal{X}$, let us indicate with $x_P$ its coordinates. Let $e^{\phantom1a}_{\mu}\left(x\right)$ be a solution of the generally covariant field equations and assume that
\begin{equation}
\left.R\right|_{P}=R\left(x_{P}\right)=0\,,
\end{equation}
where $R\left(x\right)$ is the Ricci scalar. Suppose now that we decide to change our system of coordinates in the region $\mathcal{U}$. Specifically, be $\mathcal{Y}$ the new system of coordinates and $y=F\left(x\right)$ the transformation law from one system to the other. Thus $e^{\prime\phantom1a}_{\nu}\left(y\left(x\right)\right)$, obtained from $e^{\phantom1a}_{\mu}\left(x\right)$ via the relation (\ref{covariant transformation}), is a solution of the equations of motion too. In other words, $e^{\prime\phantom1a}_{\nu}$ describes the same gravitational field as $e^{\phantom1a}_{\mu}$, but in the $\mathcal{Y}$ system of coordinates. Moreover, the Ricci scalar still vanishes around the point $P$:
\begin{equation}
\left.R^{\prime}\right|_{P}=R^{\prime}\left(y_{P}\right)=R\left(F^{-1}\left(y_{P}\right)\right)=R\left(x_{P}\right)=0\,.
\end{equation}
Let us now proceed considering the new gravitational field $E^{\phantom1a}_{\nu}$ defined as follows
\begin{equation}
E^{\phantom1a}_{\nu}\left(x\right)=e^{\prime\phantom1a}_{\nu}\left(x\right)\,,
\end{equation}
namely as the primed field in the old system of coordinates $\mathcal{X}$. It is worth stressing that the gravitational field described by $E$ is different from the one described by $e$, in particular, although the Ricci scalar constructed by $e$ is zero around the point $P$, namely $R\left(x_{P}\right)=0$, we cannot draw the same conclusion for the Ricci scalar of the field $E$. In fact we have:
\begin{equation}
\left.\mathcal{R}\right|_{P}=\mathcal{R}\left(x_{P}\right)=R^{\prime}\left(x_{P}\right)=R\left(F^{-1}\left(x_{P}\right)\right)\,.
\end{equation}
Namely, the Ricci scalar associated to the field $E$ in the space-time point $P$ is given by the Ricci scalar of $e$ calculated in the point $Q=F^{-1}\left(x_{P}\right)$ and for no any reason it must be zero.

It results that, if the gravitational field $e$ is a solution of the equations of motion, $E$ is a solution too. The reason is that the field $E$ is described in the system of coordinates $\mathcal{X}$ by the same function describing the gravitational field $e$ in the system of coordinates $\mathcal{Y}$; since the field equations do not change under a coordinates transformation, if $e$ is a solution, so is $E$.

This is in essence the content of the so called \emph{Einstein's hole argument}. The conclusion of the above described argument is that the generally covariant field equations are not deterministic, because even though $e$ and $E$ are both solutions of the same field equations, they do not determine the physics at the same space-time point $P$. For example, while the Ricci scalar associated to the gravitational field $e$ is zero around $P$, the same scalar calculated using the field $E$ is in general different from zero around the same space-time point. Since we know that classical physics is deterministic, we are at a crossroads: either the field equations cannot be generally covariant, or fixing the space-time event $P$ has no physical meaning.

Einstein had the courage to take the right road: He, in fact, understood that there is no physical meaning in fixing a particular space-time point on a generally covariant space-time. In this respect, let us consider a solution of the Einstein equations $e$ and two particles moving in this particular gravitational field. The motion of the particles is described by their respective world lines, $x_{1}\left(\tau\right)$ and $x_{2}\left(\sigma\right)$, which are determined by the gravitational field. Suppose, without any loss of generality, that the world lines of the two particles intersect at the space-time event $P$. Now consider the gravitational field $E=\phi_{*}e$, where $\phi:M\rightarrow M$ is a diffeomorphism; obviously the particles world lines $x_{1}\left(\tau\right)$ and $x_{2}\left(\sigma\right)$ are no longer solutions of the particles equations of motion in the new gravitational field. In fact, the new particles world lines, determined by the gravitational field $E$, can be easily calculated once the world lines associated with the gravitational field $e$ are known, they are:
\begin{equation}
X_{1}\left(\tau\right)=\left[\phi x_{1}\right]\left(\tau\right)\qquad\text{and}\qquad X_{2}\left(\sigma\right)=\left[\phi x_{2}\right]\left(\sigma\right)\,.
\end{equation}
In other words, a diffeomorphism, acting both on the gravitational field and on particles world lines, sends solutions to solutions. Furthermore, as a consequence of the active diffeomorphism, the particles do not intersect anymore in $P$, but in $Q=\phi\left(P\right)$. So, the fixed point $P$ loses its absolute meaning and the right physical entity is the point determined by the intersection of the world lines of the particles. In this sense the theory does not predict the value of the gravitational field around the space-time point $P$, rather around the point determined by the intersection of two world lines. Therefore, the issue contained in the hole argument is solved, the theory is deterministic because it predicts the same value of the gravitational field around the same \emph{physical} (namely determined via a diffeomorphisms invariant construction) space-time point. The characteristics of the gravitational field $e$ around the intersection of the world lines $x_{1}\left(\tau\right)$ and $x_{2}\left(\sigma\right)$, as, for example, the flatness of space-time around this point, are exactly mimed by the gravitational field $E$ around the intersection of the world lines $X_{1}\left(\tau\right)$ and $X_{2}\left(\sigma\right)$. This means that the theory has a gauge invariance in the sense of Dirac: Different solutions, correlated by gauge transformations, represent the same physical situations (see § \ref{GSC}). The gauge group is the group of diffeomorphisms, which reflects the fact that the localization of an event is not an absolute procedure, but is related to the particles and fields themselves. We will see in paragraph \ref{relationalism} that the diffeomorphisms invariance has striking consequences on the theory, which, indeed, can be considered as a partly relational theory.

Concluding, the space-time of GR is a classical structure, but it is not absolute as in classical and quantum mechanics. In other words, it is certainly possible to localize sharply a point on a generally covariant space-time, but the localization procedure requires the presence of particles and fields. In particular the diffeomorphisms invariant procedure for the localization of an event is based on the possibility to sharply recognize the points of intersection between the world lines of particles.

\subsubsection{Space-time in Quantum Gravity}

What we are going to say here must be intended by the reader as an attempt to loosely explain why the idea that QG implies an absolute limit on the localization of events has a so long tradition in the quantum gravity community. We cannot give a complete and consistent description of quantum space-time simply because the existing theories give us only a pictorial idea of how it should be; so, here, I do not pretend neither to provide new insights into the quantum gravity problem, nor to give a new pictorial description of quantum space-time. Nevertheless, I think that it could be useful and interesting to investigate the operative meaning of the localization of events in a theory that should incorporate both the general covariance and the quantum uncertainty principle.

In this respect, we want to summarize the conclusion obtained before: 
\begin{enumerate}
	\item in order to satisfy the principle of general covariance, the localization of space-time events must be realized via a diffeomorphisms invariant procedure;
	\item the quantum uncertainty does not allow to localize sharply a finite mass particle, because this would require the injection into the system of a so large quantity of energy that it is impossible to neglect the creation of copies of the analyzed particle.
\end{enumerate}
Let us now construct a physical diffeomorphisms invariant procedure of localization. In order to localize an event, we need at least two interacting particles, moving along their world lines; specifically, we consider a massive particle and a probe able to interact with the particle itself. We know from what stated above that the larger is the mass of the particle the greater is the accuracy of the localization: The relation between the energy of the probe and the uncertainty in the localization of the particle is $E=\left(\Delta x\right)^{-1}$. The physical explanation of this formula is simple. In order to determine the position of an object with a given accuracy $\Delta x$ we have to use a probe, represented by a massless particle interacting with the system under study, localized in a region at least comparable with the accuracy we require for the experiment. The localization of the probe is proportional to its Compton wavelength, so the accuracy of the experiment is proportional to the inverse of the energy of the probe. It is well known that to reveal smaller and smaller structures in particles physics we have to use higher and higher energy test particles. In this specific context, we have to consider also the presence of the gravitational field, which fixes a limit on the energy of the system. Indeed, it is necessary that the Compton wavelength of the system is larger than its Schwarzschild radius, this means that the energy during the interaction must be smaller than the Planck mass. We consider in this pictorial context the ideal scattering process, which consists in a collision between a massless probe and a particle of mass $M$, the collision lasts for a time $\Delta t$. During this period of time the system is considered frozen. We can compute the typical time of the interaction $\Delta t$ by a very simple argument: It is the time necessary to exchange information between the probe and the particle, so it is at most equal to the distance $\Delta x$ traveled by the signal during the interval $\Delta t$. The energy contained in the gravitational field during the interaction is $V\approx\frac{\ell_{Pl}^2 EM}{\Delta t}$, where $E$ is the energy of the probe. During the collision the total energy of the system must be smaller than the Planck mass $M_{Pl}=\ell^{-1}_{Pl}$, thus we can write the following relation
\begin{equation}
\frac{\ell_{Pl}^2 EM}{\Delta t}\,<M_{Pl}\,=\ell^{\,-1}_{Pl}\,.
\end{equation}
Now, remembering that the spatial displacement between the probe and the particle is at most equal to the accuracy of the localization and that the best we can do is to increase the energy of the probe up to the mass of the particle (in order to avoid the creation of copies of the particle under study), by using the relation above we obtain:
\begin{equation}
\frac{\ell_{Pl}^2}{\left(\Delta x\right)^{3}}<M_{Pl}\,.
\end{equation}
From the expression above we deduce that
\begin{equation}
\Delta x\,>\,\ell_{Pl}\,,
\end{equation}
namely the best accuracy we can obtain in determining the position of a particle is larger than the Planck length.

So space-time in Quantum Gravity is not classical. In other words, taking into account both the general covariance and the quantum uncertainty principle, we cannot sharply localize an event. An intrinsic limit given by the Planck length appears in the accuracy of a localization procedure. As a consequence, all the intuitive concepts about space and time must be abandoned in a Quantum Gravity theory, the resulting space-time structure is genuinely non classical \cite{Rov2004,Ame-Cam2006}. More radically, we could say that in Quantum Gravity space-time does not exist at all, in its place we have a fuzzy quantum structure, which fits well with the pictorial representation Wheeler gave many years ago: Quantum space-time should appear like a foam. The quantum foam is a state of the gravitational field, which, at that stage, cannot be identified with a measurable structure in a proper sense. In fact, what is generally intended as space-time fits better with its classical actualization. The quantum foam, instead, cannot be considered as a space-time, for example the existence of a minimum length suggests a fundamental discrete (non-continuous or, likely, non-commutative) structure. In order to reintroduce the common concept of space-time in Quantum Gravity, namely as a measurement of the time elapsing between two events and the spatial distances separating two disjoint events, we have to give it a different status. We mean, considering it as an ``actualization'', rather than an ``idealization''. One way to ``actualize'' it is through a relational procedure, which is the next argument we want to discuss.

\subsubsection{Relational versus Absolute space-time}\label{relationalism}

The discussion about relational or absolute space-time could appear as a pure philosophical one, in stead it regards profoundly Physics and the debate is still open and stimulating (see \cite{Barbo09,Rov09,Mar09} and references therein). This debate is a long standing one and can be traced back to the publication of the Newton's \emph{Principia Mathematica} in 1687. In his book Newton proposed an \emph{absolute} notion of space-time, according to which the geometry of space-time provides a fixed, eternal and immutable background structure on which particles move. In striking contrast with the ideas of Descartes, Leibniz, Huygens, and others who, in stead, espoused the so-called \emph{relational} philosophy, according to which space-time has to be intended as a set of ``relations'' among real objects and events \cite{Smo2005} (detailed discussions on this argument can be found in \cite{Barbo1989,Barbo2000,Sta1989}). The Newton's absolute view won against relationalism, supported by the great empirical success Newtonian mechanics had and, for a long period, no doubts were raised on which notion of space-time Physics should be based on. Nevertheless, the general covariance principle introduced by Einstein, mathematically expressed via the invariance under diffeomorphisms, seems to destroy the Newtonian absolute description of space-time. Essentially, in the Einstein's theory, fixed space-time is replaced by a dynamical structure on which the events are no longer points with assigned coordinates, but interactions between physical particles: in other words relations. So, the relational point of view deserves to be taken in serious consideration as directly suggested by the commonly accepted theory describing space-time and gravitation. 

The question we should answer is: to what extent GR is a relational theory. Obviously, to answer to this question we have to capture the differences between Newton and Einstein mechanics, as we are going to do, taking into account the ``space-time limitations'' we have.

Newton's mechanics is based on the existence and physical definition of inertial reference frames. They play a central role in all the classical theories. In particular, they allow to distinguish between accelerating and uniformly moving point particles. In fact, once fixed an inertial reference system, the distinction between what is accelerating and what is moving uniformly is a property of the geometry of the absolute space-time (background), which is completely independent of the configuration of the matter. In other words, in Newtonian physics there is a clear and absolute distinction between inertial and non-inertial motion. Furthermore, this distinction does not depend on something internal to the physical system, but only on the external geometrical properties of space-time. Physically, we can distinguish between accelerated or uniformly moving particles, by looking at the geometry of the reference frame glued to a generic particle: The presence of non-inertial forces allows us to make the distinction. This viewpoint was challenged by Mach, who proposed to eliminate absolute space-time as a cause of distinction between accelerating and non-accelerating motion, replacing it with a dynamical procedure. According to Mach, the distinction between accelerating and non-accelerating motion should be determined via the relations between all the structures which compose the entire Universe.

Mach's idea strongly influenced Einstein, who realized that acceleration should be determined with respect to a reference frame, dynamically determined by the configuration of the whole system. As a consequence, it does not exist any privileged reference frame in the Universe, and physical laws must be equivalent in all the frames. Since local reference frames are strictly connected with the geometry of space-time, then space-time itself becomes a dynamical field, no more fixed and immutable, but interacting with matter and affected by the matter content of the system. The equivalence of any reference frame, mathematically expressed by the invariance of the theory under general coordinates transformations, suggested the name General Relativity for such a theory of space-time. It is worth noting that, even though Einstein was surely influenced by Mach, we cannot naively conclude that GR is a Machian theory \cite{Rov2004}. Nevertheless, we can say that GR is, by construction, a partly relational theory as we are going to motivate.

Generally speaking, a physical theory postulates that a physical system is made up of a large collection of mutually interacting elements. The form of the physical theory is based on the properties and the specific interactions of these physical entities. The physical properties of the elements of a system, in an absolute theory, are referred to a fixed structure as, e.g., the Newtonian space-time, while their mutual interactions determine their evolution with respect to the absolute time $t$. Put differently, space-time plays the role of background to which the dynamics is referred. The same role is played by a regular lattice, often used in the framework of particle physics. In this case, the physical entities (particles or fields) are confined on the nodes of the lattice, which is fixed \emph{a priori} and not influenced by the presence of matter. 

On the contrary, the main assumption of a relational theory is that no any background exists at all; so the question is: to which structure is the dynamics referred? Looking for an answer, let us firstly digress on some aspects of the relational point of view. The relational view presumes that the fundamental properties of the elementary entities consist entirely in relations between the elementary entities themselves. Dynamics concerns with the changes of these relations. A good pictorial description of a relational theory can be obtained by considering a graph \cite{Rov2004}, characterized by some nodes representing the entities and their properties, and different classes of connections between the nodes featuring mutual interactions between adjacent entities. The state of the system is determined by the structure of the links between the nodes, whereas the dynamics modifies the structure of the connections (relations) between different nodes. It is important not to confuse the pictorial description of a relational graph with the common image everyone has of a regular lattice. In this respect, it is worth noting that the pictorial description of a relational system is completely abstract, while a regular lattice describes a precise physical situation, where a continuous space-time is substituted by a fixed discrete structure to which we refer the fields dynamics.

In relational theories time loses its usual meaning. Evolution is, in general, incorporated in the modification of the relations between the physical entities and, since time is the parameter of the evolution itself, it acquires a relational meaning as well. The concept of relational time has a great importance in QG as a large number of papers highlights (see \cite{Barbo09,Rov09,Mar09,Barbo1989,Barbo2000,Sta1989,Kuc1991,Kuc1992-1,Kuc1992-2,KucTor1991,Rov1991-2,Rov1991-3,MerMon2003,MerMon2003-2,Smo2005,Thi2006} and references therein).

Having described what we mean by relationalism, let us now focus the attention on GR. GR is a complicated theory describing all the manifestations of the gravitational interaction. In particular, it describes gravitationally dominated subsystems of the Universe as, for example, black holes or gravitationally bound systems. In some models describing subsystems of the Universe, a number of conditions on the fields and metric are imposed at the boundaries. In this case, the question whether or not GR is a relational theory is not interesting at all \cite{Smo2005}, because imposing the boundary conditions is equivalent to introducing a background. Moreover, the existence of a region of space-time which is external to the system we are modeling with our theory implies that the theory is not ``fundamental.'' But GR is widely believed to be the theory describing the whole Universe as well. In other words, GR is considered the best candidate for a cosmological theory.\footnote{In order to avoid any misunderstanding, I stress that this is absolutely not in contrast with the fact that the same theory, with suitable boundary or asymptotic conditions, can describe, as well, subsystems of the Universe. In general, it is reasonable to expect that some sort of modification can appear in the cosmological dynamical equations, as for example a non-vanishing cosmological constant. Nevertheless, the foundations of the theory remains unaffected by the specific dynamics, them being, in fact, related to the conceptual structure of the theory, which are pretty general, rather than to the peculiar dynamics it generates.} Hence, apart from the specific dynamics, if we assume that GR is, in fact, a cosmological theory, the question whether or not it provides a relational description of the dynamics of the Universe acquires a profound meaning.

GR contains a lot of structures, which are fixed \emph{a priori}, they are: dimensions, signature, topology, and differential structure. All of them belong to what is intended as ``background'', in fact, they can be varied from model to model, but they are fixed and are not subjected to dynamical laws. More precisely, they describe the manifold $\mathcal{M}$; whereas, the metric $g_{\mu\nu}$ and tensor fields $T^{(a)}$ are the dynamical entities of the theory. A space-time corresponds to a determination of the manifold, metric and fields, namely $\left(\mathcal{M},g_{\mu\nu};T^{(a)}\right)$. But, in order to define a physical space-time, we have to take into account the gauge freedom of the theory, which, as described in paragraph \ref{GRST}, is encoded in the invariance under the group of diffeomorphisms. Therefore, we define a physical space-time as an equivalence class of manifolds, metrics, and fields under the action of the group $Diff(\mathcal{M})$. We denote this equivalence class as $\left\{\mathcal{M},g_{\mu\nu};T^{(a)}\right\}$. Now, as already mentioned, the points and open sets of the manifold $\mathcal{M}$ are not preserved under the action of the diffeomorphisms group. Diffeomorphisms send points to other points, in this sense the information encoded in the physical space-time is a system of relations between the fields, rather than a collection of the values fields take in the generic points of the manifold. Then, apart from the specification of topology, signature, dimensions and differential structure, GR is a relational physical theory.

It remains to answer the question posed above. Rephrasing it, we can ask: which is the physical entity replacing the Newton's absolute space-time? The answer is now simple: it is the gravitational field! In other words, it is the gravitational field that tells objects if they are accelerating or not. This is the profound difference between Newton's and Einstein's mechanics. In the Newton's mechanics the whole dynamics is referred to an absolute structure external to the system, while in the Einstein's mechanics the dynamics is referred to the dynamical gravitational field, carrying out the relational idea.

\subsection{Possible phenomenological implications of QG}\label{phenomenology}

Generally speaking, a quantum theory of gravity is expected to describe the dynamics of the quantum \emph{space-time foam} \cite{Whe1963,Haw1978}. The specific description of the foam dynamics depends on the theory describing the quantum effects of gravity, but its phenomenological implications are shared by many different approaches to QG. In particular, a possible candidate for a quantum gravity effect due to the foamy structure of space-time is an energy dependent electromagnetic dispersion relation \emph{in vacuo}. Specifically, the modified dispersion relation is supposed to be of the following form 
\begin{equation}\label{dispersion relation}
p^2=E^2\left(1+f\left(\frac{E}{E_{QG}}\right)\right)\,,
\end{equation}
where $E_{QG}$ is an effective quantum gravity energy scale, naturally identifiable with the Planck energy. Now, let us suppose for simplicity that the Hamilton equations of motion are approximately valid in the present scenario, then the velocity of the particle follows from equation (\ref{dispersion relation}) and, for energies much lesser than the Planck scale, it turns out to be:
\begin{equation}\label{deformation velocity}
v=\frac{\partial E}{\partial p}\approx1-\xi\frac{E}{E_{QG}}\,,
\end{equation}
where $\xi$ is a positive or negative factor, which depends on the particular framework.

Before deepening into some phenomenological aspects, let us spend some words on how such modified dispersion relations have independently emerged in different QG approaches. It is commonly believed that QG effects interest too high energies to be experimented, until, some years ago, the first suggestions that quantum-gravitational fluctuations might modify particles propagation in an observable way began to appear \cite{EllHagNan1984,EllMavNan1992}. As a consequence, different classes of physical phenomena began to be studied. In particular, the effect that a modification in the particles propagation could have on the neutral kaon system \cite{EllLopMav1996,HuePes1995} was tested in laboratory experiments, fixing a lower limits on some parameters analogous to $E_{QG}$ \cite{Adl1995}. Other examples of quantum gravitational effects more related to String Theory and LQG can be found in \cite{AntBacEll1991} (see also \cite{KosSam1989} for another string motivated deformation) and \cite{GamPul99,AlfMorUrr02}. Deformed dispersion relations, consistent with the formula above, arise also in other approaches as the ``$\kappa$'' quantum deformations of the Poincar\'e symmetry \cite{LukNowRue1995,Ame-Cam1997} or quantization of point particles on a discrete space-times \cite{Hoo1996}.

Let us now deepen into the problem of finding a physical system which can be used to test such a deformed dispersion relation. According to (\ref{deformation velocity}), the deviation from the ordinary velocity of the photon is extremely small for practical purpose, nevertheless it could give a sensible effect if photons of different energies travel for a very long distance before being detected. In particular photons of different energies emitted at the same time acquire a relative time delay over a distance $L$ of the order
\begin{equation}
\Delta t\approx\xi L\frac{\Delta E}{E_{QG}}\,.
\end{equation}
So, wider the spectrum of the emitted photons and larger the distance traveled, greater is the time-delay effect. In this respect, the best candidates to observe such an effect are Gamma Ray Bursts (GRB).

GRBs are, in fact, explosive events at cosmological distances. The typical spectrum of emission is in the range $0.1 - 100$ {\rm MeV}, but it can extend up to the {\rm TeV} scale. Moreover, a time structure of the order of the millisecond is typically observed in the light curves. It should now be clear why GRB are good candidates to study the effects of deformation in the dispersion relation. By a simple calculation, in fact, it turns out that a GRB with a time structure of the order of the milliseconds, emitting photons of energy of the order of few {\rm MeV}s and exploded at a distance of $\approx10^{26}\,{\rm m}\approx\,10^{10}\,{\rm ly}$ from the Earth, could test the QG structure up to $E_{QG}\approx10^{19}\,{\rm GeV}$. Sensible sensitivities can be already obtainable from the existing GRB's data and we address the interested readers to  \cite{Ame-CamEllMav1998,Pir2004} and in particular to \cite{JacPir2006,MarPir2006}. For completeness, we remark that a small mass for the photon could produce the same time-delay effects in the arrival time of photons of different energies. But other existing experimental data fix a limit to the mass of the photon and, consequently, to such an effect well below the expected time-delay associated to QG effects.

\section{Preliminaries}\label{sec3}
This Section is dedicated to describe some fundamental arguments considered as preliminary, in the sense that they represent the foundations which the forthcoming discussion is based on.
They are pretty general and unrelated arguments, very well described in many books, in which many more details can be found. Here I collected the main results and definitions, according to my own experience as student, inspired by my own notes taken during courses and referring to my favorite textbooks. So the description is far from being complete, so I exhort the interested readers to refer to the cited Literature for a more complete treatment.

As I said in the Preface, the focus of this paper is on the canonical formulation of GR. In this respect, it is important to realize that in order to canonically formulate any theory, we have to clarify the causal structure of space-time. This is a trivial task when we treat gauge theories on a flat Minkowski fixed background, but everything becomes more involved when the dynamics of the theory directly concerns the geometrical structure of the space-time itself as in GR. In fact, the identification of the gravitational field with the geometry of space-time implies that some restrictions have to be imposed on its global (causal) structure in order to canonically formulate the theory. In particular, we have to clearly understand under which conditions it is possible to ``split'' space-time and describe the dynamics of the gravitational field as the time evolution of a geometrical spatial quantity.\footnote{It is worth noting that the word time here does not refer to the quantity measured by using clocks, which would imply that the space-time metric has already been completely determined.} It is important to understand that this is a necessary step to consistently define a canonical theory; we recall that, in fact, in classical mechanics, the Hamiltonian can be considered as the momentum conjugate to the time coordinate. This should give an idea of how much complicated the situation is in GR, since the invariance under diffeomorphisms prevents from determining a preferred time coordinate and, in general, it does not exist any global system of coordinates.  

For this reason, we consider the study of the causal structure of space-time as the natural starting point for the canonical formulation of gravity. In particular, the scope of the first part of this Section is the introduction of the Geroch theorem, which clarifies under which hypotheses a global time function can be assigned on a generic space-time. In this sense, the Geroch theorem restricts the class of space-times whose dynamics can be canonically formulated, generating a question about the resulting canonical quantum theory: Is canonical quantum gravity applicable to a restricted class of space-times as well, or the classical conditions can be relaxed in the quantum theory? Up to my understanding, this question cannot be rigorously answered until a complete canonical quantum gravity theory has been formulated; nevertheless, since the canonical quantization procedure is, generally speaking, a mathematical tool to face the problem of quantization, once the procedure has been rigorously completed, it is reasonable to expect that the resulting quantum theory will not be affected by purely classical restrictions. These are naturally relaxed by the quantization itself in a very precise and suggestive sense: In a complete theory of QG there is no room for space-time.

The discussion of the causal structure of space-time and the subsequent canonical formulation of gravity, open the way to another interesting argument, namely the initial value formulation of a theory with gauge symmetries and, in particular, of GR. It should be clear that, in fact, once GR has been rewritten as describing the ``time'' evolution of a 3-dimensional geometry, it is expectable that an initial value problem can be consistently formulated, by assigning a complete set of initial data on the initial spatial hypersurface. For physical theories formulated on a fixed Minkowski space-time, the task of the initial value problem is to extract the (unique) evolution of the system starting from a complete set of initial data, which are generally referred to the external fixed background. In this sense, the case of GR is much more involved. We cannot, in fact, refer the evolution of the initial data to a preferred background, rather, once assigned a particular starting configuration for the metric and its time derivatives, the theory describes the evolution of the background itself. From another perspective, we can say that the gauge symmetries of GR, mathematically described by the group of 4-diffs, complicates the formulation of a well posed initial value problem, not only from a mathematical perspective, but also conceptually.
Nevertheless, a suitably adapted procedure can be applied to GR in order to assign a well posed initial value problem and extract a unique evolution from the Einstein equations, even though with some limitations.

In order to clearly describe these arguments, we start by briefly recalling some elements of group theory, useful to introduce gauge theories, which will be considered as a useful example in what follows. In particular, to make the general description of the canonical formulation of a gauge theory more concrete, we will refer to the simple case of the electromagnetic field, which is a gauge theory of the abelian group $U(1)$. 

Finally, once the causal structure of space-time has been clarified and the generalities about the electromagnetic theory described, we face the problem of the initial value formulation giving a brief account of the main theorems used to study the problem. Also in facing this argument the electromagnetic theory will be useful, providing a simple, but non-trivial example to show how gauge symmetries enters in the formulation of an initial value problem for a classical system.

\subsection{Causal Structure of space-time}\label{csost}

The causal structure of flat Minkowski space-time is very simple and intuitive: Once a limit on the propagation of a signal is fixed, we can associate to any event $p$ in space-time a light cone. The future is represented by a half cone, while the past is represented by the other half. The events contained in the future half of the light cone can be reached by a matter particle leaving from $p$, all these events are generally referred as \emph{chronological future} of $p$. More generally, all the events lying in the interior of the future light cone together with those on the cone itself represent the \emph{causal future}, physically representing all the events which can be, in principle, influenced by a signal emitted from $p$.

The causal structure on a generic manifold $M$ is only locally similar to that of flat space, globally, in fact, the situation can be much more complicated. To give a complete and detailed account of the problematics correlated to the study of the causal structure of a generic space-time would require much time and space, so here we limit to give a brief sketch of this argument. In fact, non-trivial topologies or space-time singularities, in general, complicate enormously the treatment by introducing many subtleties, so we restrict the discussion by pointing out only those definitions, theorems, and lemmas we consider useful for what will be said below. For a more detailed description of the causal structure and the problem of singularities in GR, we address the reader to the book of Wald \cite{Wal1984} from which we extracted the main theorems of this Section and to the book of Hawking and Ellis \cite{HawEll1973} in which one can find complete demonstrations. 

Let me begin giving a simple, but important
\begin{definition}
The space-time $(M,g_{\mu\nu})$ is \emph{time orientable}, if $\forall p\,\in\,M$ it is possible to make a continuous designation of future and past.
\end{definition}
The simplicity of this definition stems from its intuitiveness, the importance, instead, is connected with the necessity to distinguish a particular class of space-times: in what follows we will always refer to time orientable space-times. It is easy to understand that, in general, a non-simply connected space-time cannot be time orientable; from the physical viewpoint in a non-time orientable space-time we cannot consistently distinguish between going ``forward in time'' or ``backward in time''. Time orientable space-times satisfy the property expressed by the following
\begin{lemma} 
Let $(M,g_{\mu\nu})$ a time orientable space-time, then there exists a non-unique smooth non-vanishing time-like vector field $t^{\mu}$ on $M$.
\end{lemma}
The proof of this Lemma is based on the paracompactness of $M$ and we address the reader to \cite{Wal1984} for a complete proof. It is interesting to note that the above Lemma suggests a more useful definition in order to designate time orientable space-times:
\begin{definition}
The space-time $(M,g_{\mu\nu})$ is \emph{time orientable}, if there exists on $M$ a time-like, continuous, non-vanishing vector field. 
\end{definition}
For the sake of completeness and self-consistency of the material presented in this work we recall also the following well known definitions:
\begin{definition}
A differentiable curve $\gamma(t)$ is said to be a \emph{future directed time-like curve}, if at each $p\,\in\,\gamma$ the tangent vector $t^{\mu}$ is time-like and future directed.
\end{definition}
It is simple to generalize this definition to \emph{future directed causal curve}, it is sufficient to replace the adjective ``time-like'' with ``causal''. The next definition automatically follows
\begin{definition}
The set of events that can be reached by a future directed time-like curve starting from $p$ represents the \emph{chronological future} of $p$, namely
\begin{equation}
\nonumber I^{+}(p)=\left\{q\,\in\,M:\exists\,\lambda(t)\ \text{(future directed time-like curve) with}\  \lambda(0)=p\,;\,\lambda(1)=q\right\}.
\end{equation} 
\end{definition}
Again the definition of \emph{causal future} is the same of that of chronological future except substituting the words ``future directed time-like curve'' with ``future directed causal curve''. Finally we remark that
\begin{definition}
For any subset $S\subset M$
  \begin{equation}
  \nonumber I^{+}(S)=\bigcup_{p\,\in\,S}\,I^{+}(p).
  \end{equation}
\end{definition}
It is worth noting that, even though in Minkowski space-time $I^{+}(p)$ consists of the interior of the future light cone, in an arbitrary space-time the situation could be more complicated, and the usual properties of flat spaces are in general not applicable to arbitrary space-times (it is simple to construct examples of pathological arbitrary space-times removing points from flat Minkowski space). However, at least locally, the same properties remain valid as stated by the following
\begin{theorem}
Let $(M,g_{\mu\nu})$ be an arbitrary space-time and let $p\,\in\,M$. Then there exists a convex normal neighborhood of $p$, i.e., an open set $U$ with $p\,\in\,U$ such that $\forall\,q\,,\,r\,\in\,U$ there exists a unique geodesic $\gamma$ connecting $q$ and $r$ and staying entirely within $U$. Furthermore for any such $U$, $I^{+}(p)|_{U}$ consists of all points reached by future directed time-like geodesics starting from $p$ and contained within $U$, where $I^{+}(p)|_{U}$ denotes the chronological future of $p$ in the space-time $(U,g_{\mu\nu})$. In addition $\overset{\cdot}{I}\,^{+}(p)|_{U}$ is generated by the future directed null geodesics in $U$ emanating from $p$.
\end{theorem}
The fact that all general relativistic space-times have locally the same qualitative causal structure as Minkowski space-time, does not exclude that globally remarkable differences can appear. As a consequence, a general space-time can be not \emph{causally well behaving}. In order to clarify this point consider, for instance, a space-time with topology $S^1\times \mathds{R}^3$, constructed identifying the $t=0$ and $t=1$ hyperplanes of Minkowski flat space. It is easy to realize that, in such a space-time, an observer would not have any difficulty in altering past events; in fact, the integral curves with tangent vector $t^{\mu}=(\partial/\partial t)^{\mu}$ will be closed and time-like. As a consequence, we have $\forall\,p\,\in\,M$, $I^{+}(p)=I^{+}(p)=M$ \cite{Wal1984}. Although the previous example could seem rather artificial, we stress that many general space-times with the property of allowing closed time-like curves exist, and they occur in much less artificial examples than that described above.

From a physical perspective space-times with non-trivial closed causal curves cannot be considered physically realistic, because an observer could alter past events. From a mathematical point of view, we have to ensure that time-like geodesics do not intersect themselves. But the problem is slightly more complicated than it could appear at a first glance, the reason being that we have also to consider physically unreasonable those space-times in which time-like geodesics come ``arbitrarily close'' to intersect themselves (without doing it). They could, in fact, violate the causality condition if a small perturbation of the metric occur. Then we characterize physical space-times as 
\begin{definition}
A space-time $(M,g_{\mu\nu})$ is said to be \emph{strongly causal} if $\forall\,p\,\in M$ and every neighborhood $U$ of $p$, there exists a neighborhood $V$ of $p$ contained in $U$, i.e., $V\subset U$ such that no causal curve intersects $V$ more than once.
\end{definition}
So, strongly causal space-time are characterized by the fact that causal curves cannot come arbitrarily close to themselves, but this is not sufficient to assure that one is not ``on the verge'' to violate physical causality. For this reason it is in general reasonable to give a stronger notion of causality as follows:
\begin{definition}
A space-time $(M,g_{\mu\nu})$ is said to be \emph{stably causal} if there exists a continuous non-vanishing time-like vector field $t^{\mu}$ such that the space time $(M,\widetilde{g}_{\mu\nu})$, where
\begin{equation}
\widetilde{g}_{\mu\nu}=g_{\mu\nu}-t_{\mu}t_{\nu}
\end{equation}
possesses no closed time-like curve.
\end{definition}
The definition of stable causality avoids that a strong causal space-time could violate causality by perturbing the metric. A perturbation of the metric could, in fact, ``open out'' the light cone so much that a causal curve can come arbitrarily close to itself. The light cones of space-time $(M,\widetilde{g}_{\mu\nu})$
is strictly larger than that of $(M,g_{\mu\nu})$, consequently if closed time-like curves do not exist for $(M,\widetilde{g}_{\mu\nu})$, surely they will not exist for $(M,g_{\mu\nu})$ too.

For our purposes, the content of the next theorem is particularly important.
\begin{theorem}
A space-time $(M,g_{\mu\nu})$ is stably causal if and only if there exists a differentiable function $f$ on $M$ such that $\nabla^{\mu}f$ is a past directed time-like vector field.
\end{theorem}
Or, in other words, a stably causal space-time is equivalent to the existence of a global time function. For brevity, we do not prove this theorem here, but we address the reader to \cite{Wal1984,HawEll1973} for the details of the proof. Furthermore it is important to quote the result contained in the following
\begin{corollary}
Stable causality implies strong causality.
\end{corollary}
This result is, in a certain sense, expectable and allows us to conclude that stable causality is the appropriate notion to be sure that a space-time is not going to violate causality, which is a crucial request for all physical reasonable space-times.

Above we gave the definition of causality, namely we studied the collection of events $I^{+}(S)$ which can be influenced by a set of events $S$, now we want to study the collection of events ``completely determined'' by a set of events $S$. Above all, we have to give the following two definitions:
\begin{definition}
A set $S$ is said \emph{achronal} if $I^{+}(S)\cap S\,=\,0$.
\end{definition}
\begin{definition}
Let $S$ be an achronal set, the \emph{future domain of dependence} of $S$, $D^{+}(S)$, is the collection of events $p$ such that every past inextendible causal curve passing through $p$ intersect $S$.
\end{definition}
It is worth noting that the following relations hold $S\subset D^{+}(S)$ and, $S$ being achronal, $D^{+}(S)\cap I^{-}(S)=0$. The physical importance of the future domain of dependence relies on the fact that, since no signal can travel faster than light, then any signal received in $p\,\in D^{+}(S)$ must have been registered on $S$, therefore giving suitable initial conditions on $S$, we should be able to predict what will happen in $P$. Note that if $p\,\in I^{+}(S)$, but $p\,\notin D^{+}(S)$ it is possible to reach $p$ with a signal not passing through $S$. In general, the full domain of dependence of an achronal set $S$ is defined as $D(S)=D^{+}(S)\cup D^{-}(S)$, physically representing the complete set of events which can be completely determined in future and past by fixing initial conditions on $S$.
\begin{definition}
A closed achronal set $\Sigma$ of $M$ such that $D(\Sigma)=M$ is said a \emph{Cauchy surface}.
\end{definition}
Now since the edge of an achronal set $S$ is the set of point $p\,\in S$ such that every open neighborhood $U$ of $p$ contains two point $q\,\in I^{+}(S)$ and $r\,\in I^{-}(S)$ and a time-like curve $\lambda(t)$ from $r$ to $q$ which does not intersect $S$; then it follows that the edge of a Cauchy surface is empty. Therefore, by the following
\begin{theorem}
Let $S$ be a closed achronal set with edge $(S)=0$, then $S$ is a three-dimensional, embedded, $C^{0}$ submanifold of $M$,
\end{theorem}
we can conclude that the Cauchy surface is a 3-dimensional embedded $C^{0}$ submanifold of $M$. Moreover since $\Sigma$ is achronal, it represents an ``instant of time'' of the Universe \cite{Wal1984}.
As a consequence, we give the following
\begin{definition}
\emph{(Wald \cite{Wal1984})}
A space-time $(M,g_{\mu\nu})$ which possesses a Cauchy surface is \emph{globally hyperbolic}.
\end{definition}
So, in a global hyperbolic space-time, we can predict or retrodict the entire evolution of the Universe by assigning suitable initial condition on the Cauchy surface $\Sigma$. Established the importance of a globally hyperbolic space-time, we want to give a criterion to recognize them among general space-times. In this respect, the first step is to introduce the so called Cauchy horizons:
\begin{definition} 
Let $S$ be an achronal set, the future Cauchy horizon of $S$, denoted by $H^{+}(S)$ is 
\begin{equation}
\nonumber H^{+}(S)=\overline{D^{+}(S)}-I^{-}\left[D^{+}(S)\right]
\end{equation}
\end{definition}
and let us immediately quote the following
\begin{proposition}
Being $H(S)=H^{+}(S)\cup H^{-}(S)$ and being $\overset{\bullet}{D}(S)$ the boundary of the future domain of dependence, we have
\begin{equation}
\nonumber H(S)=\overset{\bullet}{D}(S)\,.
\end{equation}
\end{proposition}
From the above proposition follows the following
\begin{corollary}
If $M$ is connected then a non-empty closed achronal set $\Sigma$ is a Cauchy surface if and only if $H(S)=0$.

\emph{Proof.} If $H(S)=0$ then by the proposition follows $\overset{\bullet}{D}(S)=0$, thus $\overline{D(S)}=\text{int}[D(S)]=D(S)$, so $D(S)$ is simultaneously closed and open, but the only sets both open and closed are the empty set and the entire set, so we conclude $D(S)=M$.
\end{corollary}
This corollary allows us to enunciate the following theorem, which represent a useful criterion to establish if a surface embedded in a manifold is a Cauchy surface:
\begin{theorem}
If $\Sigma$ is a closed achronal edgeless set, then $\Sigma$ is a Cauchy surface if and only if every inextendible null geodesic intersects $\Sigma$ and enters $I^{+}(S)$ and $I^{-}(S)$.
\end{theorem}
Now, it is easy to understand that, if a space-time is globally hyperbolic, then no closed time-like curves can exist in $M$. Indeed, either a closed time-like (or causal) curve never intersects the Cauchy surface $\Sigma$ violating global hyperbolicity, or it intersects $\Sigma$ twice violating achronality. This fact suggests that global hyperbolic space-times have a ``well causal behaviour'', as stated by the following
\begin{theorem}\label{geroch theorem}
\emph{(Geroch 1970 \cite{Ger1970}).} Let $(M,g_{\mu\nu})$ be a globally hyperbolic space-time. Then $(M,g_{\mu\nu})$ is stably causal. Furthermore, a global time function, $t$, can be chosen such that each surface of constant $t$ is a Cauchy surface. Thus $M$ can be foliated by Cauchy surfaces and the topology of $M$ is $\mathds{R}\times\Sigma$, where $\Sigma$ denotes any Cauchy surface. 
\end{theorem}
We refrain from giving the proof of this famous theorem addressing the interested reader to the original reference, nevertheless we want to stress its importance in view of the canonical approach and quantization of gravity. The Geroch's theorem, in fact, allows to recast the gravitational action in the canonical form operating a $3+1$ foliation of space-time, extracting a continuous function, which will play the role of evolution parameter. It will be clear that the dynamical degrees of freedom of the gravitational field are entirely contained in the geometry (metric modulo diffeomorphisms) of the 3-dimensional Cauchy surface. Pictorially, the evolution of the system can be described registering the changes of the 3-geometry going from one Cauchy surface to the next one, following the integral curve of the past directed time-like vector field $\nabla^{\mu}t$.

\subsection{Compact groups and gauge theories}

Let us now go on to describe some elements of group theory in view of the construction of gauge theories. Group theory, in fact, represents one of the main tool in the mathematical formalization of physical interactions and, even the briefest account of their mathematical properties would involve the interesting geometrical aspects of fiber bundles. These arguments, even though important to appreciate the general mathematical structure of this framework, would lead us away from the scope of this paper. So, I refrain from introduce the general framework of fiber bundles, giving only a brief account of the group properties we need to construct a gauge theory on a compact group. 

\subsubsection{Elements of group theory}\label{EGT}

Let us start from the following
\begin{definition}
Let $\mathcal{G}$ and $\mathcal{G}^{\prime}$ be groups, thus a map $f:\mathcal{G}\rightarrow\mathcal{G}^{\prime}$ is an homomorphism if $\forall\,g_{1}, g_{2}\in\mathcal{G}$ we have $f\left(g_{1}g_{2}\right)=f\left(g_{1}\right)f\left(g_{2}\right)$. A homomorphism $h:\mathcal{G}\rightarrow GL(V)$ is called a \emph{representation} of the group $\mathcal{G}$ and $V$ is called the \emph{representation space}. The representation is said to be of dimension $N$ if the representation space is $N$-dimensional.  
\end{definition}
A gauge theory can be constructed by using any compact group $G$, we recall that a group is said \emph{compact} if the parameters which describe the group take values in a compact set. In what follows, we will focus our attention on the compact group $SU(N)$, which is represented by the special group of the unitary $N\times N$ matrices. Its fundamental representation is $N$ dimensional, indeed, if $g\in\, SU(N)$ is an element of the group, we have that its representation $U=h\left(g\right)$ is a unitary $N\times N$ matrix, which acts on a $N$-dimensional complex space. Using the direct product, we can construct many other representations, which are in general reducible, usually called \emph{tensorial representation}. Consider, for example, the tensorial representation space $V\otimes V$; it is easy to imagine that the elements of the group will act on the representation space as follows $U_{i k}U_{j l}v_{kl}=v^{\prime}_{i j}$, which can be, conversely, used as a definition for tensors valued on the Lie algebra of a compact group. In general this representation is reducible, as one can verify constructing group invariants. Another possible representation is the conjugate one, which acts on the space $V$ as well as the fundamental representation, so they turns out to have the same dimension. Usually, the $N$-dimensional conjugate representation is denoted by the symbol $\overline{N}$. Remarkably, the conjugate representation of the group $SU(2)$ is equivalent to the fundamental one.\footnote{Two representations are said equivalent if it exists a unitary map $W:h_{1}\left(g\right)\rightarrow h_{2}\left(g\right)$, such that $W h_{1}\left(g\right)W^{\dagger}=h_{2}\left(g\right), \,\forall g$.}

Before going further introducing the so called adjoint representation, we briefly describe the parametrization of the group $SU(N)$. In order to understand which is the most convenient parametrization for this group, we observe that the generic unitary matrix $U$ can be rewritten as
\begin{equation}
U=e^{i\lambda},\qquad\text{with}\qquad\lambda=\lambda^{\dagger}\,,
\end{equation}
furthermore if $U\in\, SU(N)\Rightarrow {\rm Tr}\,\lambda=0$, so the group $SU(N)$ can be parametrized by the hermitian matrices with null trace and, as a consequence, is determined by $N^{2}-1$ parameters. Specifically, by defining a basis in the the space of the $N\times N$ hermitian matrices, $H$, formed by the $N^2-1$ hermitian matrices $\lambda^{a}$ with $a=1,2...,N^{2}-1$, satisfying the orthogonality condition 
\begin{equation}\label{orto cond}
{\rm Tr}\left(\lambda^{a}\lambda^{b}\right)=\frac{1}{2}\,\delta^{ab}\,,
\end{equation}
we can rewrite the generic element $U\in SU(N)$ as 
\begin{equation}\label{parametrization-1}
U=e^{i g^{a}\lambda^{a}},\qquad\text{where}\qquad g^{a}\in\mathds{R},\  \lambda^{a}\in H,\qquad\text{with}\qquad {\rm Tr}\,\lambda^{a}=0\,.
\end{equation}
The $\lambda^a$ matrices are said generators of the $SU(N)$ group. The fact that the commutator of two hermitian matrices is an anti-hermitian matrix, allows to write
\begin{equation}\label{commutation relation-1}
\left[\lambda^{a},\lambda^{b}\right]=if^{a b c}\lambda^{c}\,,
\end{equation}
which guarantees that (\ref{parametrization-1}) is a good parametrization for the group $SU(N)$. The structure constants of the group, $f^{a b c}$, turn out to be real and antisymmetric in the exchange of their three indexes.

Let us now consider the action of a $SU(N)$ operator infinitesimally close to the identity on the representation space. This can be parametrized by the first order expansion of the general parametrization (\ref{parametrization-1}), in symbols we have: $e^{i g^{a}\lambda^{a}}= I+i g^{a}\lambda^{a}$. The elements of the group in the connected component of the identity generate the algebra of $SU(N)$, which is fundamental for the construction of gauge theories. In this respect, we want to anticipate here that the canonical constraints we are going to calculate below are the generators of small gauge transformations only, i.e. they generate those gauge transformations in the connected component of the identity. In other words, the behavior of the states of the theory under large gauge transformations, i.e. those generated by the elements of the group characterized by a non-vanishing winding number (see the Appendix \ref{App B}), cannot be deduced by the theory. This leads to an extremely interesting issue. The observables of a gauge theory are, in fact, invariant under the full group, so they can be used to super-select states of the theory belonging to gauge sectors characterized by different winding numbers. From this perspective the global (and in general non-trivial) structure of the gauge group enters in the physical outcomes of the theory.\footnote{This fact is not only of mathematical interest, in this respect note that the solution of the so-called $U(1)_A$ puzzle in QCD is directly correlated to the topological global aspects of the gauge group. For the sake of completeness, we also stress that the extension of the theory to contain topologically non-trivial terms in the action has led to the necessity of solving the so-called strong {\it CP} problem, which contains physical predictions going ``beyond the Standard Model,'' as, for example, the existence of the axion.} 

By considering the direct product between the fundamental and the conjugate representations, we can construct the following representation, acting on $N\times N$ matrices as follows:
\begin{equation}
v^{'}_{i j}=U_{i k}U^{*}_{j l}v_{kl}\qquad\text{where}\qquad U\in SU(N),\quad v_{kl}\in V\otimes V\,.
\end{equation}
The adjoint representation is reducible, to decompose it in its irreducible components, let us identify the generic element $v_{kl}$ of the vector space as the ``$kl$'' entry of the matrix $W$, thus we have
\begin{equation}
W^{'}=UWU^{+}\,.
\end{equation}
It is worth noting that the hermitian matrices form an invariant subspace,\footnote{Considering only the hermitian matrices is not restrictive because a generic matrix $A$ can be rewritten as $A=A_{1}+i A_{2}$, with $A_{1}=A^{\dagger}_{1}$ and $A_{2}=A^{\dagger}_{2}$} as one can easily verify. The trace is an invariant too, so that we can reduce the representation considering the subspace formed by the hermitian matrices with null trace.

The representation acting on the $N\times N$ hermitian matrices with null trace is called \emph{adjoint representation} and has dimension $N^{2}-1$. 

Let us now study the algebraic structure of the adjoint representation. We consider an infinitesimal transformation, which acts on the hermitian traceless matrix $V$, we have:
\begin{equation}
V^{'}=UVU^{+}=\left(I+i g^{a}\lambda^{a}\right)V\left(I-i g^{b}\lambda^{b}\right)=V+ig^{b}\left[\lambda^{b},V\right]\,.
\end{equation}
Now, by expanding $V$ on the basis, we obtain the transformation law of the component $V^{a}$, namely:
\begin{equation}\label{adjoint representation-1}
V^{'a}=V^{a}-f^{a b c}g^{b}V^{c}=\left(\delta^{ac}-f^{a b c}g^{b}\right)V^{c}.
\end{equation}
The above expression (\ref{adjoint representation-1}) allows us to identify the generators of the adjoint representation, they are, in fact, given by the matrices of components ``$bc$'' identified with $\left(Q^{a}\right)_{bc}=f^{abc}$. It is straightforward to verify the universality of the commutation rules of the generators, in other words starting from the Jacobi identity for the $\lambda^a$ matrices and using the commutation relations (\ref{commutation relation-1}), one can demonstrate that
\begin{equation}
\left[Q^{a},Q^{b}\right]=f^{abc}Q^{c}\,.
\end{equation}
In other words, the structure constants characterize the group independently from the representation, as their name suggests. It should appear obvious that the structure constants characterize only the algebraic properties of the group, not the global ones, which cannot be deduced from the algebra.

\subsubsection{The gauge principle and physical interactions}

We are now ready to introduce $SU(N)$ Yang--Mills gauge theories. In this respect, let 
\begin{equation}
\Psi(x)=\left(\begin{array}{c}
\psi_1(x)
\\
\vdots
\\
\psi_N(x)
\end{array}\right)
\end{equation}
be a collection of $N$ Dirac spinor fields, the dynamics of which is described by the Dirac Lagrangian
\begin{equation}\label{DL}
\mathcal{L}[\Psi,\overline{\Psi}]=\overline{\Psi}(x)\left(i\gamma^{\mu}\partial_{\mu}-M\right)\Psi(x)\,,
\end{equation}
where $M$ is the $N\times N$ mass matrix.\footnote{The mass matrix $M$ can be in general neither diagonal nor hermitian. In this case, in fact, it can be easily diagonalized by a chiral transformation. This procedure, even though completely safe in the classical theory, can produce striking effects in the quantum theory, because of the chiral anomaly.} The Lagrangian above is symmetric under the action of the $SU(N)$ group, acting on the spinor field $\Psi$ and its conjugate $\overline{\Psi}$ according to the following rules:
\begin{align} 
\Psi(x)&\,\rightarrow\,\Psi^{\prime}(x)=U[g]\Psi(x)=e^{i g^a\lambda^a}\Psi(x)\,,
\\
\overline{\Psi}(x)&\,\rightarrow\,\overline{\Psi}^{\prime}(x)=\overline{\Psi}(x)U^{\dagger}[g]=\overline{\Psi}(x)\,e^{-i g^b\lambda^b}\,.
\end{align}
where $g^a$ are the $N^2-1$ constant parameters of the transformation. In other words, the collective spinor $\Psi$ transforms in the vectorial representation of the group $SU(N)$. 

The gauge principle states that the minimal coupling interaction between fermions and boson gauge fields can be obtained by requiring the local gauge invariance of the Dirac Lagrangian. In other words, we can directly extract the right minimal coupling by simply requiring that the Lagrangian (\ref{DL}) is invariant under the following local transformations
\begin{align} 
\Psi(x)\,\rightarrow\,\Psi^{\prime}(x)=U[g(x)]\Psi(x)=e^{i g^a(x)\lambda^a}\Psi(x)\,,
\\
\overline{\Psi}(x)\,\rightarrow\,\overline{\Psi}^{\prime}(x)=\overline{\Psi}(x)U^{\dagger}[g(x)]=\overline{\Psi}(x)\,e^{-i g^b(x)\lambda^b}\,.
\end{align}
It is worth noting that, in fact, the kinetic term in the Lagrangian (\ref{DL}) cannot be invariant under such a transformation for a very simple reason: The derivative operator is defined through a limiting procedure of objects transforming differently under the local $SU(N)$ group, or, equivalently, we can say that the ordinary derivative of a spinor does not transform in the vectorial representation of the $SU(N)$ group. More practically,
one can note that the action of the derivative operator on the transformed spinor field $\Psi^{\prime}(x)$ generates a term which cannot be reabsorbed unless we introduce a counter term. 

So the invariance requirement induces to modify the Lagrangian by defining a new derivative operator, $D_{\mu}$, which transforms in the adjoint representation of the $SU(N)$ group, so that 
\begin{equation}
D_{\mu}\psi\,\rightarrow\,\left(D_{\mu}\psi\right)^{\prime}=U[g(x)]D_{\mu}U^{\dagger}[g(x)]U[g(x)]\Psi=U[g(x)]D_{\mu}\Psi\,.
\end{equation}
This can be easily achieved by introducing a connection gauge field $A_{\mu}=A_{\mu}^a\lambda^a$, valued on the group $SU(N)$ and transforming according to the following equation
\begin{align}\nonumber
A_{\mu}\,\rightarrow\,A_{\mu}^{\prime}&=U[g(x)]A_{\mu}U^{\dagger}[g(x)]-i U[g(x)]\partial_{\mu}U^{\dagger}[g(x)]
\\
&=e^{i g^a(x)\lambda^a}A^c_{\mu}\lambda^c e^{-i g^a(x)\lambda^a}-i e^{i g^a(x)\lambda^a}\partial_{\mu}e^{-i g^a(x)\lambda^a}\,.
\end{align}
Consequently, for an infinitesimal gauge transformation we have:
\begin{equation}
A^a_{\mu}\,\rightarrow\,A^{\prime a}_{\mu}=A^a_{\mu}+if^{a b c}g^b A^c_{\mu}-\partial_{\mu}g^a=A^a_{\mu}-\mathcal{D}_{\mu}g^a\,,
\end{equation}
where the covariant derivative of the adjoint representation, $\mathcal{D}_{\mu}$, has been defined. In order to rigorously introduce the mathematical concept of connections and study their properties, we should digress on the geometry of fiber bundles, but this is far from the scope of this paper. So let us just remark that the replacement
\begin{equation}
\partial_{\mu}\Psi(x)\,\rightarrow\,D_{\mu}\Psi(x)=\partial_{\mu}\Psi(x)+i A_{\mu}\Psi\,,
\end{equation}
is a consequence of the fact that the global symmetry group has been promoted to be a local group and this introduces a fiber bundle structure which motivates the introduction of connections. 

Now, it is a simple exercise writing a Dirac Lagrangian which features the required properties of symmetry, i.e.
\begin{align}\nonumber
\mathcal{L}[\Psi,\overline{\Psi},A]&=\overline{\Psi}(x)\left(i\gamma^{\mu}D_{\mu}-M\right)\Psi(x)
\\
&=\overline{\Psi}(x)\left(i\gamma^{\mu}\partial_{\mu}-M\right)\Psi(x)-A_{\mu}^i\overline{\Psi}(x)\lambda^i\Psi(x)=\mathcal{L}[\Psi,\overline{\Psi}]+\mathcal{L}_{\rm int}[\Psi,\overline{\Psi},A]\,,
\end{align}
Remarkably, a new interaction term denoted by $\mathcal{L}_{\rm int}$ has appeared in the full Lagrangian, representing the minimal interaction of the $SU(N)$ gauge connection and the spinor matter fields. 

So, by fulfilling the requirement of the gauge principle, we have been able to automatically extract the correct minimal interaction between fermions and boson gauge fields. But, the gauge principle does not give us any hint on how to construct a kinetic term for the new gauge field $A_{\mu}$. From the general theory of fiber bundles, we know that once a connection has been defined, a natural curvature tensor can be constructed. Usually, in physics books, the curvature tensor is defined as follows:
\begin{equation}\label{definition curvature}
\left[D_{\mu},D_{\mu}\right]\Psi=i F_{\mu\nu}\Psi\,,
\end{equation}
where $F_{\mu\nu}=F_{\mu\nu}^a\lambda^a$ is said curvature or strength tensor. By the above definition it is easy to extract the explicit expression of the curvature tensor as function of the connection, i.e.
\begin{equation}
F_{\mu\nu}=\partial_{\mu}A_{\nu}-\partial_{\nu}A_{\mu}+\left[A_{\mu},A_{\mu}\right]\,.
\end{equation}
The curvature tensor satisfies the so-called Bianchi identity, 
\begin{equation}\label{bI}
D_{[\mu}F_{\nu\rho]}=0\,,
\end{equation}
which, remembering the definition (\ref{definition curvature}), can be considered as a consequence of the Jacobi identity associated to the covariant derivative operator $D_{\mu}$.

It is worth noting that if the gauge symmetry is represented by the $U(1)$ group, the structure constants vanish as a consequence of the vanishing of the commutator between the group generators (Abelian group) and, in this specific case, the connection physically describes the electromagnetic potential, while the curvature tensor is the electromagnetic field strength.

The curvature tensor transforms as a proper element of the adjoint representation (in contrast with the connection), specifically we have:
\begin{equation}
F_{\mu\nu}\,\rightarrow F^{\prime}_{\mu\nu}=U[g(x)]F_{\mu\nu}U^{\dagger}[g(x)]\,.
\end{equation}
This fact, and the analogy with the electromagnetic field, suggests the expression of the action for the field $A_{\mu}$. Specifically, requiring that the dynamics be described by second order partial differential equations of motion, a good action is:
\begin{equation}
S[A]=-\frac{1}{2}\int d^4x{\rm Tr}F_{\mu\nu}F^{\mu\nu}=-\frac{1}{4}\int d^4x\sum_a F^a_{\mu\nu}F^{a\,\mu\nu}\,,
\end{equation}
where in the second equality we used the orthogonality condition (\ref{orto cond}). The trace acts on the $N^{2}-1$ gauge internal indexes and makes the action invariant under the action of the $SU(N)$ group as can be easily demonstrated. 

As a final remark, note that we can add a term to the above action without affecting the classical equations of motion, i.e. the action
\begin{equation}\label{topological sector}
S_{\theta}[A]=-\frac{1}{4}\int d^4\!x\sum_a F^a_{\mu\nu}F^{a\,\mu\nu}+\frac{\theta}{64\pi^2}\int d^4\!x\,\epsilon^{\mu\nu\rho\sigma}\sum_b F^b_{\mu\nu}F^{b}_{\rho\sigma}\,
\end{equation} 
is dynamically equivalent to $S[A]$, since the $\theta$-term contribution to the classical equations of motion vanishes identically according to the Bianchi identity (\ref{bI}). Even though such a term does not affect the classical theory, it produces striking effects in the non-perturbative quantum theory (further details are given in Appendix \ref{App B}). Keep in mind the form of the action $S_{\theta}[A]$, since an analog situation will characterize the Ashtekar--Barbero formulation of canonical gravity.

\subsection{Gauge Symmetries and Constraints}\label{GSC}

Let us now start describing the canonical formulation of theories with gauge symmetries. This argument is particularly important for us, since it is the classical starting point of the canonical quantization of gauge systems. At the end of the section we will calculate the canonical Hamiltonian of the electromagnetic theory, which will serve as a description of the formalism in a simple practical example. 

\subsubsection{General formalism}

Let us consider a physical system described by a Lagrangian
\begin{equation}\label{sl}
L=L(q_i,\dot q_k)\,,
\end{equation}
where Latin indexes denotes the different generalized coordinates and velocities which determine the classical motion of the system on the configuration space. The Lagrangian equations of motion are
\begin{equation}\label{emot}
\frac{d}{dt}\left(\frac{\partial L}{\partial\dot{q}_i}\right)=\frac{\partial L}{\partial q_i}\,.
\end{equation}
By suitably defining the momenta and performing the so-called Legendre transformation, it is possible to refer the dynamics of a physical system to the generalized coordinates and their conjugate momenta. In this respect, we define the momentum $p_k$ as 
\begin{equation}
p_k=\frac{\partial L}{\partial\dot{q}_k}\,.
\end{equation}
The couple of canonical variables $(q_i,p_k)$ are the coordinates of the so-called phase space. Usually, in classical mechanics, we introduce the hypothesis that the momenta are independent functions of the velocities. Even though this hypothesis is fulfilled in many interesting classical macroscopic systems, it is too restrictive to be applied to more fundamental physical theories, e.g. those based on the gauge principle.

In general, the Lagrangian could be singular, namely 
\begin{equation}
\det\left[\frac{\partial^2 L}{\partial\dot{q}_i\partial\dot{q}_k}\right]=0\,.
\end{equation}
If this is the case, the velocities cannot be all inverted as functions of the generalized coordinates and momenta.\footnote{It is easy to show, starting from the equations of motion (\ref{emot}), that the vanishing of the above determinant implies also that the accelerations at a generic time cannot be uniquely determined as functions of the positions and velocities at the same time.} As a consequence, the momenta are not all independent, rather some relations among them crop up directly from their definitions. Specifically, we have
\begin{equation}\label{pc}
\phi_m(q_i,p_k)=0\,,\qquad\, m=1,2,\cdots,M\,.
\end{equation}
These relations are called \emph{primary constraints}, emphasizing that they result from the very definition of the momenta. The submanifold in the phase space determined by the conditions (\ref{pc}) is called primary constraint surface.

At this point, let us define the following function
\begin{equation}
H=\sum_k p_k\dot{q}_k-L(q_i,\dot{q}_k(p_j))\,,
\end{equation}
and calculate its variation:
\begin{align}\nonumber
\delta H&=\sum_i \delta p_i\dot{q}_i+\sum_k p_k\delta\dot{q}_k-\sum_j\left(\frac{\delta L(q_i,\dot{q}_k)}{\delta q_j}\right)\delta q_j-\sum_l\left(\frac{\delta L(q_i,\dot{q}_k)}{\delta \dot{q}_l}\right)\delta\dot{q}_l
\\
&=\sum_k\dot{q}_k \delta p_k-\sum_j\left(\frac{\delta L(q_i,\dot{q}_k)}{\delta q_j}\right)\delta q_j\,,
\end{align}
this is called canonical Hamiltonian and its variation depends only on the positions and canonical momenta. However, the Hamiltonian defined above is not uniquely determined because we can add to it any linear combination of the primary constraints. In other words, the theory cannot distinguish between the Hamiltonian $H$ defined above and the new Hamiltonian $H^{\star}=H+\sum_m c_m\phi_m$.
Nevertheless, since the above equation has to hold for any variation, provided that the variation preserves the conditions (\ref{pc}), we can obtain the following Hamiltonian equations of motion:
\begin{subequations}
\begin{align}\label{em}
\dot{q}_i&=\frac{\partial H}{\partial p_i}+\sum_m u_m\frac{\partial\phi_m}{\partial p_i}\,,
\\
-\dot{p}_k&=\frac{\partial H}{\partial q_k}+\sum_m u_m\frac{\partial\phi_m}{\partial q_k}\,,
\end{align}
\end{subequations}
which are in accordance with the general method of the calculus of variations applied to a system with constraints. We stress that the symbol $u_m$ denotes a completely arbitrary set of functions. Before going on, it is convenient to introduce a formalism that allows to write the canonical equations of motion in a compact way. We are referring to the Poisson brackets. Let $f$ and $g$ be two generic functions of the canonical variables, then we define:
\begin{equation}
\left[f,g\right]=\sum_i\left(\frac{\partial f}{\partial q_i}\frac{\partial g}{\partial p_i}-\frac{\partial f}{\partial p_i}\frac{\partial g}{\partial q_i}\right)\,.
\end{equation}
It immediately follows that 
\begin{equation}
\left[q_i,p_k\right]=\delta_{ik}\,,
\end{equation}
where $\delta_{ik}$ is the Kronecker symbol. Moreover, it is easy to demonstrate that the equations of motion (\ref{em}) can be easily rewritten as
\begin{align}
\dot{q}_i&=\left[q_i,H\right]+\sum_m u_m\left[q_i,\phi_m\right]\,,
\\
\dot{p}_k&=\left[p_k,H\right]+\sum_m u_m\left[p_k,\phi_m\right]\,.
\end{align}
more generically, we have 
\begin{equation}
\dot{g}=\left[g,H\right]+\sum_m u_m\left[g,\phi_m\right]\,,
\end{equation}
where $g$ is a generic functions of the canonical variables. The Poisson brackets are defined only for those quantities that are functions of the canonical variables, nonetheless, the above definition for the time derivative of the generic function $g$ can be rewritten more concisely as:
\begin{equation}\label{pbt}
\dot{g}=\left[g,H_T\right]\,,
\end{equation}
where the total Hamiltonian $H_{T}$ is defined as
\begin{equation}
H_T=H+\sum_m u_m\phi_m\,.
\end{equation}
One may immediately wonder about the acceptability of this definition, because one of the term that come out from Poisson bracket (\ref{pbt}) is $\sum_m\left[g,u_m\right]\phi_m$. This is badly defined because the $u_m$ are arbitrary functions not depending on the canonical variables. But, the correctness of this definition stems from the fact that the Poisson brackets multiply the vanishing functions $\phi_m$. So, we have to carefully consider the fact that in dealing with the Poisson formulation of classical mechanics in theories with constraints, these have to be imposed only after having calculated all the Poisson brackets, otherwise we would affect the consistency of the construction. That is the reason why the weakly vanishing symbol ``$\approx$'' is widely used; it emphasizes the fact that the constraints have to be imposed at the end of the canonical analysis, limiting the evolution of the system to a restricted region of the phase space.

As a consistency requirement, we have to impose another condition on the dynamics. Namely, we have to require that the primary constraints surface be preserved by the Hamiltonian flow, i.e.
\begin{equation}\label{cc}
\dot{\phi}_n=\left[\phi_n,H\right]+\sum_m u_m\left[\phi_n,\phi_m\right]\approx 0\,.
\end{equation}
If the above equation is automatically satisfied, namely the time derivative of the primary constraints vanishes on the primary constraints surface, then no other consistency check is necessary. But, in general, different cases may occur. In particular, if $\dot{\phi}_n\neq 0$ for certain values of $n$, then some \emph{secondary constraints} $\chi(q_i,p_k)\approx 0$ have to be imposed, further restricting the available region of the phase space. Obviously, the same consistency condition in Eq. (\ref{cc}) has to be satisfied by the secondary constraints, otherwise tertiary constraints have to be imposed and so on.

Since there is no fundamental difference between primary and secondary constraints, we collect all of them in the same symbol, i.e. 
\begin{equation}
\phi_m(q_i,p_k)\approx 0\,,\qquad\, m=1,2,\cdots,M+K\,, 
\end{equation}
where $K$ is the number of secondary constraints. In general, the consistency equation contains interesting information about the arbitrary functions entering in the Hamiltonian, so a closer look at it is in order. We have:
\begin{equation}
\left[\phi_n,H\right]+\sum_m u_m\left[\phi_n,\phi_m\right]\approx 0\,,
\end{equation}
generally, if $\det\left\{\left[\phi_n,\phi_m\right]\right\}\neq 0$, we can extract a solution for the $u_m$ by inverting the matrix $C_{nm}=\left[\phi_n,\phi_m\right]$; for convenience let us indicate this set of solutions as $U_m(q_i,p_k)$. Moreover, in order to write the general solution of the equation above, we must add the solution of the homogeneous equation to the particular solutions $U_m(q_i,p_k)$, namely 
\begin{equation}
\sum_m V_m\left(\left[\phi_n,\phi_m\right]\right)\approx 0\,.
\end{equation}
In general, there can be a certain number of independent solution of the equation above, which are denoted as $V_{a m}$. So that the most general solution of the consistency condition is $u_n=U_n+\sum_a v_a V_{a n}$, where the components of the functions $u_m$ that can be fixed by the consistency conditions have been separated from that which remains arbitrary. In terms of this new expression for $u_m$, the total Hamiltonian can be rewritten as
\begin{equation}\label{TH}
H_T=H+\sum_m U_m\phi_m+\sum_a v_a\phi_a=H^{\prime}+\sum_a v_a\phi_a\,,
\end{equation} 
where we have defined $H^{\prime}=H+\sum_m U_m\phi_m$ and $\phi_a=V_{a m}\phi_m$. The canonical equations of motion can be calculated through the total Hamiltonian by using the Poisson brackets formalism. Obviously, even though some of the arbitrariness has been eliminated via the consistency condition, they still contain some arbitrary functions, $v_a$. Nevertheless, the canonical equations of motion are equivalent, by construction, to the Lagrangian ones.

At this point, we have to introduce some useful terminology. We call \emph{first class} all those functions of the canonical variables which have weakly vanishing Poisson brackets with all the constraints. By remembering that the $\phi$'s are the only independent quantities which are weakly zero, then the function $S=S(q_i,p_k)$ is first class if
\begin{equation}
\left[S,\phi_m\right]=\sum_n c_{m n}\phi_n\,,\qquad \forall\, m=1,\cdots,M+K\,.
\end{equation}
All the other functions of the canonical variables are said to be \emph{second class}. It is easy to demonstrate that the Poisson brackets of two first class functions is first class as well as the Hamiltonian $H^{\prime}$ defined in (\ref{TH}). This terminology applies to constraints as well: We call \emph{first class constraints} the set of $\phi_m$ with $m\leq K+M$ such that
\begin{equation}
\left[\phi_m,\phi_n\right]=\sum_p c_{m n p}\phi_p\,,\qquad \forall\, m=1,\cdots,M+K\,,
\end{equation}
while the others are referred as \emph{second class constraints}.

Finally, we want to digress on the transformations induced by the first class constraints. In order to do that, let us stress that in a theory with constraints, because of the presence of the arbitrary functions $v_a$, the evolution of the generalized coordinates and momenta are not uniquely determined by the initial state. This means that there are many choices of the fundamental variables that characterize the same physical state. In this respect, it is interesting to consider particular values of phase space variables (let us call them $g$) at an initial time, e.g. $g_0=g(t=0)$, and look at their values after an infinitesimal temporal lapse $\delta t$. By using the Poisson brackets, we obtain:
\begin{equation}
g({\delta t})=g_0+\dot{g}\delta t=g_0+\left[g,H_{T}\right]\delta t=g_0+\delta t\left(\left[g,H^{\prime}\right]+\sum_a v_a\left[g,\phi_a\right]\right)\,.
\end{equation}
Imagine that we had initially taken different functions $v^{\prime}_a$, then we would have obtained:
\begin{equation}
g^{\prime}({\delta t})=g_0+\dot{g}\delta t=g_0+\left[g,H_{T}\right]\delta t=g_0+\delta t\left(\left[g,H^{\prime}\right]+\sum_a v^{\prime}_a\left[g,\phi_a\right]\right)\,.
\end{equation} 
In other words, during the infinitesimal time $\delta t$, the difference $\Delta v_a$ of the two functions $v_a$ and $v^{\prime}_a$ (i.e. $\Delta v_a=v_a-v^{\prime}_a$) generates a difference between $g$ and $g^{\prime}$ given by
\begin{equation}
\Delta g({\delta t})=\delta t\Delta v_a\left[g,\phi_a\right]=\epsilon_a\left[g,\phi_a\right]\,,
\end{equation} 
where $\epsilon_a$ is a small quantity, being proportional to $\delta t$. So, according to the above rule, the variables describing a particular physical configuration of the system can be arbitrarily changed without affecting the state of the system. In other words, many different sets of canonical variables, related each other by the above transformation, equivalently describe the same physical state. Hence, the functions $\phi_a$ result to be the generators of gauge transformations.

Concluding, we have found that first class primary constraints generate gauge transformations, but it is, in general, expectable that also secondary first class constraints are generators of gauge transformations, and this is, in fact, the case in many mechanical systems. It is commonly believed that all the first class constraints generate gauge transformations, even though, this belief is not supported by a rigorous proof and is sometimes referred as Dirac's conjecture.

\subsubsection{Electromagnetic canonical theory}

The electromagnetic theory provides a very simple, but non-trivial and practically interesting example to operatively use the procedure just described. In order to be definite, we start from the Lagrangian for the Maxwell's theory, i.e.
\begin{equation}
L(A,\partial A)=-\frac{1}{4}\int d^3\!\!x\,F_{\mu\nu}F^{\mu\nu}\,.
\end{equation}
Here the fundamental variable is the electromagnetic potential $A_{\mu}$, which, as is well known, is the connection field associated to the $U(1)$ gauge symmetry; namely, acting with a gauge transformation on $A_{\mu}$, we have that 
\begin{equation}\label{gd}
A_{\mu}\,\rightarrow\,A_{\mu}^{\prime}=A_{\mu}+\partial_{\mu}\lambda\,,
\end{equation}
where $\lambda$ is a generic function of the space-time points. According to what said above, the gauge symmetry will reveal its presence in the canonical theory through the appearance of first class constraints, associated with the generators of the gauge transformation. Hence, let us start the canonical analysis of the electromagnetic theory by calculating the conjugate momenta to the variables $A_{\mu}$. Namely,
\begin{equation}
E^{\alpha}=\frac{\partial L}{\partial A_{\alpha}}=-\partial^0 A^{\alpha}+\partial^{\alpha}A^0=F^{\alpha0}\,,\qquad\text{and}\qquad P^0=\frac{\partial L}{\partial A_{0}}=0\,,
\end{equation}
are respectively the momenta conjugate to $A_{\alpha}$ and $A_0$ (the Greek indexes from the beginning of the alphabet indicate purely spatial components, while the index $0$ indicates the time component). The resulting phase space is 8-dimensional with coordinates $(A_{\alpha},A_0,E^{\alpha},P^0)$, and can be equipped with the following symplectic structure:
\begin{equation}
\left[A_{\alpha}(t,x),E^{\beta}(t,x^{\prime})\right]=\delta^{\beta}_{\alpha}\delta(x,x^{\prime})\,,\qquad\left[A_0(t,x),P^0(t,x^{\prime})\right]=\delta(x,x^{\prime})\,,
\end{equation}
the other brackets vanishing. 

The fact that the right hand side of the momentum $P^0$ does not contain any velocity implies that the Lagrangian is singular, according to definition (\ref{sl}); moreover it generates a primary constraint, i.e.
\begin{equation}
\phi:= P^0\approx 0\,.
\end{equation}
Now, the canonical Hamiltonian can be calculated by performing the Legendre transformation, obtaining:\footnote{It is worth noting that the velocity $\dot{A}_{\alpha}$ can be rewritten as function of the momentum as $\dot{A}_{\alpha}=E_{\alpha}+\partial_{\alpha}A_0$, where we have taken into account the signature of the metric, i.e. $-,+,+,+$.}
\begin{equation}
H=\int d^3\!\!x\left[\frac{1}{2}E^{\alpha}E_{\alpha}+\frac{1}{4}F_{\alpha\beta}F^{\alpha\beta}-E^{\alpha}\partial_{\alpha}A_0\right]\,.
\end{equation}
So, the primary total Hamiltonian results to be:
\begin{equation}\label{Th}
H_T=\int d^3\!\!x\left[\frac{1}{2}E^{\alpha}E_{\alpha}+\frac{1}{4}F_{\alpha\beta}F^{\alpha\beta}+A_0\partial_{\alpha}E^{\alpha}+u\phi\right]\,,
\end{equation}
where we integrated by parts the third term. As a consistency check, we calculate the time derivative of the primary constraint $P^0$, we have:
\begin{equation}
\dot{\phi}=\left[\phi,H\right]=-\partial_{\alpha}E^{\alpha}\,,
\end{equation}
so that a secondary constraint has to be imposed, namely
\begin{equation}
\chi:=\partial_{\alpha}E^{\alpha}\approx 0\,.
\end{equation}
The next consistency check does not generate any tertiary constraint, indeed
\begin{equation}
\dot{\chi}=\left[\chi,H\right]=0\,.
\end{equation}
So, the theory generates one primary and one secondary constraint which form a set of first class constraints, i.e.
\begin{equation}
\left[\phi,\phi\right]=0\,,\quad\left[\phi,\chi\right]=0\,,\quad\left[\chi,\chi\right]=0\,.
\end{equation}
Now, in order to get the most general physically possible motion, we write the extended Hamiltonian containing the secondary constraint as well, we have:
\begin{equation}\label{eh}
H_E=\int d^3\!\!x\left[\frac{1}{2}E^{\alpha}E_{\alpha}+\frac{1}{4}F_{\alpha\beta}F^{\alpha\beta}+A_0\partial_{\alpha}E^{\alpha}+u\phi+v\chi\right]\,.
\end{equation}
The equations of motion can be easily obtained by calculating the Poisson bracket of the canonical variables with the extended Hamiltonian above; but, interestingly enough, given the structure of $H_E$, some simplifications are possible. In this respect, let us firstly, calculate the equation of motion of $A_0$, obtaining
\begin{equation}
\dot{A_0}=\left[A_0,H_E\right]=u\,.
\end{equation}
This reveals the nature of one of the ambiguities of the theory, which turns out to be the time derivative of $A_0$. More importantly, the above equations states that the evolution of  $A_0$ is completely arbitrary. So, remembering the expression of the secondary constraint $\chi$, we can reabsorb the variable $A_0$ in the definition of the arbitrary function $v$. Moreover, the momentum $P^0$ is constrained to vanish along all the evolution and its presence in the Hamiltonian only ensures that the variable $A_0$ is an arbitrary function. So, in order to simplify the expression of the extended Hamiltonian and reduce the number of unphysical degrees of freedom, we can drop both the variables $A_0$ and $P^0$. It is worth stressing that the dynamics of the physically relevant degrees of freedom is not affected by this reduction. Finally, the total Hamiltonian turns out to be:
\begin{equation}
H_T=\int d^3\!\!x\left[\frac{1}{2}E^{\alpha}E_{\alpha}+\frac{1}{4}F_{\alpha\beta}F^{\alpha\beta}+v\partial_{\alpha}E^{\alpha}\right]\,,
\end{equation}
which can be used to calculate the canonical equations of motion, or to quantize the system by implementing the Dirac procedure described in § \ref{DQP}. We remark that the only survived constraint is the first class Gauss constraint $\partial_{\alpha}E^{\alpha}\approx 0$. A Gauss constraint appears in Yang--Mills gauge theories of non-abelian groups as well; in that case the ordinary derivative is replaced by a covariant derivative, so the connection field enters in the Gauss constraint, complicating its mathematical structure (see, e.g. (\ref{Gauss Y-M})). 

It is interesting to note that the Gauss constraint, in fact, generates gauge transformations. Let us, for example, consider the action of the smeared Gauss constraint, i.e.
\begin{equation}
G(f)=\int d^3\!\!x\,f(t,x)\partial_{\alpha}E^{\alpha}(t,x)\,,
\end{equation}
on the fundamental variable $A_{\alpha}$, we have:
\begin{equation}
\delta A_{\alpha}(t,x)=\left[A_{\alpha}(t,x),G(f)\right]=\partial_{\alpha}f(t,x)\,,
\end{equation}
where the generic smearing function $f$ plays the role of $\lambda$ in (\ref{gd}). 

\subsection{Initial Value Formulation}

We conclude this section giving an account of the fundamental theorems for establishing a well posed initial value formulation for gauge theories and GR. The results expressed by these theorems are pretty general, but as we will show below, they can be easily applied to specific cases. As an example, we will address the simple problem regarding the initial value formulation of Maxwell's electromagnetism. The case of electromagnetism is interesting because, being a gauge theory, shares with GR the presence of first class constraints and, as a consequence, the necessity to make a proper ``gauge choice'' in order to write the equations in a suitably form to face the initial value problem.

\subsubsection{Some important theorems}

First of all we enunciate the following theorem without demonstrating it, addressing the reader to \cite{HawEll1973} for a complete proof (a partial proof can be found in Wald's book \cite{Wal1984}):
\begin{theorem}\label{theorem1}
Let $(M,g_{\mu\nu})$ be a globally hyperbolic space-time (or a globally hyperbolic region of an arbitrary space-time) and let $\nabla_{\mu}$ be any derivative operator. Let $\Sigma$ be a smooth, space-like Cauchy surface. Consider the system of $n$ linear equations for $n$ unknown functions $\phi_1,\dots,\phi_n$ of the form
\begin{equation}\label{eq theo}
g^{\mu\nu}\nabla_{\mu}\nabla_{\nu}\phi_i+\sum_{j}A^{\mu}_{\phantom1i j}\nabla_{\mu}\phi_j+\sum_{j}B_{i j}\phi_j+C_i=0\,,
\end{equation}
namely a linear, diagonal, second order hyperbolic system. Then equation (\ref{eq theo}) has a well posed initial value formulation on $\Sigma$. More precisely, given arbitrary smooth initial data, $(\phi_i,n^{\mu}\nabla_{\mu}\phi_i)$ for $i=1,\dots,n$ on $\Sigma$, there exists a unique solution of equation (\ref{eq theo}) throughout $M$. Furthermore, the solutions depend continuously on the initial data. Finally, a variation of the initial data outside of a closed subset, $S$, of $\Sigma$ does not affect the solution in $D(S)$.
\end{theorem}
It is worth noting that the theorem explicitly refers to linear systems of equations, moreover, although there are few results concerning the initial value formulation for non-linear systems of equations, an important result exists concerning the so called quasi-linear, diagonal second order hyperbolic equations due to Leray (1952) and is contained in the following theorem: 
\begin{theorem}\label{theo 2}
Let $(\phi_0)_1,\dots,(\phi_0)_n$ be any solution of a quasi-linear hyperbolic system of equations below
\begin{equation}\label{eq theo 2}
g^{\mu\nu}(x;\phi_j,\nabla_{\mu}\phi_j)\nabla_{\mu}\nabla_{\nu}\phi_i=F_i(x;\phi_j,\nabla_{\mu}\phi_j)\,,
\end{equation}
on a manifold $M$ and let $(g_0)^{\mu\nu}=g^{\mu\nu}(x;(\phi_0)_j,\nabla_{\mu}(\phi_0)_j)$. Suppose $(M,(g_0)_{\mu\nu})$ is globally hyperbolic (or alternatively consider a globally hyperbolic region of this space-time). Let $\Sigma$ be a smooth space-like Cauchy surface for $(M,(g_0)_{\mu\nu})$. Then, the initial value formulation of equation (\ref{eq theo 2}) is well posed on $\Sigma$ in the following sense: for initial data on $\Sigma$ sufficiently close to the initial data for $(\phi_0)_1,\dots,(\phi_0)_n$, there exists an open neighborhood $O$ of $\Sigma$ such that equation (\ref{eq theo 2}) has a solution $\phi_1,\dots,\phi_n$, in $O$ and $(O,g_{\mu\nu}(x;\phi_j,\nabla_{\mu}\phi_j))$ is globally hyperbolic. The solution is unique in $O$ and propagates causally in the sense that if the initial data for $\phi_{1}^{\prime},\dots,\phi_{n}^{\prime}$ agree with that of $(\phi_0)_1,\dots,(\phi_0)_n$ on a subset $S$ of $\Sigma$, then the solution agree on $O\cap D^{+}(S)$. Finally the solutions depend continuously on the initial data.
\end{theorem}
As before we do not give the demonstration of the theorem, which can be find in \cite{HawEll1973}, together with some other interesting properties of the solutions. Let us remark that equation (\ref{eq theo 2}) differs from (\ref{eq theo}), because the Lorentzian metric field $g_{\mu\nu}$ is now allowed to depend on the unknown variables and their first derivatives, while the smooth functions $F_i$ may have a non-linear dependence on these variables. An interesting recent application of Leray's theorem is the demonstration that a well posed initial value formulation can be formulated for the scalar-gravity coupled system, as showed in \cite{Cho-BruIsePol2006}.

\subsubsection{Initial value formulation for the electromagnetic field}

The strategy we want to follow in order to show that the Maxwell system of equations has a well posed initial value formulation should be now clear: It basically consists in recasting the equations in a form which can be traced back to those in line (\ref{eq theo}).

So let us start recalling the expression of the Maxwell system of equations in Minkowski space-time:
\begin{equation}\label{max eq}
\partial_{\mu}\left(\partial^{\mu}A^{\nu}-\partial^{\nu}A^{\mu}\right)=0\,.
\end{equation}
We can easily split the background, fixing a one parameter family of hypersurfaces $\Sigma_t$ parametrized by constant values of the inertial time $t$, in particular let us assume that $\Sigma_0=\Sigma_{t=t_0}$ be our initial hypersurface. This procedure allows to emphasize a central feature of the Maxwell system of equations, namely the appearance of the so called Gauss constraint, due to the fact that the time component of the equation above does not contain any second time derivative term. Indeed,
\begin{equation}\label{em gauss}
\partial_{\alpha} E^{\alpha}=0,
\end{equation}
where
\begin{equation}
\qquad E^{\alpha}=F^{\alpha0}=\partial^{\alpha}A^0-\partial^0 A^{\alpha}=\partial^{\alpha}A^0+\dot{A}^{\alpha}\,,
\end{equation}
is the electric field. Thus equation (\ref{em gauss}) represents a constraint for the initial data $(A^{\mu},\dot{A}^{\nu})$ on $\Sigma_0$. In other words, the choice of the initial data is not free, they must, in fact, satisfy the Gauss constraint, otherwise they cannot generate solutions of the Maxwell's equations. One could expect that differentiating the Gauss constraint with respect to time, an equation containing a second time derivative of the scalar potential $A_0$ can be obtained. But, as can be easily verified, the Bianchi identity $dF=0$ (where $F=dA$ is the curvature 2-form associated to the electromagnetic field $A$ and $d$ is the exterior derivative operator) prevents from generating second order time derivatives of $A_0$. As a side remark, we stress that in the opposite case, the initial value problem  would have a simple solution, at least in the sense expressed by the Cauchy-Kowalewski theorem:
\begin{theorem}
\emph{(Cauchy-Kowalewski theorem):} Let $(t,x^1,\dots,x^m-1)$ be coordinates of $\mathds{R}^m$. Consider a system of $n$ partial differential equations for $n$ unknown functions $\phi_1,\dots,\phi_n$ in $\mathds{R}^m$, having the following form
\begin{equation}\label{cauchy-kowalewski}
\frac{\partial^2\phi_{i}}{\partial t^2}=F_i\left(t,x^a;\phi_j;\frac{\partial\phi_{j}}{\partial t};\frac{\partial\phi_{j}}{\partial x^a};\frac{\partial^2\phi_{j}}{\partial t\partial x^a};\frac{\partial^2\phi_{i}}{\partial x^a\partial x^b}\right)\,,
\end{equation}
where each $F_{i}$ is an analytic function of its variables. Let $f_{i}(x^a)$ and $g_{i}(x^a)$ be analytic functions. Then there is an open neighborhood $O$ of the initial hypersurface $t=t_0$ such that within $O$ there exists a unique analytic solution of equation (\ref{cauchy-kowalewski}) such that $\phi_{i}(t_0,x^a)=f_{i}(x^a)$ and $\frac{\partial\phi_{i}}{\partial t}(t_0,x^a)=g_{i}(x^a)$.
\end{theorem}
Unfortunately, this is not the case. The Maxwell system of equations (\ref{max eq}) is under-determined, being equivalent to the dynamical equations for the spatial components of the electromagnetic field, which contain second time derivatives, plus a constraint equation for four unknown functions. But this feature is absolutely not unexpected. It simply reflects the presence of the $U(1)$ gauge freedom. In other words, the existing gauge freedom prevents the Maxwell equations from completely determining the potential $A_{\mu}$, namely we should expect that they uniquely determine the potential up to a gauge transformation. From this perspective, it is easy to understand that the system of Maxwell equations admit a well posed initial value formulation only for the \emph{physical} states of the theory. In fact, once realized how the gauge transformations enter into the game, the solution becomes simple. 

In order to clarify this point, let us fix the Lorentz gauge, i.e.
\begin{subequations}
\begin{equation}\label{max system 2}
\partial_{\mu}A^{\mu}=0\,,
\end{equation}
which is particularly useful to treat this problem. The gauge choice allows to simplify the structure of Eq. (\ref{max eq}), i.e.
\begin{equation}\label{max system 1}
\partial_{\nu}\partial^{\nu}A^{\mu}=0\,.
\end{equation}
\end{subequations}
The system of equations are physically equivalent to (\ref{max eq}). More precisely, solutions of the Eq. (\ref{max eq}) can differ from those obtainable by solving the system of equations in (\ref{max system 2}) and (\ref{max system 1}) only by a gauge transformation. So, from a physical perspective, the dynamics is well described by Eqs. (\ref{max system 2}) and (\ref{max system 1}), with the remarkable advantage that now we can use the result of theorem (\ref{theorem1}).

Specifically, let us suppose that the initial data are chosen in such a way that they satisfy the Lorentz gauge condition in Eq. (\ref{max system 2}) on the initial hypersurface $\Sigma_0$ (if they don't we can operate on them by a suitable gauge transformation), then, using equation (\ref{max system 1}) and the Schwarz theorem, we can write:
\begin{equation}
\partial_{\mu}\partial^{\mu}\left(\partial_{\nu}A^{\nu}\right)=\partial_{\nu}\left(\partial_{\mu}\partial^{\mu}A^{\nu}\right)=0\,.
\end{equation}
Hence, provided that equation $\partial_{\mu}\partial^{\mu}A^{\nu}=0$ is satisfied everywhere, according to theorem (\ref{theorem1}) also the gauge condition will be satisfied everywhere if and only if $\partial\partial_{\mu}A^{\mu}/\partial t=0$ on $\Sigma_0$. It is worth explaining the role played here by the Gauss constraint. At a first glance the Gauss constraint seems to have disappeared; actually, it has only been written in a different form. In fact, we have required that $\partial\partial_{\mu}A^{\mu}/\partial t=0$ on $\Sigma_0$, namely
\begin{align}\nonumber
0&=\frac{\partial}{\partial t}\,\partial_{\mu}A^{\mu}=\partial_{0}^2A^0+\partial_{\alpha}\partial_{0}A^{\alpha}
\\
&=\partial_{0}^2A^{0}+\partial_{\alpha}\left(E^{\alpha}-\partial^{\alpha}A^0\right)
=\partial_{\alpha}E^{\alpha}\,,
\end{align}
where we used equation (\ref{max system 1}). It is not surprising at all that the Gauss constraint turns out to be encapsulated in the initial conditions, because the initial field configuration must satisfy not only the gauge condition, but also the constraint. In particular, its role is to assure that, if equation (\ref{max system 1}) holds everywhere and, provided that $\partial_{\alpha}E^{\alpha}=0$ on $\Sigma_0$, the Lorentz gauge condition remains valid throughout all the evolution for the gauge transformed initial data. 

It remains only to solve the dynamical equations (\ref{max system 1}), with the given, suitably chosen, initial data. But now the problem is simple, the equations have in fact the desired form, i.e. the form required to apply the result of theorem (\ref{theorem1}), which indeed establishes the existence of a well posed initial data formulation. The last question is: can we conclude that the solutions are unique? We can briefly answer to this question supposing that the original systems of Maxwell equations (\ref{max eq}) provides two solutions, $S_1$ and $S_2$, with the same initial conditions. By making a gauge transformation, it is possible to recast them into the solution of the equation (\ref{max system 1}) with the same initial condition. But, since the solution of the system (\ref{max system 1}) is unique, once assigned suitably initial conditions, then we conclude that $S_1$ and $S_2$ can differ at most by a gauge transformation. In other words they represent the same physical field configuration. Concluding, physically the solution is, in fact, unique.

\section{Canonical General Relativity}\label{sec4}
As we have clarified above, the canonical constraints play a crucial role in the initial value formulation of a theory with gauge freedom. Furthermore, once the canonical analysis has been completed and the second class constraints eliminated through the Dirac prescription, the system can be canonically quantized, by requiring that the operators corresponding to the first class constraints annihilates the state functional. The same procedure and considerations are, in general, valid as far as we regard GR, which features a gauge symmetry correlated with the invariance under diffeomorphisms. So in this section, in view of discussing the problem of quantum gravity, we canonically reformulate GR, digressing on the main aspects of its initial value formulation. 

Specifically, starting from the description of the so-called 3+1 splitting procedure, we finally arrive at the canonical equations of motion of GR, initially studied by Dirac \cite{Dir1964} and then by Arnowitt, Deser, and Misner \cite{ArnDesMis1959,ArnDesMis1960-1,ArnDesMis1960-2,ArnDesMis1962}. This will also allow us to address the initial value problem, which will be only briefly sketched, emphasizing the role of gauge symmetries.

\subsection{3+1 splitting of space-time}\label{3+1SOST}

The splitting procedure is a tool which allows to sort an evolution parameter out of the covariant general relativistic space-time. It is worth stressing that covariance is not lost in this formulation, even though it is no longer manifest as in the Lagrangian approach. As suggested by Theorem (\ref{geroch theorem}), by using a gauge transformation, which we refer to as embedding diffeomorphism, it is possible to ``slice'' a globally hyperbolic space-time, representing it as the evolution in ``time'' of 3-dimensional space. As any gauge transformation, the embedding diffeomorphisms does not affect the dynamical content of the theory. In other words, the canonical Hamiltonian equations plus constraints (obviously!) are completely equivalent to the usual Einstein equations.

Technically, what we are going to do is the following: Starting from the Hilbert--Einstein action,
\begin{equation}\label{H-E action}
S_{HE}(g)=\frac{1}{2}\int\limits_{M}d^4x\sqrt{-g}\,R\,,
\end{equation}
where $R=g_{\mu\rho}g_{\nu\sigma}R^{\mu\nu\rho\sigma}$ is the Ricci scalar curvature ($R_{\mu\nu\rho}^{\ \phantom1\phantom1\phantom1\sigma}v_{\sigma}=\left[\nabla_{\mu},\nabla_{\nu}\right]v_{\rho}$ being the Riemann curvature tensor), we restrict $M$ to be a globally hyperbolic space-time, $\mathcal{M}$, so that, according to Theorem (\ref{geroch theorem}), $\mathcal{M}=\mathds{R}\times\Sigma^3$; $\Sigma^3$ being a compact three-dimensional manifold without boundary.

It is interesting to note that the restriction to globally hyperbolic space-times is a quite strong requirement, especially in view of the formalization of a quantum theory of gravity. Being aware of this strong hypothesis, necessary to canonically quantize the theory, we expect that it could be relaxed once a rigorous formulation of the quantum theory will be at hand, as suggested in \cite{Thi2001}. So far, up to the author's knowledge, no rigorous prescription exists that allows to get rid of this hypothesis or consider different topologies directly in the quantum theory.\footnote{The case of Loop Quantum Gravity is quite different. In Loop Quantum Gravity, in fact, the continuum space-time is replaced by a discrete genuinely quantum structure. This implies that in this theory no room is left for (the classical concept of) space-time, which, in this sense, is not restricted by the hypothesis of the Geroch's theorem. In other words, in Loop Quantum Gravity the quantization procedure itself relaxes the classical restrictions on the space-time structure.}

However, once the hypotheses of the Geroch theorem are satisfied, we can foliate $\mathcal{M}$ by Cauchy hypersurfaces $\Sigma_{t}^{3}\stackrel{def}{=}y_{t}(\sigma)$, in other words $\forall\, t\in\mathds{R}, \exists$ a globally injective immersion (embeddings) $y_{t}:\Sigma^3\rightarrow M$, defined by $y_{t}(x^i)=y(t,x^i)$, where $x^i\in\sigma$ are the coordinates over the hypersurfaces $\Sigma^3$. Hence, $\Sigma_{t}^{3}$ represents a foliation of the manifold $M$ parametrized by the continuous function $t$. Now let us denote the unit normal vector to the hypersurfaces $\Sigma^{3}_{t}$ as $n^{\mu}$, so that the four-dimensional metric $g_{\mu\nu}$ induces a three-dimensional Riemannian metric $h_{\mu\nu}$ on each hypersurface:
\begin{equation}\label{h=g+nn}
h_{\mu\nu}=g_{\mu\nu}+n_{\mu}n_{\nu}\,.
\end{equation}
The above relation is often referred as \emph{first fundamental form} of $\Sigma^3$. Now, consider a vector field $t^{\mu}$, called ``deformation vector'', satisfying the the following relation $t^{\mu}\nabla_{\mu}t=1$. It generates a 1-parameter family of diffeomorphisms, $\phi_{t}:\mathds{R}\times\Sigma^3\rightarrow M$, defined as $(t,x)\rightarrow y^{\mu}(t,x):=y^{\mu}_t(x)$, called embedding diffeomorphisms. Geometrically, the deformation vector represents the ``flow of time'' throughout space-time, in other words it is the tangent vector to the ``time line'', namely the directional derivative it generates corresponds to an increment in label time $t$. Remarkably, the label time $t$ does not correspond to physical time, the measurement of which would imply the knowledge of the space-time metric; rather, it is a mere label denoting the different Cauchy hypersurfaces.\footnote{Note that if we do not fix the metric, the lapse of time, say $\Delta t$, dividing two different spatial hypersurfaces, $\Sigma^{3}_{t}$ and $\Sigma^{3}_{t+\Delta t}$, is completely general and correlated with the integral curve of $t^{\mu}$, which generates the embedding diffeomorphism. So it is not referable to any real physical measurement, being correlated to gauge transformations.}  

So, the embedding diffeomorphisms is completely arbitrary and can be usefully parametrized by decomposing the deformation vector in its normal and tangential components with respect to $\Sigma_t$. Specifically, by defining the ``Lapse function'' $N$ and the ``Shift vector'' $N^{\mu}$ as
\begin{subequations}
\begin{align}
N&=-n_{\mu}t^{\mu}=\left(n^{\mu}\nabla_{\mu}t\right)^{-1}\,,
\\
N_{\mu}&=h_{\mu\nu}t^{\nu}\,,
\end{align}
\end{subequations}
we have:
\begin{equation}\label{lap shi dec}
t^{\mu}(y)=\left.\left(\frac{\partial y^{\mu}(t,x)}{\partial t}\right)\right|_{y(t,x)=y_{t}(x)}=N(y)n^{\mu}(y)+N^{\mu}(y)\,.
\end{equation}
It is important to note that in order to generate a consistent foliation the Lapse function has to be monotonic.

At this point a brief digression is in order. It should be easy for the reader familiar with the canonical formalism to imagine that the lapse function and shift vector, being two completely arbitrary functions which parametrize the time flow, will turn out to be Lagrange multipliers. As regarding the 4-metric, four of its entries directly depend on them. So, only six out of the ten component of the metric are dynamical variables. This fact nicely reflects in the canonical theory, where the 3-metric, with its six components, turns out to be the fundamental dynamical variable. In other words, a globally hyperbolic space-time represents the time evolution of a Riemannian 3-metric field on a 3-dimensional abstract manifold \cite{Wal1984}, while the other four components express only the arbitrariness we have in choosing the reference system \cite{LanLif1962}. It is worth recalling that the Einstein equations in vacuum 
\begin{equation}
R_{\mu\nu}=0
\end{equation}
are, in fact, six equations (not ten!), because the Bianchi identity $\nabla_{\mu}R^{\mu}_{\ \nu}=\frac{1}{2}\partial_{\nu}R$ relates four of the ten components of the Ricci tensor. 

Let us now enter in the technical details of the canonical formulation of gravity, enunciating the following 
\begin{lemma}
Let $(M,g_{\mu\nu})$ be a space-time and let $\Sigma$ be a smooth space-like hypersurface in $M$. Let $h_{\mu\nu}$ denote the induced metric on $\Sigma$ and let $D_{\mu}$ denote the covariant derivative operator associated with the metric $h_{\mu\nu}$. Then the action of $D_{\mu}$ is given by the formula
\begin{equation}
D_{\rho}T^{\nu_1\cdots\nu_n}_{\phantom1\phantom1\phantom1\phantom1\phantom1\phantom1\mu_1\cdots\mu_m}=h^{\nu_1}_{\phantom1\alpha_{1}}\cdots h^{\nu_n}_{\phantom1\alpha_{n}} h^{\phantom1\beta_1}_{\mu_{1}}\cdots h^{\phantom1\beta_m}_{\mu_{m}}h_{\rho}^{\phantom1\sigma}\nabla_{\sigma}T^{\alpha_1\cdots\alpha_n}_{\phantom1\phantom1\phantom1\phantom1\phantom1\phantom1\beta_1\cdots\beta_m}\,,
\end{equation}
where $\nabla_{\sigma}$ is the derivative operator associated with $g_{\mu\nu}$.

\emph{Proof.} It is simple to verify that $D_{\mu}$ satisfies the following properties: Linearity, Leibniz rule, Commutativity with contraction, Torsion free ($\forall\,f\in\mathcal{F},\left[D_{\mu},D_{\nu}\right]f=0$) and acts as a directional derivative on scalar functions $f$. Moreover we have:
\begin{equation}\label{3-covariant}
D_{\mu}h_{\rho\sigma}=h_{\rho}^{\phantom1\alpha}h_{\sigma}^{\phantom1\beta}h_{\mu}^{\phantom1\gamma}\nabla_{\gamma}\left(g_{\alpha\beta}+n_{\alpha}n_{\beta}\right)\,,
\end{equation} 
because $\nabla_{\mu}g_{\rho\sigma}=0$ and $h_{\mu\nu}n^{\nu}=0$. Thus $D_{\mu}$ is the unique derivative operator associated with $h_{\mu\nu}$. $\Box$
\end{lemma}

Let us define the \emph{second fundamental form} of $\Sigma^3$, called extrinsic curvature:
\begin{equation}
K_{\mu\nu}=h_{\mu}^{\phantom1\rho}h_{\nu}^{\phantom1\sigma}\nabla_{(\rho}n_{\sigma)}=\frac{1}{2}\,h_{\mu}^{\phantom1\rho}h_{\nu}^{\phantom1\sigma}£_{n}h_{\rho\sigma}=\frac{1}{2}\left(£_{n}h\right)_{\mu\nu}\,,
\end{equation}
where the symbol $£_v$ denotes the Lie derivative with respect to the vector field $v^{\mu}$.
The extrinsic curvature is a spatial vector by definition and represents the parallel transport of the normal vector along the hypersurface $\Sigma$, or the variation of the three-metric along the integral line of $n^{\mu}$. We can easily rewrite the extrinsic curvature in order to make explicit the Lie time derivative of the 3-metric; namely
\begin{align}
\nonumber K_{\mu\nu}&=\frac{1}{2}\left(£_{n}h\right)_{\mu\nu}=\frac{1}{2}h_{\mu}^{\phantom1\rho}h_{\nu}^{\phantom1\sigma}\left(n^{\alpha}\nabla_{\alpha}h_{\rho\sigma}+h_{\rho\alpha}\nabla_{\sigma}n^{\alpha}+h_{\alpha\sigma}\nabla_{\rho}n^{\alpha}\right)
\\\nonumber 
&=\frac{1}{2N}h_{\mu}^{\phantom1\rho}h_{\nu}^{\phantom1\sigma}\left(N n^{\alpha}\nabla_{\alpha}h_{\rho\sigma}+h_{\rho\alpha}\nabla_{\sigma}(N n^{\alpha})+h_{\alpha\sigma}\nabla_{\rho}(N n^{\alpha})\right)
\\
&=\frac{1}{2N}h_{\mu}^{\phantom1\rho}h_{\nu}^{\phantom1\sigma}£_{(t-N)}h_{\rho\sigma}=\frac{1}{2N}\left(\dot h_{\mu\nu}-2D_{(\mu}N_{\nu)}\right)\,,
\end{align}
where we defined $\dot h_{\mu\nu}=h_{\mu}^{\phantom1\rho}h_{\nu}^{\phantom1\sigma}£_{t}h_{\rho\sigma}$ and we used equation (\ref{3-covariant}).

Once defined the covariant derivative operator (\ref{3-covariant}), we can define the curvature tensor of the Cauchy surface $\Sigma$ as usual:
\begin{equation} \ ^{(3)}\!R_{\mu\nu\rho}^{\phantom1\phantom1\phantom1\sigma}\omega_{\sigma}=D_{\mu}D_{\nu}\omega_{\rho}-D_{\nu}D_{\mu}\omega_{\rho}\,.
\end{equation}
Now using the prescription to express the covariant derivative on the 3-dimensional manifold as projection of the 4-dimensional derivative operator we have:
\begin{align}
\nonumber D_{\mu}D_{\nu}\omega_{\rho}&=D_{\mu}\left(h_{\nu}^{\phantom1\sigma}h_{\rho}^{\phantom1\tau}\nabla_{\sigma}\omega_{\tau}\right)=h_{\nu}^{\phantom1\alpha}h_{\rho}^{\phantom1\beta}h_{\mu}^{\phantom1\gamma}\nabla_{\gamma}\left(h_{\alpha}^{\phantom1\sigma}h_{\beta}^{\phantom1\tau}\nabla_{\sigma}\omega_{\tau}\right)
\\\nonumber
&=h_{\nu}^{\phantom1\sigma}h_{\rho}^{\phantom1\tau}h_{\mu}^{\phantom1\gamma}\nabla_{\gamma}\nabla_{\sigma}\omega_{\tau}+h_{\nu}^{\phantom1\alpha}h_{\rho}^{\phantom1\beta}h_{\mu}^{\phantom1\gamma}\nabla_{\gamma}\left(h_{\alpha}^{\phantom1\sigma}h_{\beta}^{\phantom1\tau}\right)\nabla_{\sigma}\omega_{\tau}
\\
&=h_{\nu}^{\phantom1\sigma}h_{\rho}^{\phantom1\tau}h_{\mu}^{\phantom1\gamma}\nabla_{\gamma}\nabla_{\sigma}\omega_{\tau}+h_{\rho}^{\phantom1\tau}K_{\nu\mu}n^{\sigma}\nabla_{\sigma}\omega_{\tau}+h_{\nu}^{\phantom1\sigma}K_{\mu\rho}n^{\tau}\nabla_{\sigma}\omega_{\tau}\,,
\end{align}
where we used the expression of the 3-metric as function of the 4-metric and the normal vector (\ref{h=g+nn}). 

At last, we have all the elements to write down the relation between the 3-dimensional curvature tensor as function of the 4-dimensional Riemann tensor and extrinsic curvature:
\begin{align}
\nonumber\ ^{(3)}\!R_{\mu\nu\rho}^{\phantom1\phantom1\phantom1\sigma}\omega_{\sigma}&=D_{\mu}D_{\nu}\omega_{\rho}-D_{\nu}D_{\mu}\omega_{\rho}=2D_{[\mu}D_{\nu]}\omega_{\rho}
\\\nonumber
&=2h_{[\mu}^{\phantom1\gamma}h_{\nu]}^{\phantom1\sigma}h_{\rho}^{\phantom1\tau}\nabla_{\gamma}\nabla_{\sigma}\omega_{\tau}+2h_{\rho}^{\phantom1\tau}K_{[\nu\mu]}n^{\sigma}\nabla_{\sigma}\omega_{\tau}-2K_{[\mu|\rho|}K_{\nu]}^{\phantom1\sigma}\omega_{\sigma}
\\
&=h_{\mu}^{\phantom1\gamma}h_{\nu}^{\phantom1\beta}h_{\rho}^{\phantom1\tau}h^{\sigma}_{\phantom1\alpha}R_{\gamma\beta\tau}^{\phantom1\phantom1\phantom1\alpha}\omega_{\sigma}-K_{\mu\rho}K_{\nu}^{\phantom1\sigma}\omega_{\sigma}+K_{\nu\rho}K_{\mu}^{\phantom1\sigma}\omega_{\sigma}\,,
\end{align}
where, in the second line, we used $h_{\nu}^{\phantom1\sigma}K_{\mu\rho}n^{\tau}\nabla_{\sigma}\omega_{\tau}=h_{\nu}^{\phantom1\sigma}K_{\mu\rho}\nabla_{\sigma}\left(n^{\tau}\omega_{\tau}\right)-h_{\nu}^{\phantom1\sigma}K_{\mu\rho}\omega_{\tau}\nabla_{\sigma}n^{\tau}=K_{\mu\rho}K_{\nu}^{\phantom1\tau}\omega_{\tau}$ and $K_{[\mu\nu]}=0$. Now, considering that $\omega_{\mu}$ is a common factor, we can write down the \emph{first Gauss-Codacci relation}:
\begin{equation}\label{first G-C}
\ ^{(3)}\!R_{\mu\nu\rho}^{\phantom1\phantom1\phantom1\sigma}=h_{\mu}^{\phantom1\gamma}h_{\nu}^{\phantom1\beta}h_{\rho}^{\phantom1\tau}h^{\sigma}_{\phantom1\alpha}R_{\gamma\beta\tau}^{\phantom1\phantom1\phantom1\alpha}-K_{\mu\rho}K_{\nu}^{\phantom1\sigma}+K_{\nu\rho}K_{\mu}^{\phantom1\sigma}\,.
\end{equation}
With an analogous procedure, we can obtain the \emph{second Gauss-Codacci relation} as well:
\begin{equation}\label{second G-C}
D_{\mu}K^{\mu}_{\phantom1\nu}-D_{\nu}K^{\mu}_{\phantom1\mu}=R_{\mu\rho}n^{\rho}h^{\mu}_{\phantom1\nu}\,.
\end{equation}
At this point, once we have realized that
\begin{equation}
R_{\mu\nu\rho\sigma}h^{\mu\rho}h^{\nu\sigma}=R_{\mu\nu\rho\sigma}\left(g^{\mu\rho}+n^{\mu}n^{\rho}\right)\left(g^{\nu\sigma}+n^{\nu}n^{\sigma}\right)=R+2R_{\nu\sigma}n^{\nu}n^{\sigma}\,,
\end{equation}
we can write the Ricci scalar as
\begin{equation}
R=R_{\mu\nu\rho\sigma}h^{\mu\rho}h^{\nu\sigma}-2R_{\nu\sigma}n^{\nu}n^{\sigma}
\end{equation}
and from the first Gauss-Codacci relation we have:
\begin{equation}\label{R3-R4}
R=\ ^{(3)}\!R+(K^{\mu}_{\phantom1\mu})^2-K_{\mu\nu}K^{\mu\nu}-2R_{\nu\sigma}n^{\nu}n^{\sigma}\,.
\end{equation}
Moreover,
\begin{align}\label{Rnn-kk}
\nonumber R_{\nu\sigma}n^{\nu}n^{\sigma}&=R_{\nu\rho\sigma}^{\phantom1\phantom1\phantom1\rho}n^{\nu}n^{\sigma}=-n^{\nu}\left(\nabla_{\nu}\nabla_{\rho}-\nabla_{\rho}\nabla_{\nu}\right)n^{\rho}
\\
&=(K^{\mu}_{\phantom1\mu})^2-K_{\mu\nu}K^{\mu\nu}-\nabla_{\mu}\left(n^{\mu}\nabla_{\rho}n^{\rho}\right)+\nabla_{\rho}\left(n^{\mu}\nabla_{\mu}n^{\rho}\right)\,.
\end{align}
Therefore, being $\sqrt{-g}=N\sqrt{h}$, from equations (\ref{R3-R4}) and (\ref{Rnn-kk}) we finally obtain the following action for the gravitational field (having dropped the last term in the last line above, which being a total divergence, does not affect the equations of motion):
\begin{equation}
S_{3+1}(N,N^{\mu},h_{\mu\nu})=\frac{1}{2}\int\limits_{\mathds{R}\times\sigma}d t d^3x N\sqrt{h}\left(^{(3)}\!R+K_{\mu\nu}K^{\mu\nu}-(K^{\mu}_{\phantom1\mu})^2\right)\,.
\end{equation}
It is particularly useful to pull back spatial tensors to the hypersurface $\Sigma^3$. This can be easily done by suitably defining the projectors $X^{\mu}_{\alpha}=\left(\partial y^{\mu}(t,x)/\partial x^{\alpha}\right)_{|y_{t}(x)=y(t,x)}$, where $x^{\alpha}$ are spatial coordinates on $\Sigma^3$. In this respect, let us define
\begin{subequations}
\begin{align}
h_{\alpha\beta}&=h_{\mu\nu}X^{\mu}_{\alpha}X^{\nu}_{\beta}=g_{\mu\nu}X^{\mu}_{\alpha}X^{\nu}_{\beta}\,,\label{3-d metric}
\\
K_{\alpha\beta}&=K_{\mu\nu} X^{\mu}_{\alpha}X^{\nu}_{\beta}\,,\label{3-d ex curv}
\end{align}
\end{subequations}
so that we can rewrite the extrinsic curvature as
\begin{equation}\label{extrinsic curvature}
K_{\alpha\beta}=\frac{1}{2N}\left(\dot h_{\alpha\beta}-2D_{(\alpha}N_{\beta)}\right)\,,
\end{equation}
where only spatial indexes appear.

The Lagrangian contains the time derivatives of the 3-metric field through the terms depending on the extrinsic curvature, while no time derivatives of the lapse function and the shift vectors appear. Hence, the Lagrangian is singular, so we expect that the theory generates four primary constraints as we are going to demonstrate.

The next step in the canonical analysis is the definition of the conjugate momenta. Once the space-time has been split, we recall that the fundamental variables are $N$, $N^{\alpha}$ and $h_{\beta\gamma}$, the conjugate momenta of which respectively are
\begin{subequations}
\begin{align}
p^{(N)}&=\frac{\partial\mathcal{L}_{3+1}}{\partial\dot N}=0\,,\label{p=0}
\\
p^{(N)}_{\alpha}&=\frac{\partial\mathcal{L}_{3+1}}{\partial\dot N^{\alpha}}=0\,,\label{pk=0}
\\
p^{\alpha\beta}&=\frac{\partial\mathcal{L}_{3+1}}{\partial\dot h_{\alpha\beta}}=\sqrt{h}\left(K^{\alpha\beta}-h^{\alpha\beta}K\right)\,.
\end{align}
\end{subequations}
So, the phase space is twenty dimensional and coordinatized by the set $N, N^{\alpha}, h_{\beta\gamma}, p^{(N)}, p_{\alpha}^{(N)}, p^{\beta\gamma}$ and equipped with the following symplectic structure:
\begin{subequations}
\begin{align}
\left\{N(t,x),p^{(N)}(t,x^{\prime})\right\}&=\delta(x,x^{\prime})\,,
\\
\left\{N^{\alpha}(t,x),p_{\beta}^{(N)}(t,x^{\prime})\right\}&=\delta^{\alpha}_{\beta}\delta(x,x^{\prime})\,,
\\
\left\{h_{\alpha\beta}(t,x),p^{\gamma\delta}(t,x^{\prime})\right\}&=\delta_{\alpha\beta}^{\gamma\delta}\delta(x,x^{\prime})\,,\label{Ste}
\end{align}
\end{subequations}
where the symbol $\left\{\cdots,\cdots\right\}$ denotes the Poisson brackets, while $\delta_{\alpha\beta}^{\gamma\delta}=\frac{1}{2}\left(\delta^{\gamma}_{\alpha}\delta^{\delta}_{\beta}-\delta^{\gamma}_{\beta}\delta^{\delta}_{\alpha}\right)$.

From the definition of conjugate momenta, we immediately obtain four primary constraints as expected, i.e.
\begin{subequations}\label{primary constraints}
\begin{align}
\mathcal{C}^{(N)}\mathrel{\mathop:}=p^{(N)}&\approx 0\,,\label{p=00}
\\
\mathcal{C}_{\alpha}^{(N)}\mathrel{\mathop:}=p^{(N)}_{\alpha}&\approx 0\,.\label{pk=00}
\end{align}
\end{subequations}

Now, we can perform the Legendre transformation. Because of the presence of primary constraints, we cannot re-express all the velocities as functions of the fundamental variables and their conjugate momenta. This implies that the Hamiltonian, usually defined as
\begin{equation}
H=\sum_i p_i\dot q_i-\mathcal{L}\,,
\end{equation}
where $q_i$ and $p_k$ are the generalized coordinates on the phase space, is not uniquely determined as function of the fundamental variables and momenta. In other words, because of the presence of the primary constraints, the Hamiltonian is well defined only on a restricted region of the phase space determined by the primary constraints. So that,
in order to take into account the restriction of the phase space implied by the four constraints (\ref{primary constraints}), we have to introduce four Lagrange multipliers $\lambda$ and $\lambda^{\alpha}$, which have to be varied independently in the action. We have,
\begin{align}\label{split action}
\nonumber S_{3+1}(N,p_{N},N^{\alpha},p_{\beta}^{(N)},h_{\gamma\delta},p^{\alpha\beta})=&\,\frac{1}{2}\int\limits_{\mathds{R}\times\sigma}d t d^3x\left[p^{\alpha\beta}£_{t}h_{\alpha\beta}+p_{N}\dot N+p_{\gamma}^{(N)}\dot N^{\gamma}\right.
\\
&\left.-NH-N^{\alpha}H_{\alpha}-\lambda\mathcal{C}^{(N)}-\lambda^{\beta}\mathcal{C}^{(N)}_{\beta}\right]\,,
\end{align}
where the super-Hamiltonian $H$ and super-momentum $H^{\alpha}$ are defined as:
\begin{subequations}
\begin{align}
H&=\frac{1}{\sqrt{h}}\left[p_{\alpha\beta}p^{\alpha\beta}-\frac{1}{2}\left(h_{\alpha\beta}p^{\alpha\beta}\right)^2\right]-\sqrt{h}\,^{(3)}\!R=\frac{1}{2}\,G_{\alpha\beta\gamma\delta}p^{\alpha\beta}p^{\gamma\delta}-\sqrt{h}\,^{(3)}\!R\,,
\\
H^{\gamma}&=-2\,D_{\delta}p^{\gamma\delta}\,.
\end{align}
\end{subequations}
Above we have introduced the so called super-metric $G_{\alpha\beta\gamma\delta}=\frac{1}{\sqrt{h}}\left(h_{\alpha\gamma}h_{\beta\delta}+h_{\alpha\delta}h_{\beta\gamma}-h_{\alpha\beta}h_{\gamma\delta}\right)$. 

Finally, we write down the canonical Hamiltonian for the gravitational field, 
\begin{equation}
\mathcal{H}_{\rm can}=\int\limits_{\sigma^3} d^3x\left[NH+N^{\alpha}H_{\alpha}+\lambda \mathcal{C}_{(N)}+\lambda^{\beta}\mathcal{C}_{\beta}^{(N)}\right]\,,
\end{equation}
and go on to discuss the dynamics.

\subsection{Canonical constrained dynamics}\label{CCD}

This Section is devoted to the study of the constrained dynamics of the gravitational field. Namely, starting from the split action (\ref{split action}), we are going to calculate the Hamiltonian equations of motion. At the end, a discussion about the formulation of a well posed initial value problem will follow, in correlation also with a well known issue of canonical quantum gravity, the so-called problem of time. In GR, in fact, once suitable initial conditions have been assigned and the gauge fixed, the time evolution is uniquely determined and depends continuously on the initial data; but, as remarked more than once the $t$ parameter appearing in the equations of motion does not represent physical time, rather it is a label denoting the different Cauchy hypersurfaces. The fact that the Hamiltonian is constrained to weakly vanish in General Relativity, implies that the observables, which must commute with all the constraints, are ``frozen'', namely they do not evolve. The concept of time evolution can be reintroduced in such a ``frozen'' formalism through a physical procedure, relating the evolution to the dynamics of other fields coupled to gravity. We consider this aspect of GR extremely natural, because the theory is describing the dynamics of the space-time itself and not the dynamics of a field on a background. Moreover, from a quantum perspective, the frozen formalism does not affect the interpretation of the outcomes of the theory, which describes the actual quantum state of space-time and, eventually, the transition from one quantum configuration to another. From this perspective, the classical idea of evolution in time has to be completely abandoned, as the concept of classical trajectories has to be abandoned in describing the quantum transitions of an electron in an atom. 

Having in mind the above premise, let me start the canonical analysis by varying action (\ref{split action}) with respect to the Lagrange multipliers $\lambda$ and $\lambda^k$, thus obtaining the primary constraints (\ref{primary constraints}). In order to guarantee that the dynamics of the system is consistent we have to require that the constraints be preserved during the evolution, namely that the Poisson brackets of the constraints with the Hamiltonian vanish. We have,
\begin{subequations}
\begin{align}
\dot{\mathcal{C}}^{(N)}(t,x)&=\left\{\mathcal{H}_{\rm can},\mathcal{C}^{(N)}(t,x)\right\}=H(t,x)\,,
\\
\dot{\mathcal{C}}^{(N)}_{\alpha}(t,x)&=\left\{\mathcal{H}_{\rm can},\mathcal{C}^{(N)}_{\alpha}(t,x)\right\}=H_{\alpha}(t,x)\,,
\end{align}
\end{subequations}
hence a set of secondary constraints have to be imposed, i.e.
\begin{subequations}
\begin{align}
H(t,x)&\approx 0 \label{sup-ham}\,,
\\
H_{\alpha}(t,x)&\approx 0 \label{sup-mom}\,,
\end{align}
\end{subequations}
for all $x\,\in\Sigma$. The above weak equations are called super-Hamiltonian and super-momentum constraints and generate the following algebra
\begin{subequations}\label{algebra}
\begin{align}
\left\{H_{\alpha}\left(t,x\right),H_{\beta}\left(t,x^{\prime}\right)\right\}
&=H_{\beta}(t,x)\partial_{\alpha}\delta\left(x,x^{\prime}\right)-H_{\alpha}(t,x)\partial_{\beta}\delta\left(x^{\prime},x\right)\,,
\\
\left\{H_{\alpha}\left(t,x\right),H\left(t,x^{\prime}\right)\right\}
&=H\left(t,x\right)\partial_{\alpha}\delta\left(x,x^{\prime}\right)\,,
\\
\left\{H\left(t,x\right),H\left(t,x^{\prime}\right)\right\}
&=H^{\beta}\left(t,x\right)\partial_{\beta}\delta\left(x,x^{\prime}\right)-H^{\beta}\left(t,x^{\prime}\right)\partial_{\beta}\delta\left(x^{\prime},x\right)\,.
\end{align}
\end{subequations}
The algebra above reveals that the super-Hamiltonian and super-momentum constraints are the generators of diffeomorphisms: It is worth noting that differently from the usual Yang-Mills gauge theories, the algebra of the constraints has structure functions instead of structure constants. Furthermore, the above relations prevent from the emergence of tertiary constraints. 

The fact that the Poisson brackets between the whole set of constraints weakly vanish indicates that the super-Hamiltonian and super-momentum together with the primary constraints form a set of first class constraints. Interestingly enough, this set can be easily reduced, by taking into account that the primary constraints can be strongly satisfied by considering the lapse function $N$ and the shift vector $N^{\alpha}$ themselves as Lagrange multipliers. In this respect, let us consider the dynamical equations for $N$ and $N^{\alpha}$, i.e.
\begin{subequations}
\begin{align}
\dot{N}(t,x)&=\lambda(t,x)\,,
\\
\dot{N}^{\alpha}(t,x)&=\lambda^{\alpha}(t,x)\,,
\end{align}
\end{subequations}
so, as we can immediately understand, the evolution of the Lapse function and the Shift vector is completely arbitrary, being their time derivatives related to the Lagrange multipliers $\lambda$ and $\lambda^{\alpha}$, which are unspecified functions. Henceforth, the system will be described on the phase space by the remaining variables $h_{\alpha\beta}$ and $p^{\gamma\delta}$, while the Lapse function and the Shift vector are treated as Lagrange multipliers. This automatically solves the primary constraints (\ref{primary constraints}), since the momenta conjugated to $N$ and $N^{\alpha}$ vanish strongly. The dynamical equations for $h_{\alpha\beta}$ and $p^{\gamma\delta}$ can be directly calculated from the reduced Hamiltonian
\begin{equation}\label{reduced split action}
\mathcal{H}=\int\limits_{\Sigma^3}d^3x\left[NH+N^{\alpha}H_{\alpha}\right]\,,
\end{equation}
obtaining:
\begin{subequations}
\begin{align}
\dot{h}_{\alpha\beta}=&2N\,G_{\alpha\beta\gamma\delta}p^{\gamma\delta}+2\nabla_{(\alpha}N_{\beta)}\,, \label{eq per h}
\\[12pt]
\nonumber\dot{p}^{\alpha\beta}=&\frac{1}{2\sqrt{h}}\,N h^{\alpha\beta}\left(p^{\gamma\delta}p_{\gamma\delta}-\frac{1}{2}\,p^{2}\right)-\frac{2N}{\sqrt{h}}\,\left(p^{\alpha\gamma}p_{\gamma}^{\phantom1\beta}-\frac{1}{2}\,pp^{\alpha\beta}\right)
-N\sqrt{h}\left(^{(3)}\!R^{\alpha\beta}-\frac{1}{2}\,^{(3)}\!Rh^{\alpha\beta}\right)
\\[8pt]
&+\sqrt{h}\left(\nabla^{\alpha}\nabla^{\beta}N-h^{\alpha\beta}\nabla^{\gamma}\nabla_{\gamma}N\right)
-2p^{\gamma(\alpha}\nabla_{\gamma}N^{\beta)}+\nabla_{\gamma}\left(N^{\gamma}p^{\alpha\beta}\right)\,,
\label{eq per p}
\end{align}
\end{subequations}
where we ignored all the boundary terms and used equation (\ref{sup-mom}). The system of equations (\ref{sup-ham}), (\ref{sup-mom}), (\ref{eq per h}) and (\ref{eq per p}) is equivalent to the vacuum Einstein equations $R_{\mu\nu}=0$. As remarked above, the super-Hamiltonian and super-momentum constraints are first class (see Eqs. (\ref{algebra})) and reflect the gauge invariance of the theory. Provided that the spatial equations of motion are satisfied, they, in fact, generate a diffeomorphisms flow on the phase space, according to the following identifications:
\begin{subequations}
\begin{align}
\left\{H(N),\dots\right\}&=£_{Nn^{\mu}}(\dots)\,,
\\
\left\{\overset{\rightarrow}{H}(\overset{\rightarrow}{N}),\dots\right\}&=£_{N^{\mu}}(\dots)\,,
\end{align}
\end{subequations}
where we used the following notation:
\begin{equation}
H(N)=\int\limits_{\sigma}d^3x\,N(t,x)H(t,x)\quad\text{and}\quad\overset{\rightarrow}{H}(\overset{\rightarrow}{N})=\int\limits_{\sigma}d^3x\,N^i(t,x)H_i(t,x)\,.
\end{equation} 

In particular, the super-momentum or vectorial constraint is clearly correlated with spatial diffeomorphisms and can be satisfied by introducing the Wheeler superspace.\footnote{The Wheeler superspace is the space of the 3-metrics modulo 3-diffeomorphisms. Namely, two metric fields related by a spatial diffeomorphisms represent the same point on the Wheeler superspace.} The super-Hamiltonian or scalar constraint, instead, represents a serious obstacle toward the canonical quantization of the gravitational field. As we remarked before, it generates diffeomorphisms along the normal vector to the Cauchy hypersurfaces, provided that the spatial equations of motion are satisfied. Interestingly enough, as noted by Wald and Kucha\v{r}, the scalar constraint is strictly analogous to the constraint coming out when one tries to parametrize an original non-constrained theory on fixed background. More specifically, an analogous constraint crops up when one introduces within the Lagrangian a time function which labels the hypersurfaces $\Sigma_t$, starting from an initial hypersurface $\Sigma_0$ and then treats this ``time function'' as a dynamical variable \cite{Wal1984,Kuc1973}. But, as one can easily verify in the case of the point particle in flat space-time \cite{MerMon2004-2}, the parametrized theory is linear in the momentum conjugate to the time function, thus the theory can be easily deparametrized by solving the constraint with respect to this momentum. The scalar constraint of GR is, in stead, quadratic in the momenta, therefore such a deparametrization seems not to be possible, at least in pure gravity.\footnote{As soon as matter fields are considered, a deparametrization is, in fact, possible \cite{BroKuc1995} (see also \cite{MerMon2003,MerMon2003-2,Thi2006}). In this framework, the evolution of the physical system can be interpreted in terms of relational variables \cite{Rov1991-1,Rov1991-2} and the so-called problem of time can be solved, by referring the evolution to the dynamical ``relations'' between distinguished fields.}

In order to understand how many physical degrees of freedom are described by the theory, let us count the number of independent variables. Obviously, the presence of first class constraints associated with the gauge freedom of the theory, indicates that there is more than one set of canonical variables which correspond to a particular physical state. In other words, a physical state is well described by two or more different sets of canonical variables if they are correlated by a gauge transformation. Or, we can rephrase saying, a physical state is represented by the class of equivalence of canonical variables under the symmetries of the theory. Usually, in order to eliminate such an ambiguity in the description of the physical system, a set of gauge conditions are imposed on the canonical variables. The gauge conditions are chosen \emph{ad hoc} and are, sometimes, suggested by the mathematical or physical structure of the theory; but they are not a consequence of the theory. They are external conditions. Nevertheless, they are completely admissible, since they only remove the unphysical degrees of freedom of the theory. Obviously, they have to fulfill some consistency requirements: Firstly, they have to be accessible, i.e. it must exist a transformation which maps the original set of variables to the set satisfying the gauge conditions; secondly, we have to require that the gauge conditions be preserved by the symmetry flow. The two requirements above imply that the number of gauge conditions has to be equal to the number of first class constraints in order to completely fix the gauge. Moreover, the determinant of the matrix constructed by the Poisson brackets between the gauge conditions and the first class constraints has to be different from zero. Remarkably, this is exactly the definition of a set of second class constraints, which can be in principle solved through the Dirac procedure \cite{Dir1964}. At this point, the count of the number of physical degrees of freedom is easy. We have to subtract to the number of canonical variables the total number of second class constraints, or, according to what said before, the number of first class constraints plus the number of gauge conditions plus the second class constraints not coming from the gauge fixing. In other words, we have to subtract twice the number of first class constraints plus the number of independent second class constraints to the number of canonical variables. It is worth noting that in the present case the number of physical degrees of freedom in the phase space is four, which correspond to the two polarization of the graviton in the configuration space.
 
Concluding, we can say that the symmetry group of GR is well implemented in the canonical formalism, which is in this sense generally covariant, even though the covariance of the theory is not manifest as in the Lagrangian formulation. We want also to stress the importance of the invariance under diffeomorphisms, pointing out that every possible observable for this theory must be invariant under this group of symmetries of the action. But the meaning of this statement goes over the usual meaning it has in Yang-Mills gauge theories, because the request of 4-diffeomorphisms invariance involves also the dynamics, therefore the definition of an observable is not only a kinematical problem, it necessarily implies to solve the dynamics. In other words in GR kinematics and dynamics are inextricably bound. Quoting a famous sentence by J. Stachel, we can say that in GR: ``\emph{There is no kinematics without dynamics}''.

\subsection{Initial value formulation for gravity}

As we have shown above, the Hamiltonian of GR is not a true Hamiltonian, but a linear combination of constraints. In particular, if we assume that the spatial components of the Einstein equations are satisfied, then it generates a flow along the integral curve of the deformation vector, namely a gauge flow.\footnote{It is worth stressing that the lapse function and the shift vector entering in the Hamiltonian are arbitrary functions of the space-time points.} Nevertheless, the Cauchy problem can be well posed until the appearance of a singularity affects the consistency of the theory itself.

In order to construct a parallel with the case of electromagnetism, it is worth recalling that the Gauss constraint comes out from the time component of the Maxwell equations, which does not contain second time derivatives. The same considerations are valid for Einstein equations in vacuum. In fact, the equations $G_{\mu\nu}n^{\nu}=0$ do not contain second time derivatives of anyone of the metric component, namely, as in the electromagnetic case, these equations depend only on the initial data: In other words, they represent a restriction on the possible acceptable initial data set. So, we expect that the canonical constraints be contained in the equations $G_{\mu\nu}n^{\nu}=0$, indeed we have:
\begin{subequations}\label{constraints II}
\begin{align}
G_{\mu\nu}n^{\mu}n^{\nu}&=-\frac{H}{2\sqrt{h}}=0\,,
\\
G_{\mu\nu}n^{\mu}e^{\nu}_{\phantom1i}&=\frac{H_{i}}{2\sqrt{h}}=0\,.
\end{align}
\end{subequations}
Therefore, the constraints equations are actually equivalent to four of the Einstein dynamical equations; furthermore the Bianchi identity $\nabla_{\mu}G^{\mu\nu}=\nabla_{\mu}R^{\mu\nu}-1/2\partial^{\nu}R=0$, together with the equations of motion for the spatial components, implies that the constraints are \emph{involutive}. Namely, provided that the super-Hamiltonian and super-momentum constraints are satisfied on the initial Cauchy surface and the equations of motion are satisfied everywhere, then also the constraints are satisfied along the evolution. A very simple argument allows to show what just claimed. Assuming that we have already solved the equations for the spatial components of the gravitational field, then the Bianchi identity represents a relation between the time derivative of the normal components of the Einstein tensor $G_{\mu\nu}n^{\nu}$ and the non-time differentiated components of $G_{\mu\nu}$ and their spatial derivatives. Now, by pulling back equation $\nabla_{\mu}G^{\mu\nu}=0$ on the solution for the spatial components of the Einstein equations and realizing that the spatial part of $G_{\mu\nu}$ vanishes, the Bianchi identity becomes a linear homogeneous system of four first order partial equations for the four unknown functions $G_{\mu\nu}n^{\nu}$. Then, it follows that if $G_{\mu\nu}n^{\nu}$ vanish on the initial slice, they must vanish on any slice.

From the Lagrangian point of view the vanishing of four of the Einstein vacuum equations could appear as an under-determination of the components of the metric field, instead it just reflects the invariance of the theory under reparametrization, as we have already explained before. In other words, the apparent under-determination is not a physical one, exactly as in the electrodynamics case. In fact, if $\phi:M\rightarrow M$ is a diffeomorphisms and $(M,g_{\mu\nu})$ is a solution of the Einstein equations, then $(M,\phi^{*}g_{\mu\nu})$ is a solution too. So, the metric contains four arbitrary components, corresponding to a free choice of reference system. So that the remaining components are exactly six, as the number of the dynamical equations. 

Now, in analogy with the electromagnetic case, we fix the gauge: since the gauge freedom of GR regards the general coordinates transformation, then choosing a gauge means to fix a particular system of coordinates. A suitable choice is the ``harmonic'' gauge, characterized by a system of coordinates satisfying the following equations:
\begin{equation}\label{harm coord}
\nabla_{\mu}\nabla^{\mu}y^{\rho}=0\,.
\end{equation}
This choice does not affect the generality of the procedure. Specifically, in a neighborhood of a portion of space-time covered by an original set of coordinates, say $x^{\mu}$, we can proceed as follows: Firstly, let us note that equation (\ref{harm coord}) has the form of equation (\ref{eq theo}), therefore, assuming as initial data the old set $x^{\mu}$ and its derivative $\nabla_{\mu}x^{\nu}$, we can uniquely solve equation (\ref{harm coord}) in a neighborhood of the portion of space-time covered by the old set of coordinates. Since $\nabla_{\mu}x^{\nu}$ are linearly independent, then also $\nabla_{\mu}y^{\rho}$ are linearly independent and consequently the set $y^{\rho}$ will provide a local coordinates system. The choice of the harmonic coordinates will result in the following equation:
\begin{equation}\label{Stef1}
0=\nabla_{\mu}\nabla^{\mu}y^{a}=\frac{1}{\sqrt{-g}}\partial_{\mu}\left(\sqrt{-g}g^{\mu\nu}\partial_{\nu}y^{a}\right)=\frac{1}{\sqrt{-g}}\left(\partial_{\mu}\sqrt{-g}g^{\mu a}\right)=\partial_{\mu}g^{\mu a}+\frac{1}{2}g^{\mu a}g^{\rho\sigma}\partial_{a}g_{\rho\sigma}\,.
\end{equation}
Furthermore, we have that the Ricci tensor $R_{\mu\nu}$ can be written as
\begin{equation}\label{Stef2}
R_{\mu\nu}=-\frac{1}{2}g^{\rho\sigma}\left[-2\partial_{\sigma}\partial_{(\nu}g_{\mu)\rho}+\partial_{\sigma}\partial_{\rho}g_{\mu\nu}+\partial_{\mu}\partial_{\nu}g_{\rho\sigma}\right]+F_{\mu\nu}\left(g,\partial g\right)\,.
\end{equation}
The above expression emphasizes that the Ricci tensor is, in fact, linear in the second derivatives of the metric tensor, where $F_{\mu\nu}\left(g,\partial g\right)$ contains the non-linear dependence on the metric and its derivatives. So that, considering both Eq. (\ref{Stef1}) and (\ref{Stef2}), we can isolate, in the Einstein equations, the non-linear dependence on the metric and its derivatives, obtaining \cite{Cho-Bru1962}:
\begin{equation}\label{red ein eq}
^{(h)}\!R_{\mu\nu}=R_{\mu\nu}+g_{\rho(\mu}\partial_{\nu)}\nabla_{\sigma}\nabla^{\sigma}y^{\rho}=-\frac{1}{2}g^{\rho\sigma}\partial_{\rho}\partial_{\sigma}g_{\mu\nu}+\tilde{F}_{\mu\nu}(g,\partial g)=0\,.
\end{equation}
Therefore, the Einstein equations in vacuum are equivalent to the system of equations (\ref{red ein eq}) (generally referred as reduced Einstein equations) and (\ref{harm coord}). The reduced Einstein equations are suitable to apply the result of the Leray's theorem (\ref{theo 2}).

In this respect, let $h_{\alpha\beta}$ and $K_{\gamma\delta}$ be the metric and extrinsic curvature of the hypersurface $\Sigma^3$ and let us assume that they satisfy the constraints (\ref{constraints II}). Then, after having chosen a suitable system of coordinates over a portion of the hypersurface $\Sigma^3$, assign as initial data for the metric and its time derivative the set $(h_{\alpha\beta}^0,\dot{h}_{\alpha\beta}^0)$, such that the extrinsic curvature $K_{\alpha\beta}$ on $\Sigma^3$ results from this choice via equation (\ref{extrinsic curvature}). Since the Einstein equations involve all the ten components of the metric field, we have to give the initial values of $g_{00}$ and $g_{0\alpha}$ too. A very simple choice could be $g_{00}=-1$, $g_{0\alpha}=0$ and, as a consequence $K_{\alpha\beta}=\frac{1}{2}\dot{h}_{\alpha\beta}=\frac{1}{2}\dot{g}_{\alpha\beta}$. Now the time derivative of the ``$0$'' components of the metric tensor $\partial g_{0\mu}/\partial t$ remains undetermined by this choice, but they can be fixed via the gauge fixing condition $\nabla_{\mu}\nabla^{\mu}y^{a}=0$ on $\Sigma^3$. In the canonical formalism this means that we initially choose the value of the Lapse function $N=-1$ and Shift vector $N^i=0$, then via the gauge fixing we assign the time evolution of these two geometrical objects. Interestingly enough, by assigning the functional form of the Lapse and Shift one can completely fix the gauge, i.e. the reference system.

Now, let us suppose that the chosen initial conditions are sufficiently near that of flat space-time, then, according to theorem (\ref{theo 2}), we can solve equations (\ref{red ein eq}) in a neighborhood of that portion of $\Sigma^3$ covered by the original set of coordinates. Moreover, $\Sigma^3$ will be a Cauchy surface of the globally hyperbolic space-time generated by the solution.

Furthermore, it is possible to demonstrate that a solution of equations (\ref{red ein eq}) will be a solution of the vacuum Einstein equations in a neighborhood of a portion of space-time where the condition (\ref{harm coord}) holds. This concludes this brief digression about the initial value problem in gravity. The interested reader can find the demonstration of the last statement in \cite{Wal1984}.

Concluding we can say that we can give a prescription to demonstrate that it exists, at least locally, a solution of the Einstein equations, moreover the solution depends continuously on the initial data and the space-time it generates is globally hyperbolic. This demonstration is based on the assumption that the set of initial data is near to that of flat space, but this requirement can be relaxed using a trick. The idea is that any curved space if observed from a sufficiently small scale appears nearly flat, so the trick consists simply in rescaling the initial data metric function via a coordinate transformation if they did not appear sufficiently flat (for details see \cite{Wal1984}).

\section{Ashtekar Canonical Gravity}\label{sec5}
As is well known, the program of canonical quantization is not a rigid algorithm and can be slightly adapted to the theory one is going to construct. In fact, the general program of quantization of classical systems requires to make choices in different steps of the quantization procedure, as is briefly described in the next section. Generally speaking, one could say that the achievement of the desired result, namely the construction of a rigorous quantum theory, depends on the choices made in the different steps of the canonical procedure. The structure of the theory could result remarkably simplified if a smart choice of variables were done, allowing to consistently reduce the difficulties one has to face in the quantization procedure. In other words, a smart choice of fundamental variables could make the theory manageable in view of quantization. 

To be more specific, the introduction of the Ashtekar self-dual $SL(2,C)$ connections \cite{Ash86-87} allows to reduce the phase space of GR to that of a Yang--Mills gauge theory, which can be non-perturbatively quantized by formulating the theory using holonomies and fluxes as fundamental variables.\footnote{The loop formalism does not work properly in Yang--Mills gauge theories, remarkably, the reason of this failure can be traced back to the basic assumption of a continuum space-time. This fact suggests that it may work well in QG, where a discrete space-time naturally emerges.} But, let me follow the natural order of things, clarifying one thing at the time. 

It is possible to demonstrate, in fact, that by introducing the Ashtekar self-dual $SL(2,\mathbb{C})$ connections in the framework of canonical GR, a Gauss constraint, which incorporates the generators of the local Lorentz boosts and rotations in a complex combination, appears besides the vectorial and scalar constraints, both connected with the diffeomorphisms gauge invariance of the theory. Simultaneously, the high non-linearity of the Arnowitt--Deser--Misner (ADM) canonical formulation of GR disappears: the new canonical constraints depend polynomially on the fundamental variables, both in vacuum and in the presence of matter \cite{AshRomTat89}.\footnote{It is worth noting that the standard ADM formulation of GR requires that the metric field is non-degenerate, since it contains the three-dimensional Ricci scalar, which is constructed by the Ricci tensor saturating the indexes with the inverse metric field. In the Ashtekar formulation the same requirement is not mandatory, since the constraints are polynomial. So, we can say that the Ashtekar self-dual formulation of Gravity represents a possible extension of GR allowing the presence of degenerate metrics. Whether or not this extension has any physical relevance, up to my knowledge, is not completely understood yet.} By using the Ashtekar formulation of GR, a background independent quantum theory of gravity was later formulated \cite{Ash91}. But the use of complex fundamental variables generates a serious difficulty connected with the implementation of the \emph{reality conditions} in the quantum theory, which are strictly necessary to ensure that the evolution of the system is real. This difficulty has not been overcome so far and, basically, it can be considered the technical motivation which led to the adoption of the real Ashtekar-Barbero (AB) connections \cite{Ash1987-88,Bar1995} as fundamental variables, instead of the complex ones. The link existing between real and complex variables can be clarified by observing that both are obtainable from the ADM canonical pair via a contact transformation. In particular, a suitable canonical transformation allows to introduce a finite complex number, $\beta\neq 0$, namely the Barbero--Immirzi (BI) parameter, in the definition of the new variables, so that they correspond to the (anti)self-dual ones when $\beta=\pm\,i$ and to the real ones for any real value of $\beta$.

Geometrically, the main difference between these two sets of possible new variables for GR is the following: while the complex connections are the projection over the 3-space of the self-dual part of the Ricci spin connections, the real ones are non-trivially related to them, complicating their reconstruction \cite{Sam00}. In fact, the real $SU(2)$ valued connections contain only half of the necessary information for reconstructing the Lorentz valued connections of GR \cite{Thi2001}, motivating also the necessity of fixing the temporal gauge in order to avoid the appearance of second class constraints.\footnote{The temporal gauge fixing consists in rotating the local basis by using a suitable Wigner boost so that, at every instant of ``time'', its zeroth component is parallel to the normal vector to the instantaneous Cauchy hypersurface $\Sigma^3_t$. This condition reduces the local $SO(3,1)$ gauge group to the subgroup of spatial rotations, $SO(3)$, by fixing the boost component of the Lorentz symmetry.} By fixing the temporal gauge, the accessible part of the phase space is determined by first class constraints only \cite{BarSa00} and the system can be quantized through the Dirac procedure. The result is a non-perturbative background independent quantum theory of gravity called \emph{Loop Quantum Gravity} (LQG) \cite{AshLew2004,Rov2004,Thi2001,AshRovSmo1992}.\footnote{LQG besides providing interesting physical predictions as the quantization of areas and volumes \cite{Rov2004} (see also \cite{DitThi07,Rov07-1}), has been able to cure the inevitable singular behavior of classical GR in symmetric spacetimes \cite{Mod04,AshPawSin06,AshBoj06}. Furthermore, the recently obtained results about the graviton propagator have strengthened the physical content of the theory, providing new insights into its non-singular behavior \cite{Rov06,BiaModRov06,ChrLivSpe07}.}

Since the BI parameter has been introduced via a canonical transformation, one can naively believe that different values of $\beta$ correspond to unitary equivalent quantum theories. Strangely enough, this is not the case. In fact, $\beta$ enters in the spectrum of the main geometrical observables of the theory, e.g. the spectra of the area and volume operators, revealing that a one parameter family of non-equivalent quantum theories exists.
As argued by Rovelli and Thiemann \cite{RovThi98}, two dynamically equivalent $SO(3)$-valued connections exist and, as a consequence, an ambiguity appears in the theory, which is essentially expressed by the presence of the BI parameter.

Immirzi suggested that the appearance of the BI parameter in the quantum theory was a consequence of the temporal gauge fixing \cite{Imm97}, so that it would have disappeared in a fully Lorentz covariant theory. But this expectation was not completely confirmed by the so-called \emph{Covariant Loop Quantum Gravity} (CLQG), which is a fully Lorentz covariant quantum theory of gravity, constructed \emph{\`a la} Dirac relaxing the time gauge condition \cite{Ale02}.\footnote{It is worth remarking that the complicated form of the Dirac brackets, necessary to solve the second class constraints (see \cite{Dir1964,HenTei1992}), prevents the fully Lorentz covariant theory from being rigorously formalized.} This approach, in fact, revealed a correlation between the choice of the fundamental variables and the appearance of the BI ambiguity in the quantum theory. In other words, in CLQG different choices of the fundamental variables are possible. In particular, for a geometrically well motivated specific choice of variables the resulting area spectrum no longer depends on the BI parameter \cite{AleVas01}. But, choosing different fundamental variables considered as a direct generalization of the AB connections, the resulting area spectrum turns out to depend on the BI ambiguity again \cite{AleLiv03}, reproducing the result of the gauge fixed theory (see also the interesting paper \cite{CiaMon09}).

Recently, it has been proposed the idea that the BI parameter is, in fact, analogous to the $\theta$-angle of the topological sector of Yang--Mills gauge theories \cite{Mer08,Mer09p} (for a brief description of the topological sector of Yang--Mills gauge theories see Appendix B). This idea, initially proposed by Gambini, Obregon and Pullin \cite{GamObrPul1998},
has been lately reconsidered in relation to the proposal to generalize the action for gravity to contain a topological term \cite{Mer09p,Mer06,Mer06p}. This argument will be better described below in \ref{HACFAG}, but it is worth anticipating that the presence of a topological term, called Nieh--Yan density \cite{NieYan82}, which further generalize the so-called Holst modification \cite{Hol1996}, allows, in fact, to construct a precise analogy between the BI parameter and the $\theta$-angle. Furthermore, by clarifying the large structure of the gauge group involved in gravity through the Nieh-Yan density, it is possible to demonstrate its supposed topological origin and, as a consequence, the existence of non-unitary equivalent quantum theories associated to different values of $\beta$.

Having briefly outlined the AB formulation of GR and some recent aspects concerning the interpretation of the BI parameter, let me now enter in more technical details, starting from the tetrad formulation of gravity and the consequent generalization of the 3+1 splitting procedure.

\subsection{3+1 splitting again}\label{3+1SA}

Let us introduce a one to one map $e:M^{4}\rightarrow TM^{4}_{x}$, which sends tensor fields on $M^{4}$ in tensor fields in the Minkowskian tangent space $TM^{4}_{x}$. The fields $e_{\mu}^{\phantom1a}$ are commonly called tetrads or vierbein (or, more physically, gravitational field! \cite{Rov2004}) and represent a local reference system for space-time. They satisfy the following relations with the metric field:
\begin{equation}\label{tetrametric}
g_{\mu\nu}=\eta_{a b}e_{\mu}^{\phantom1a}e_{\nu}^{\phantom1b}\,,\qquad e_{\mu}^{\phantom1a}\,e^{\mu}_{\phantom1b}=\delta^{a}_{b}\,,\qquad e_{\mu}^{\phantom1a}\,e^{\nu}_{\phantom1a}=\delta_{\mu}^{\nu}\,, 
\end{equation}
where Greek and Latin indexes run from $0$ to $3$, and transform respectively under general coordinates transformations and local Lorentz transformations. The symbol $\eta_{ab}$ denotes the metric tensor in the local Minkowski frame. So, the tetrad fields incorporate all the metric properties of $M$. It is worth noting that the converse is not true. In fact, there are infinitely many choices of the local basis which reproduce the same metric tensor: This is clearly a consequence of the local Lorentz gauge invariance, manifestly present in this formalism. This is also the reason why there are more components in $e_{\mu}^{\phantom1a}$ than in the metric $g_{\mu\nu}$, the difference being exactly six, that is the number of degrees of freedom of the group $SO(3,1)$ representing the number of independent parameters of a Lorentz transformation in the tangent space-time.

As we briefly described in § \ref{EGT}, the presence of a local gauge freedom requires the introduction of a covariant derivative $D_{\mu}$ transforming in the adjoint representation of the gauge group. This implies the introduction of a Lorentz valued connection (often referred as spin connection), here denoted as $\omega^{ab}(x)$ and satisfying the following property $\omega^{ab}(x)=-\omega^{b a}(x)$. The covariant derivative operator $D_{\mu}$ acts on Lorentz valued tensor fields as follows:
\begin{align}\nonumber
D_{\mu}T_{\nu_1\dots\nu_n}^{\ \ \ \ \ \ a_1\dots a_m}&=\partial_{\mu}T_{\nu_1\dots\nu_n}^{\ \ \ \ \ \ a_1\dots a_m}-\sum_{k=1}^{n}\Gamma_{\mu\nu_{k}}^{\rho}T_{\nu_1\dots\nu_{k-1}\rho\nu_{k+1}\dots\nu_n}^{\ \ \ \ \ \ \ \ \ \ \ \ \ \ \ \ \ \ \ \ \ a_1\dots a_m}+\sum_{l=1}^{m}\omega^{a_l} _{\phantom1\phantom2b\mu}T_{\nu_1\dots\nu_n}^{\ \ \ \ \ \ a_1\dots a_{l-1}b a_{l+1}\dots a_m}
\\
&=\nabla_{\mu}T_{\nu_1\dots\nu_n}^{\ \ \ \ \ \ a_1\dots a_m}+\sum_{l=1}^{m}\omega^{a_l} _{\phantom1\phantom2b\mu}T_{\nu_1\dots\nu_n}^{\ \ \ \ \ \ a_1\dots a_{l-1}b a_{l+1}\dots a_m}\,,
\end{align}
where $\Gamma_{\mu\nu}^{\rho}$ denotes the affine connection. Now, requiring the compatibility of the above defined covariant derivative operator with the tetrad basis, we can extrapolate the expression of the spin connection as function of the local basis vectors:
\begin{equation}\label{spin connections}
D_{\mu}e_{\nu}^{\phantom1a}=0\quad\Longrightarrow\quad\omega^{a}_{\phantom1b\mu}=e^{\phantom1a}_{\rho}\nabla_{\mu}e^{\rho}_{\phantom1b},
\end{equation}
where $\nabla_{\mu}$ as usual satisfies the metric compatibility condition, i.e. $\nabla_{\mu}g_{\rho\sigma}=0$. 

As we are going to show, the same conclusion can be derived from the solution of the second Cartan structure equation, which will be extensively used below. In this respect, from the expression given above for the four dimensional spin connection as function of the tetrad fields, it is easy to derive the following equation:
\begin{equation}\label{tor1}
\nabla_{[\mu}e_{\rho]}^{\phantom1a}=-\omega^{\,\,a b}_{[\mu}\,e_{\rho] b}\,.
\end{equation}
If the torsion-less condition holds, namely
\begin{equation}
\nabla_{\mu}\nabla_{\nu}f=\nabla_{\nu}\nabla_{\mu}f\,,
\end{equation}
and remembering that the $\nabla$ operator is compatible with the metric, we can easily deduce
\begin{equation}\label{tor2}
\partial_{[\mu}e_{\rho]}^{\phantom1a}=-\omega^{\,\,a b}_{[\mu}\,e_{\rho] b}\,,
\end{equation} 
which recalling the definition of exterior derivative of n-forms can be rewritten as:
\begin{equation}\label{torsionless condition}
d e^{a}+\omega^{a}_{\phantom1 b}\wedge e^{b}=0\,.
\end{equation}
The above equation is called homogeneous second Cartan structure equation and completely determines the spin connection as function of the gravitational field.\footnote{For the reader's convenience, we collected a description of the forms formalism in Appendix \ref{App A}, where he/she can also find the Hilbert-Palatini and matter actions translated in the forms language.} It is worth noting that in the case of non-vanishing torsion, the affine connection $\Gamma^{\rho}_{\mu\nu}$ has also an antisymmetric part, then the second Cartan structure equation generalizes to:
\begin{equation}\label{II Cartan structure}
d e^{a}+\omega^{a}_{\phantom1 b}\wedge e^{b}=T^a\,,
\end{equation}
where the torsion 2-form $T^a$ is defined as
\begin{equation}
T^a_{\mu\nu}=e_{\rho}^{\phantom1a}\left(\Gamma^{\rho}_{\mu\nu}-\Gamma^{\rho}_{\nu\mu}\right)\,.
\end{equation}
In this more general case, the solution of the second Cartan structure equation provides the full spin connection as a sum of its torsion-less component $\omega^{ab}[e]$ plus a contortion term $K^{ab}$, namely
\begin{equation}
\Omega^{ab}[e,\dots]=\omega^{ab}[e]+K^{ab}[e,\dots]\,,
\end{equation}
where the dots indicate that the contortion component can depend on matter fields as in the case of spinors coupled to gravity.

The spin connection 1-form generates, in the usual way, the curvature 2-form $R^{a}_{\phantom1b}$ through the following identification (in what follows let us use the compact notation $d^{(\omega)}(\dots)\equiv d(\dots)+\omega\wedge(\dots)$):
\begin{equation}\label{Riemann}
d^{(\omega)}\circ d^{(\omega)}v^a=R^{a}_{\phantom1b}\wedge v^b\,.
\end{equation}
Explicitly we have
\begin{equation}\label{Riemann tensor}
R^{a}_{\phantom1b}=d\omega^{a}_{\phantom1b}+\omega^{a}_{\phantom1c}\wedge\omega^{c}_{\phantom1b}\,,
\end{equation}
which is called first Cartan structure equation. The curvature tensor satisfies the Bianchi identity
\begin{equation}\label{bianchi1}
d^{(\omega)}R^{a}_{\phantom1b}=0\,,
\end{equation}
which is a consequence of the Jacobi identity applied to the covariant exterior derivatives operator, i.e. $d^{(\omega)}\circ d^{(\omega)}\circ d^{(\omega)}=0$. Another identity can be obtained applying the exterior covariant derivative operator on the left and right hand sides of the second Cartan structure equation (\ref{II Cartan structure}), i.e.
\begin{equation}\label{bianchi ciclic general}
R_{\phantom1b}^{a}\wedge e^{b}=d T^a+\omega^{a}_{\phantom1b}\wedge T^b\,,
\end{equation}
which in the torsion-less case reduces to the Bianchi cyclic identity:
\begin{equation}\label{bianchi cyclic}
R_{\phantom1b}^{a}\wedge e^{b}=0\,.
\end{equation}

It is possible to rewrite the action for gravity in the tetrad formalism by using the relations given in (\ref{tetrametric}), specifically we have
\begin{equation}\label{tetrad action}
S[e]=\frac{1}{2}\int d^4 x\det[e] e^{\mu}_{\ a}e^{\nu}_{\ b}R_{\mu\nu}^{\ \ \ \ ab}\,,
\end{equation}
which can be used as starting point to canonically reformulate the gravitational theory.
Actually, the canonical formulation of tetrad gravity can be straightforwardly deduced from the canonical theory described in Sections \ref{3+1SOST} and \ref{CCD}, by taking into account the presence of an additional local symmetry in this new framework. 

So, in order to construct the canonical theory, let us assume that the space-time is a globally hyperbolic metric manifold $(M=\mathds{R}\times\Sigma^3,g_{\mu\nu}=\eta_{a b}e_{\mu}^{\ a}e_{\nu}^{\ b})$. Let $h_{\mu\nu}=g_{\mu\nu}+n_{\mu}n_{\nu}$ and $K_{\mu\nu}=1/2£_n h_{\mu\nu}$ be respectively the first and second fundamental form of the Cauchy hypersurface $\Sigma^3$, so that, by using the tetrad fields, we can write:
\begin{equation}
h_{\mu\nu}=\eta_{a b}e_{\mu}^{\phantom1a}e_{\nu}^{\phantom1b}+n_{\mu}n_{\nu}=-e_{\mu}^{\phantom1 0}e_{\nu}^{\phantom1 0}+\delta_{i j}e_{\mu}^{\phantom1i}e_{\nu}^{\phantom1j}+n_{\mu}n_{\nu}\,.
\end{equation}
The expression of the first fundamental form above suggests a very simple gauge choice to eliminate \emph{ab initio} some non-dynamical degrees of freedom. Specifically, by operating a Wigner boost it is possible to rotate the local basis in such a way that the component $e_{\mu}^{\phantom1 0}$ results to coincide with the normal vector $n_{\mu}$. This gauge choice, called temporal gauge, reduces the local gauge symmetry to the group of spatial rotations, since the boost component of the original Lorentz group is obliged to vanish by having chosen the direction of one of the component of the local basis.\footnote{It is worth stressing that the fixation of the temporal gauge remarkably simplifies the canonical theory. Nevertheless, we have to say that, the temporal gauge is not a mandatory choice to construct the canonical theory. In fact, we can canonically formulate the theory as well without fixing the gauge, but the technical difficulties one would have to face in solving the second class constraints are far from the scope of this paper.}

By implementing the temporal gauge and projecting the 3-metric on the hypersurface, we have
\begin{equation}\label{3-metric tetrad}
h_{\alpha\beta}=\delta_{i j}e_{\alpha}^{\phantom1i}e_{\beta}^{\phantom1j}\,,
\end{equation}
where Greek indexes $\alpha,\beta,\gamma\dots$ run over $1,2,3$, while Latin indexes $i,j,k,\dots$ are related to the $SO(3)$ (or, via an isomorphism, to the $SU(2)$) local symmetry.
Note that the orthogonality condition $n^{\mu}h_{\mu\nu}=0$ is automatically fulfilled according to the properties of the tetrad basis. This allows us to easily write the components of the tetrad fields and their inverse in the coordinates system $(t,x)$, we have \begin{equation}\label{vierbein}
e_{\mu}^{\phantom1a}=\left(
\begin{array}{cc}
N & N^{\alpha}e_{\alpha}^{\phantom1i}
\\
0 & e_{\alpha}^{\phantom1i}
\end{array}
\right)
\qquad\text{and}\qquad
e^{\nu}_{\phantom1b}=\left(
\begin{array}{cc}
\displaystyle\frac{1}{N} & 0
\\
-\displaystyle\frac{N^{\beta}}{N} & e^{\beta}_{\phantom1j}
\end{array}
\right),
\end{equation}
as before $N$ denotes the Lapse function, while $N^{\alpha}$ the Shift vector. This identification of some of the components of the local basis with the Lapse and Shift is possible by considering that the line element
\begin{equation} ds^2=e_{\mu}^{\phantom1a}e_{a\nu}dx^{\mu}dx^{\nu}=-N^2dt^2+h_{\alpha\beta}\left(N^{\alpha}dt+dx^{\alpha}\right)(N^{\beta}dt+dx^{\beta})\,,
\end{equation}
corresponds to the ADM decomposition of the 4-metric. It is important to note that the 3-metric (\ref{3-metric tetrad}) is invariant under a local $SO(3)$ rotations of the tetrad basis, namely the dreibein carries three degrees of freedom more with respect to the three-metric $h_{\alpha\beta}$. As a consequence also the number of constraints of the canonical theory written in tetradic formalism must increase. In particular, we should have three (first class) constraints more in order to reabsorb the local gauge degrees of freedom connected with the $SO(3)$ symmetry. 

In this respect, consider the second fundamental form. Define the 1-form $K_{\alpha}^{\phantom1 i}$ on $\Sigma^3$ and contract the internal index with the tetrads, i.e.
\begin{equation}
\mathcal{K}_{\alpha\beta}=\delta_{i k}K_{\alpha}^{\phantom1 i}e^k_{\beta}\,.
\end{equation}
Now, it is easy to realize that the symmetric part of the tensor $\mathcal{K}_{\alpha\beta}$ is the extrinsic curvature, namely
\begin{equation}
\mathcal{K}_{(\alpha\beta)}=K_{\alpha\beta}\,,
\end{equation}
while its antisymmetric part, corresponding to the antisymmetric part of the extrinsic curvature, which is naturally symmetric, has to vanish. Then the 1-form $K_{\alpha}^{\phantom1 i}$ must satisfies the following constraint:
\begin{equation}\label{rough rotational}
\mathcal{K}_{[\alpha\beta]}=K_{[\alpha}^{\phantom1 i}e_{\beta]i}\approx 0\,.
\end{equation}
$\mathcal{K}_{[\alpha\beta]}$ is a $3\times 3$ antisymmetric matrix constrained to vanish, so it reabsorbs exactly the three degrees of freedom more that we have introduced with the choice of the tetrads as elementary variables.\footnote{The reader may wonder about the legitimacy of the  weakly vanishing symbol in (\ref{rough rotational}). In this respect, we have to say that $K^i_{\alpha}$ has been generically defined, but only when it satisfies the condition expressed by (\ref{rough rotational}), it can be safely related to the extrinsic curvature. Since the final goal will be to change variables and describe the canonical dynamics through the canonical couple $(K^i_{\alpha},E^{\beta}_k)$, the second being defined below, such a canonical system describes ordinary gravity when Eq. (\ref{rough rotational}) is satisfied. In other words, condition (\ref{rough rotational}) plays exactly the role of a constraint (weak equation) limiting the physically relevant evolution to a restricted region of the enlarged phase space.} 

As a next step, let me introduce the weighted triad fields
\begin{equation}\label{elec field}
E^{\alpha}_i=e\,e^{\alpha}_{\phantom1 i}\,, 
\end{equation}
where $e=\det[e_{\alpha}^{\phantom1 i}]$.\footnote{Notice that the inverse of the weighted triad is divided by the determinant of $e_{\alpha}^{\phantom1 i}$.} We can rewrite the constraint (\ref{rough rotational}) as
\begin{equation}\label{rot constraint}
\mathcal{R}_{i j}=e e_{i}^{\alpha}e_{j}^{\beta}\mathcal{K}_{\alpha\beta}=e e_{i}^{\alpha}e_{j}^{\beta} K_{[\alpha}^{\phantom1 k}e_{\beta]k}=K_{\alpha [i} E^{\alpha}_{j]}\approx 0\,.
\end{equation}

At this point we are ready for changing variables, specifically we can easily rewrite the canonical ADM variables as:
\begin{subequations}
\begin{align}
h_{\alpha\beta}&=\det(E^{\delta}_l)\delta_{i j} E_{\alpha}^{\phantom1i}E_{\beta}^{\phantom1j}\,,\label{new metric}
\\
p^{\alpha\beta}&=2\delta^{(\alpha}_{\delta}E^{\beta) i}E_{\gamma}^{\phantom1k}E^{[\gamma}_{\phantom1\phantom1i}K^{\delta]}_{\phantom1k}\,.\label{new momentum}
\end{align}
\end{subequations}
The new couple of fundamental variables can be introduced in the scalar and vectorial constraints too, we obtain 
\begin{subequations}
\begin{align}
H_{\alpha}&=-2D_{\beta}\left[K_{\alpha}^{\phantom1i}E_{\phantom1i}^{\beta}-\delta^{\beta}_{\alpha}K_{\gamma}^{\phantom1i}E_{\phantom1i}^{\gamma}\right]\,,\label{new vector}
\\
H&=\frac{1}{\left(\det[E_{\alpha}^{\phantom1 i}]\right)^{1/2}}E^{\alpha}_{\phantom1i}E^{\beta}_{\phantom1j}\left(K_{\alpha}^{\phantom1j}K_{\beta}^{\phantom1i}-K_{\alpha}^{\phantom1i}K_{\beta}^{\phantom1j}\right)-\left(\det[E_{\alpha}^{\phantom1 i}]\right)^{1/2}\ ^{(3)}R(E)\,,\label{new scalar}
\end{align}
\end{subequations}
where the Ricci scalar curvature$^{(3)}\!R(E)$ is considered as function of the weighted tetrads. At this point, it is important to demonstrate that the canonical dynamics, described by the new variables on the extended phase space,\footnote{The adjective ``extended'' refers to the fact that the phase space associated to the new couple of variables has six dimensions more with respect to the ADM one.} is equivalent to that described by the usual ADM variables. In this respect, we can easily demonstrate that once the extended phase space is equipped with the natural symplectic structure
\begin{subequations}\label{new symplectic}
\begin{align}
\left\{E^{\phantom1i}_{\gamma}(t,x),K_{\phantom1j}^{\delta}(t,y)\right\}&=\delta_{j}^{i}\delta^{\delta}_{\gamma}\delta(x-y)\,,
\\
\left\{E^{\phantom1i}_{\gamma}(t,x),E^{\phantom1j}_{\delta}(t,y)\right\}&=\left\{K_{\phantom1i}^{\gamma}(t,x),K_{\phantom1j}^{\delta}(t,y)\right\}=0\,,
\end{align}
\end{subequations}
the constraint
\begin{equation}
\mathcal{R}(\alpha)=\int\limits_{\sigma}d^3x\alpha^{i k}K_{\alpha i}E^{\alpha}_{\phantom1j},
\end{equation}
where $\alpha^{i k}$ is an arbitrary antisymmetric matrix function, generates $SO(3)$ rotations. In fact, calculating the Poisson algebra generated by the rotational constraint, we obtain:
\begin{equation}
\left\{\mathcal{R}(\alpha),\mathcal{R}(\alpha^{\prime})\right\}=\mathcal{R}\left(\left[\alpha,\alpha^{\prime}\right]\right)\,,
\end{equation}
which is exactly the algebra of spatial $SO(3)$ rotations. Any Poisson bracket between the rotational constraint and the ADM canonical variables vanishes, simply because the latter are manifestly rotations invariant. The 3-metric tensor $h_{\alpha\beta}$, being a function only of the weighted tetrad fields (see relation (\ref{new metric})), simply satisfies the following relation:
\begin{equation}
\left\{h_{\alpha\beta}(t,x),h_{\gamma\delta}(t,y)\right\}=0\,.
\end{equation}
The case of the canonical momenta is more complicated, finally we have \cite{Thi2001}:
\begin{equation}
\left\{p^{\alpha\beta}(t,x),p^{\gamma\delta}(t,y)\right\}=-\frac{\sqrt{h}}{8}\left[h^{\alpha\gamma}\mathcal{R}^{\beta\delta}+h^{\alpha\delta}\mathcal{R}^{\beta\gamma}+h^{\beta\gamma}\mathcal{R}^{\alpha\delta}+h^{\beta\delta}\mathcal{R}^{\alpha\gamma}\right](t,x)\ \delta(x,y)\,.
\end{equation}
Namely, the above brackets vanish as soon as the rotation constraint is satisfied. In other words, as soon as $\mathcal{R}(\alpha)=0$, the ADM canonical variables, written as functions of the extended phase space variables, $(K_{\alpha}^i,E^{\gamma}_k)$, generate the usual Poisson brackets (\ref{Ste}). At last, with few algebraic passages, we also obtain
\begin{equation}
\left\{p^{\alpha\beta}(t,x),h_{\gamma\delta}(t,y)\right\}=\delta^{\alpha}_{(\gamma}\delta^{\beta}_{\delta)}\delta(x,y)\,.
\end{equation}
Summarizing, the new extended phase space elementary variables $E^{\phantom1i}_{\alpha}$ and $K_{\phantom1i}^{\alpha}$ reduces through the definitions in lines (\ref{new metric}) and (\ref{new momentum}) to the ADM ones, moreover their Poisson brackets mimic the ADM ones as soon as the rotational constraint is satisfied. So we can conclude that the Hamiltonian system described by the action
\begin{align}\label{new split action}
\nonumber S_{3+1}(E,K)=\frac{1}{2}\int\limits_{\mathbb{R}\times\sigma}d t d^3x&\left(K_{\phantom1i}^{\alpha}\overset{\cdot}{E}\,^{\phantom1i}_{\alpha}-NH-N^{\alpha}H_{\alpha}+O^{i k}\mathcal{R}_{i k}\right),
\end{align}
once solved the rotational constraint, $\mathcal{R}_{i k}=0$, is equivalent to that described by the ADM action (\ref{reduced split action}). 

The constraints (\ref{rough rotational}), (\ref{new vector}) and (\ref{new scalar}) are first class, as can be demonstrated with some algebra. They reflect the gauge structure of the theory, indeed they are correlated to the automorphisms of the tangent bundle, namely the $SO(3)$ rotational symmetry, and to the space-time diffeomorphisms.\footnote{It should be clear from what stated above that this statement does not mean that the flow generated by the ``new'' super-Hamiltonian and super-momentum constraints on the extended phase space is a diffeomorphisms. Rather, once the rotational constraint is satisfied, i.e. on the constraint surface determined by the condition $\mathcal{R}_{i j}=0$, it still exists a representation of the diffeomorphisms group. There, in fact, the Hamiltonian flow of the rotational invariant variables $h_{\alpha\beta}$ and $p^{\alpha\beta}$ is equivalent to that generated on the ADM phase space.}

\subsection{Canonical transformations and new variables for gravity}

Let us begin noting that, given the symplectic structure (\ref{new symplectic}), the transformations $E^{\alpha}_{\phantom1j}\ \rightarrow\ E^{\alpha}_{\phantom1j}/\beta$ and $K_{\alpha}^{\phantom1j}\ \rightarrow\ \beta K_{\alpha}^{\phantom1j}$ are canonical, in fact they do not change the symplectic structure. The parameter $\beta$ is called Barbero--Immirzi (BI) parameter and is, in general, a complex number. It is worth noting that the rotational constraint remains invariant under this rescaling, so we can rewrite it as
\begin{equation}
\mathcal{R}_{k}=\epsilon_{k i j}\mathcal{R}^{i j}=\epsilon_{k i j}\,^{(\beta)}\!K_{\alpha}^{\phantom1i}\,^{(\beta)}\!E^{\alpha j}\approx 0,
\end{equation}
where we used the properties of the total antisymmetric symbol $\epsilon_{k i j}$ and we indicated as $^{(\beta)}\!K_{\alpha}^{\phantom1i}$ and $^{(\beta)}\!E^{\alpha}_{\phantom1i}$ the canonical variables rescaled by the BI parameter.

Let us now introduce the connection associated with the $SO(3)$ symmetry (or equivalently $SU(2)$). Namely, we introduce the connection 1-form $\Gamma^{i j}_{\phantom1\phantom2\alpha}$, in order to covariantly derive $SO(3)$ valued tensors. In particular, let us define the connection via its action on the generic tensor $T_{\alpha_1\dots\alpha_n}^{i_1\dots i_m}$ containing both vectorial and $SO(3)$ indexes, i.e.
\begin{equation}   
D_{\beta}T_{\alpha_1\dots\alpha_n}^{\ \ \ \ \ \ \ i_1\dots i_m}=\partial_{\beta}T_{\alpha_1\dots\alpha_n}^{\ \ \ \ \ \ \ i_1\dots i_m}-\sum_{k=1}^{n}\Gamma_{\beta\alpha_{k}}^{\gamma}T_{\alpha_1\dots\alpha_{k-1}\gamma\alpha_{k+1}\dots\alpha_n}^{\ \ \ \ \ \ \ \ \ \ \ \ \ \ \ \ \ \ \ \ \ \ i_1\dots i_m}+\sum_{l=1}^{m}\Gamma^{i_l}_{\phantom1j \beta}T_{\alpha_1\dots\alpha_n}^{\ \ \ \ \ \ \ i_1\dots i_{l-1}j i_{l+1}\dots i_m}\,.
\end{equation}
It can be verified by direct analysis that the generalized covariant derivative above, $D_{\beta}$, sends each smooth $SO(3)$ valued tensor field of type $(p,q)$ to a smooth $SO(3)$ valued tensor of type $(p,q+1)$ on $\Sigma$. As usual, we require that the covariant derivative operator is compatible with the tetrad basis, i.e. it annihilates the field $e_{\alpha}^{\phantom1i}$, namely we have
\begin{equation}\label{spin connection}
D_{\beta}e_{\alpha}^{\phantom1i}=0\quad\Longrightarrow\quad \Gamma^{i j}_{\phantom1\phantom2\alpha}=e^{\beta i}\nabla_{\alpha}e_{\beta}^{\phantom1j}\,.
\end{equation}
The curvature 2-form, $R_{\alpha\beta}^{\phantom1\phantom2\phantom3ij}$, is defined by considering the commutators of two covariant derivatives on an $SO(3)$-valued scalar, i.e.
\begin{equation}\label{curvature 3-dim} 
R_{\alpha\beta\!\phantom1j}^{\phantom1\phantom2\phantom3i}v^{j}=\left[D_{\alpha},D_{\beta}\right]v^i\,\longrightarrow\,R_{\alpha\beta}^{\ \ \ \ i k}=2\partial_{[\alpha}\Gamma_{\ \ \beta]}^{i k}+\Gamma^{i}_{\ j[\alpha}\Gamma^{j k}_{\ \ \beta]}\,.
\end{equation}
The following relation holds:
\begin{equation}
D_{\alpha}E^{\beta}_{j}=\nabla_{\alpha}\left(e\,e^{\beta}_{j}\right)-\Gamma^{k}_{j\alpha}e\,e^{\beta}_{k}=e D_{\alpha}e^{\beta}_{j}=0\,, 
\end{equation}
where we have taken into account the compatibility condition (\ref{spin connection}). Consequently we have,
\begin{equation}\label{geo con}
D_{\alpha}E^{\alpha}_{j}=0\,,
\end{equation}
moreover, being $\partial_{\alpha}E^{\alpha}_{\phantom1j}=\partial_{\alpha}\left(e\,e^{\alpha}_{\phantom1j}\right)=\nabla_{\alpha}E^{\alpha}_{\phantom1j}$, we finally obtain the following important relation:
\begin{equation}\label{rel ab}
D_{\alpha}E^{\alpha}_{j}=\partial_{\alpha}E^{\alpha}_{j}-\Gamma^{k}_{\ j\alpha}E^{\alpha}_{k}=\partial_{\alpha}E^{\alpha}_{j}+\epsilon_{j k}^{\phantom1\phantom2l}\Gamma^{k}_{\alpha}E^{\alpha}_{\phantom1j}=0\,,
\end{equation}
where we defined $\Gamma^{k}_{\alpha}\overset{def}{=}-1/2\,\epsilon^{k}_{\phantom1i j}\Gamma^{i j}_{\phantom1\phantom2\alpha}$. It is worth noting that the above defined $\Gamma^{i}_{\alpha}$ as function of the dreibeins fields can be recasted as function of the weighted triads $E^{\alpha}_{\phantom1j}$, we give below the expression:\footnote{The expression of the 3-dimensional spin connection as function of the densitized triad is the one that we should use in (\ref{new scalar}) to rewrite the Ricci scalar as function of the $E^{\alpha}_i$. This can be easily done considering that $R(E)=\frac{E^{\alpha}_i E^{\beta}_k}{\det[E]}R_{\alpha\beta}^{\ \ \ \ i k}[\Gamma(E)]$, where $R_{\alpha\beta}^{\ \ \ \ i k}[\Gamma(E)]$ is the curvature tensor associated with the connection $\Gamma$ and defined in (\ref{curvature 3-dim}).}
\begin{align}\label{gamma e}
\nonumber\Gamma^{i}_{\alpha}&=\frac{1}{2}\,\epsilon^{i}_{\phantom1j k}E^{\beta k}\left[\partial_{\beta}E_{\alpha}^{\phantom1j}-\partial_{\alpha}E_{\beta}^{\phantom1j}+E^{\gamma j}E_{\alpha l}\partial_{\beta}E_{\gamma}^{\phantom1l}\right]
\\
&+\frac{1}{4}\,\epsilon^{i}_{\phantom1j k}E^{\beta k}\left[2E_{\alpha}^{\phantom1j}\frac{\partial_{\beta}\left(\det[E_{\gamma}^{\phantom1 i}]\right)}{\det[E_{\gamma}^{\phantom1 i}]}-E_{\beta}^{\phantom1j}\frac{\partial_{\alpha}\left(\det[E_{\gamma}^{\phantom1 i}]\right)}{\det[E_{\gamma}^{\phantom1 i}]}\right]\,,
\end{align}
which can be calculated simply by substituting the definition (\ref{elec field}) in the expression (\ref{spin connection}) for the spatial spin connection. We also note that $\Gamma^{k}_{\alpha}$ is not affected by the rescaling $E^{\alpha}_{\phantom1j}\ \rightarrow\ ^{(\beta)}E^{\alpha}_{\phantom1j}=E^{\alpha}_{\phantom1j}/\beta$ of the weighted tetrad fields. By using the strong equation (\ref{rel ab}), we can now replace the rotational constraint (\ref{rot constraint}) with a Gauss constraint, $G_k$, of an $SO(3)$ or $SU(2)$ Yang-Mills gauge theory, as follows 
\begin{align}\label{Gauss}
\nonumber G_{k}&=D_{\alpha}E^{\alpha}_k+\mathcal{R}_k=\partial_{\alpha}\,^{(\beta)}\!E^{\alpha}_{\phantom1k}+\epsilon_{k i}^{\phantom1\phantom2j}\Gamma^{i}_{\alpha}\,^{(\beta)}\!E^{\alpha}_{\phantom1j}+\epsilon_{k i}^{\phantom1\phantom2j}\,^{(\beta)}\!K_{\alpha}^{\phantom1i}\,^{(\beta)}\!E^{\alpha}_{\phantom1j}
\\
&=\partial_{\alpha}\,^{(\beta)}\!E^{\alpha}_{\phantom1k}+\epsilon_{k i}^{\phantom1\phantom2j}\,^{(\beta)}\!\mathcal{A}^{i}_{\alpha}\,^{(\beta)}\!E^{\alpha}_{\phantom1j}\approx 0\,,
\end{align}
where the Ashtekar-Barbero connection, defined as $^{(\beta)}\!\mathcal{A}^{i}_{\alpha}=\Gamma^{i}_{\alpha}+\beta K_{\alpha}^{\phantom1i}$, has been introduced. For $\beta=\pm\,i$ we obtain the original definition of the Ashtekar self-dual variables for gravity. 

Again a change of variables is in order. Specifically, we can replace the canonical couple $(K^{i}_{\alpha},E_{k}^{\beta})$ with the new couple $(^{(\beta)}\!\mathcal{A}^{i}_{\alpha},^{(\beta)}\!E_{k}^{\beta})$ and, once checked that the replacement does not affect the symplectic structure (canonical transformation),\footnote{The only non-trivial check is that the Poisson bracket $\left\{^{(\beta)}\!A_{\phantom1i}^{\gamma}(t,x),\,^{(\beta)}\!A_{\phantom1j}^{\delta}(t,y)\right\}$ in fact vanishes. This is a non-trivial result, mainly based on the fact that $\left\{\Gamma^{i}_{\alpha}(E),K^{j}_{\beta}\right\}=0$.} i.e. 
\begin{subequations}\label{Ash symplectic}
\begin{align}
\left\{^{(\beta)}\!E^{\phantom1i}_{\gamma}(t,x),\,^{(\beta)}\!A_{\phantom1j}^{\delta}(t,y)\right\}&=\delta_{j}^{i}\delta^{\delta}_{\gamma}\delta(x,y)\,,
\\
\left\{^{(\beta)}\!E^{\phantom1i}_{\gamma}(t,x),\,^{(\beta)}\!E^{\phantom1j}_{\delta}(t,y)\right\}&=\left\{^{(\beta)}\!A_{\phantom1i}^{\gamma}(t,x),\,^{(\beta)}\!A_{\phantom1j}^{\delta}(t,y)\right\}=0\,,
\end{align}
\end{subequations}
we can go on to rewrite the canonical constraints as functions of the new variables. 

So, let us define the curvature of the new connections
\begin{equation}\label{def}
\mathcal{F}^{k}_{\alpha\beta}=2\partial_{[\alpha}\ ^{(\beta)}\!\mathcal{A}^{k}_{\beta]}+\epsilon^{k}_{\phantom1i j}\,^{(\beta)}\!\mathcal{A}_{\alpha}^{i}\,^{(\beta)}\!\mathcal{A}_{\beta}^{j},
\end{equation}
satisfying the following relation
\begin{equation}\label{rel1}
\mathcal{F}^{k}_{\alpha\beta}=R_{\alpha\beta}^{\,k}+2D_{[\alpha}\,^{(\beta)}\!K^{\phantom1k}_{\beta]}+\epsilon^{k}_{\phantom1i j}\,^{(\beta)}\!K^{\phantom1i}_{\alpha}\,^{(\beta)}\!K^{\phantom1j}_{\beta}.
\end{equation}
As a useful formula, we rewrite the 3-dimensional Bianchi cyclic identity, $R_{[\alpha\beta}^{\phantom1\phantom1\phantom1i j}e_{\gamma]j}=0$, as 
\begin{equation}
\epsilon^{i\! j}_{\phantom1\, k}R_{[\alpha\beta}^{\,k}E_{\gamma]j}=0\,,
\end{equation} 
which, after some algebra and considering the definition (\ref{elec field}), can be further reduced to 
\begin{equation}\label{rel2}
R^j_{\alpha\beta}E^{\beta}_j=0\,.
\end{equation} 
By using the definition (\ref{def}) and the equations (\ref{rel1}) and (\ref{rel2}), the canonical constraints can be rewrite as follows:
\begin{subequations}\label{can con}
\begin{align}
H_{\alpha}&=\,^{(\beta)}\!E_{i}^{\gamma}\mathcal{F}^{i}_{\alpha\gamma}-\,^{(\beta)}\!K^{i}_{\alpha}G_i\,,
\\
H&=\frac{\beta^2}{\left(\det[E_{\alpha}^{\phantom1 i}]\right)^{1/2}}\,^{(\beta)}\!E^{\alpha}_{i}\,^{(\beta)}\!E^{\gamma}_{j}\,\left[\epsilon^{i j}_{\phantom1\phantom1k}\mathcal{F}^{k}_{\alpha\gamma}-2\left(\beta^{2}+1\right)K_{[\alpha}^{i}K_{\gamma]}^{j}\right]+\frac{\beta^2}{\left(\det[E_{\alpha}^{\phantom1 i}]\right)^{1/2}}\,^{(\beta)}E^{\gamma}_{j}
D_{\gamma}G^j\,.
\end{align}
\end{subequations}
It is interesting to note that both the scalar and vectorial constraints (\ref{can con}) involve the Gauss constraint. Obviously, since the new expressions of the canonical constraints (\ref{Gauss}) and (\ref{can con}) are the consequence of a well defined canonical transformation, the structure of their Poisson brackets is unaffected. In other words, they still represent a set of first class constraints, correlated with the gauge symmetry of the system; they limit the dynamics of the system to a restricted region of the extended phase space. The same dynamics can be obtained directly working with the following set of first class constraints:  
\begin{subequations}\label{new ash constraints}
\begin{align}
&G_i=\mathcal{D}_{\alpha}E^{\alpha}_i=\partial_{\alpha}E^{\alpha}_i+\epsilon_{i j}^{\phantom1\phantom1k}\ ^{(\beta)}\!\mathcal{A}^{j}_{\alpha}E^{\alpha}_{k}\approx 0\,, \label{gauss law}
\\
&C_{\alpha}=E_{i}^{\beta}\mathcal{F}^{i}_{\alpha\beta}\approx 0\,, \label{vector constraint}
\\
&C=\frac{1}{\left(\det[E_{\alpha}^{\phantom1 i}]\right)^{1/2}}E^{\alpha}_{i}E^{\beta}_{j}\,\left[\epsilon^{i j}_{\phantom1\phantom1k}\mathcal{F}^{k}_{\alpha\beta}-2\left(\beta^{2}+1\right)K_{[\alpha}^{i}K_{\beta]}^{j}\right]\approx 0\,, \label{scalar constraint}
\end{align}
\end{subequations}
which are dynamically equivalent to the previous ones. 

It is worth noting that the Gauss law (\ref{gauss law}) and the vector constraint (\ref{vector constraint}) do not depend on the BI parameter $\beta$, while the scalar constraint (\ref{scalar constraint}) is $\beta$-dependent, implying that the physical predictions of the quantum theory will in general depend on the Immirzi parameter. A weird fact is that even physical quantities not directly depending on the Hamiltonian, for example the area operator come out to be $\beta$-dependent. 

As a final remark, the values $\beta=\pm\,i$ corresponding to the complex Ashtekar (anti)self-dual variables are pretty special: The constraints become polynomial, provided that we can someway reabsorb the determinant of the densitized triad in the denominator of the scalar constraint, e.g., by defining a densitized Lapse function \cite{AshRomTat89}.

\subsection{Holst action as Lagrangian formulation of Ashtekar gravity}\label{HACFAG}

So far, we have never introduced the action which correspond to the AB constraints calculated previously. We dedicate this last part of the section to this argument, describing also a recent proposal for a possible generalization of the so-called Holst action.

The Holst action represents an important contribute in understanding the geometrical content of the Ashtekar-Barbero formalism. In \cite{Hol1996}, Holst showed that the AB canonical constraints of GR \cite{Ash1987-88,Bar1995} can be derived by splitting a generalized Hilbert-Palatini action. The Hilbert-Palatini action in tetrad formalism is:\footnote{Let us use differential forms in order to be more concise in formulating this argument. Some details are given in Appendix \ref{App A}.} 
\begin{equation}
S[e,\omega]=\frac{1}{2}\int e_{a}\wedge e_{b}\wedge \star R^{a b}(\omega)\,.
\end{equation}
The Riemann tensor is a function of the spin-connection, $\omega$, which is considered as a separate distinct variables with respect to the gravitational 1-form $e^a=e^a_{\mu}dx^{\mu}$. In other words, the action as to be varied with respect to both the gravitational field and the spin connections to write down the full set of equations of motion. Specifically, by varying the action with respect to $\omega$ and $e$ respectively, we obtain
\begin{subequations}
\begin{align}\label{se}
de^a+\omega^{a}_{\ b}\wedge e^b&=0\,,
\\
\epsilon^{a}_{\ b c d}e^b\wedge R^{c d}(\omega)&=0\,.\label{EE}
\end{align}
\end{subequations}
The first one of the above equations is the second Cartan structure equation (\ref{II Cartan structure}), containing the information about the relation between the spin connection and the gravitational field. This can be easily solved and the solution can be put in (\ref{EE}), which are the Einstein equations in vacuum. 

It is worth noting that the presence of matter, in general, affects the Einstein equations by generating a source in the right hand side of (\ref{EE}). Remarkably spinor fields, which interact with gravity through both the gravitational field and the spin connection, generate also a source for the second Cartan structure equation, namely they represent a source for torsion \cite{Mer06}.

We claim that the Holst action
\begin{equation}\label{holst action}
S\left(e,\omega\right)=S_{\rm HP}\left(e,\omega\right)+S_{\rm Hol}\left(e,\omega\right)
=\frac{1}{2}\int e_{a}\wedge e_{b}\wedge\left(\star R^{a b}-\frac{1}{\beta} R^{ab}\right)\,,
\end{equation}
where $\beta$ is the BI parameter, is the starting point to formulate canonical gravity in the AB variables. The Holst action is made up of two parts, the first one is the usual Hilbert-Einstein action, while the second one is an ``on (half-)shell'' vanishing term. Classically, the Holst action is dynamically equivalent to the HP one, indeed, by varying it with respect to the spin connection we get the unmodified second Cartan structure equation, while the variation with respect to the gravitational field gives
\begin{equation}\label{complete einstein equations}
\epsilon^{a}_{\ b c d}\,e^{b}\wedge R^{c d}\left(\omega\right)-\frac{1}{\beta}\,e_{b}\wedge R^{a b}\left(\omega\right)=0\,.
\end{equation}
But as previously demonstrated, the homogeneous second Cartan structure equation (\ref{se}) implies the cyclic Bianchi identity (\ref{bianchi cyclic}), which ensures that the Einstein dynamics is preserved from the Holst modification. 

We note that in the Holst formulation the BI parameter turns out to be a multiplicative constant in front of a on (half-)shell vanishing term, this clarifies why it does not affect the classical dynamics, while it has important effects in the quantum regime as remarked previously. This behavior is reminiscent of the parameter characterizing the topological sector of Yang--Mills gauge theories (see action (\ref{topological sector}) and the comment below; see also Appendix \ref{App B}). If the $\theta$-angle and the BI parameter $\beta$ have an analogous origin, then it must exist a classical framework where the analogy between the two parameters can be made manifest. In the pure gravitational case, in fact, the argument proposed fails to be completely convincing. The Holst modification, in fact, is not a topological density. It does not reduce to a total divergence, rather it is an on-shell identically vanishing term. But the action (\ref{holst action}) can be further generalized to include in the picture also the interesting case of torsional space-times. In particular, in \cite{Mer06,Mer06p}, by introducing spinor matter fields, an interesting hint was given to complete the Holst picture; specifically, the presence of spinors can generate the necessary torsion contribution to generalize the Holst modification and construct a topological term. In other words, by using a non-minimal coupling between spinors and gravity,\footnote{See \cite{Kau08} for the extension to supergravity theories.} it has been,
indirectly, demonstrated that the EC action can be generalized without modifying the classical dynamics by adding the Nieh--Yan topological density \cite{NieYan82}, i.e. 
\begin{align}\label{new action for gravity}
\nonumber S&_{\rm Grav}=S_{\rm HP}\left[e,\omega\right]+S_{\rm NY}\left[e,\omega\right]
=\frac{1}{2}\int e_a\wedge e_b\wedge\star R^{ab}
+\frac{1}{2\beta}\int\left(T^a\wedge T_a-e_a\wedge e_b\wedge R^{a b}\right)\,.
\end{align}
By remembering the definition of the torsion 2-form $T^a=d e^a+\omega^a_{\ b}\wedge e^b$, the NY term can be easily rewritten as a total divergence, i.e.
\begin{equation}\label{Nieh--Yan}
\int\left(T^a\wedge T_a-e_a\wedge e_b\wedge R^{a b}\right)=\int d\left(e_a\wedge T^a\right)\,.
\end{equation}
The modification is now a true topological term related to the Chern--Pontryagin classes \cite{ChaZan97}. This generalization is quite natural \cite{DatKauSen09,Mer09} and has been the starting point to construct a precise analogy between the BI parameter and the $\theta$-angle of Yang--Mills gauge theories, presented in \cite{Mer08}, as well as other recent works \cite{CalMer09,MerTav09}. The structure of the large gauge group, which is supposed to be at the base of the proposed interpretation of the BI parameter is quite subtle; a possible framework is described in \cite{Mer09p}.

Finally, we want to briefly digress on an interesting possibility which has lately attracted much interest. The analogy existing between the $\theta$-angle of Yang--Mills gauge theories and the BI parameter in gravity suggest a further generalization, namely the idea that the BI parameter is actually a field \cite{TavYun09,CalMer09,TorKra09}. Initially, this idea was considered just as a possible generalization of the theory, but recently it has been demonstrated that promoting the BI parameter to be a field could be necessary in order to reabsorb a divergence coming from the chiral anomaly on space-time with torsion \cite{Mer09}. This proposal has an interesting outcome, indeed, once the BI field is coupled to gravity via the Nieh--Yan density, it generates a torsion contribution in the second Cartan structure equation. In the case of pure gravity interacting with the BI field, the net effect of the presence of such a torsion contribution is the appearance of a kinetic term for the BI field, which turns out to behave like a decoupled pseudo-scalar field. A more interesting dynamics appears as soon as we consider the presence of fermion fields. Indeed, the BI field couples to the fermion axial current and, through the chiral anomaly, it interacs with boson fields as well \cite{Mer09,MerTav09,MerTom09}. As it happens for the QCD axion, instantonic effects can provide an extremely small mass to the BI field, which can be easily evaluated \cite{Mer09,MerTav09}, allowing one to extract some interesting cosmological implications \cite{LatMer09}. 

We conclude this section saying that the nature of the BI parameter is still debated; it is, in fact, still argument of active discussion from both the purely classical and quantum perspective. The idea that it is the expectation value of a super-weakly interacting pseudo-scalar field is particularly fascinating and rich from the theoretical point of view.

\section{Quantization Program}\label{sec6}
Previously in this paper, precisely in § \ref{WDWNQTG?}, we discussed some general and very well known arguments which motivates the attempts to formulate a consistent QG. Remarkably, the necessity of a quantum theory of gravity was pointed out by Einstein himself in 1916: More than ninety years later a fully consistent and complete formulation of a quantum gravity theory still lacks.

On the one hand, it appeared immediately obvious from the pioneering works of Dirac, Wheeler, and DeWitt, that the problem of quantizing gravity was much more conceptually involved than other analogous problems regarding the other interactions. This led to the idea that the problem of QG could not be solved separately from the other interactions, namely, that it was inextricably bound to the issue of unification. So, for many years, the problem of quantizing pure gravity marginally interested physicists, more attracted by the attempt of unifying the other interactions or, more recently, by the idea of supersymmetry and extra-dimensions. Another interesting aspect that is worth mentioning is the belief that the main question to answer to construct a consistent quantum gravity was the disappearance of time. In other words, initially it seemed that the so called \emph{frozen formalism} was the main obstacle to obtain a physically consistent formulation of QG. But the peculiar role played by time in ordinary quantum mechanics is mainly correlated with the, let's say, evolutionary interpretation of physical theories that we have been used to by classical mechanics. The presence of an evolutionary parameter is neither a fundamental request of the quantization procedure, nor a fundamental ingredient for the physical interpretation of the theory. A quantum theory without time can be, in fact, perfectly consistent.

On the other hand, the physical situation a theoretical physicist is called to face in constructing a quantum theory of gravity should appear as the best he/she can imagine. At present, in fact, there is no strong experimental constraint on the quantum gravity regimes. Naively, one could expect that a rich variety of different consistent theories has been formulated so far, on the contrary, we do not have anyone. Most likely the reason is the double nature of GR, namely it is the field theory describing gravity and, simultaneously, it is the theory describing the structure of space-time. Any quantum gravity theory, in fact, has to put together three fundamental dynamical elements, i.e. geometry, gravity and quantum laws. In this perspective, the ordinary quantum theory of field cannot provide any insightful hint as one of the fundamental ingredients lacks, i.e. the dynamical nature of the space-time geometry.

We know, in fact, that as soon as we treat quantum gravity perturbatively, namely neglecting the full dynamics of space-time, as one would do following the prescriptions of quantum field theory, the result we get is a non-renormalizable theory. So, it seems pretty natural to incorporate the full dynamics of space-time in the theory through a non-perturbative approach. But one may wonder if a non-perturbative quantum theory of gravity can actually be a consistent theory. This question is often suggested by a naive analogy based on the behavior at high energy of the Fermi theory for the weak interaction. As is well known, the Fermi model contains a point-like four fermions interaction, which is non-renormalizable. Fermi's model works well at low energy, but it is doomed to fail at high energies. A striking progress was done in this sense by completely reformulating the theory through the introduction of the massive bosons $W^{\pm}$ and $Z^0$ carrying the weak interaction. 

It is often argued that an analogous procedure has to be applied to GR, since its perturbative non-renormalizability points in a direction similar to that of the Fermi model of weak interaction. Nevertheless, this argumentation completely fails in getting an essential difference between the weak and the gravitational interaction, namely the fact that perturbative expansions presuppose that the space-time is a smooth continuum at all the energy scales. But, there is no reason to believe that the classical concept of continuum space-time has to survive at scales of the order of the Plank energy. That is why a non-perturbative approach, able to incorporate the complete dynamics of the geometry of space-time, may safely describe quantum general relativity.    

Furthermore, the failure of the standard perturbation expansion in gravity may well reflect the fact that GR is characterized by a non-trivial fixed point of the renormalization group flow. This extremely fascinating aspect of perturbative QG has been well described during this school by Roberto Percacci, who, in his two lectures, has pointed out that there is a growing evidence that this is exactly the case. Furthermore the requirement that the fixed point should continue to exist also in presence of matter fields constrains the possible couplings in an interesting physical way \cite{PerPer03}.

In general, it is expectable that a consistent quantum theory of gravity is able to remove singularities, replacing them with a well defined quantum state of the gravitational field. Initially, this was just a hope, but now encouraging results exist. They are mainly due to the general quantization program of Loop Quantum Gravity, which faces the problem of quantum gravity merging together the three main ingredients said above. They are still partial results, in the sense that it is possible to remove singularities at least in symmetric space-times, both of cosmological origin, as the Big Bang singularity \cite{AshPawSin06,AshPawSinVan07,Boj08,AshCorSin08,Ash08,Ash09,Boj09,AshWil09} and those resulting from a complete gravitational collapse \cite{AshBoj06,Mod2006,AshTavVar08}. Of course, a theorem establishing a general result about the avoidance of singularities in LQG still lacks.

However, the existing results make us confident that a suitable background independent quantization of the gravitational field can solve the problem of classical singularities; the program of canonical quantization of gravity has exactly this task. But, it is a matter of fact, that so far the complicated structure of the canonical constraints has prevented from making progresses in the full theory. 

Below, we describe the main features of canonical quantization, starting from a brief account of the prescriptions of the Dirac quantization procedure, then we digress on the old Wheeler-DeWitt quantum gravity and finally we go on to briefly introduce some aspects of Loop Quantum Gravity.

\subsection{Dirac quantization procedure}\label{DQP}

The Dirac quantization procedure is a set of prescriptions aiming to consistently face the quantization of constrained physical system. One useful example to understand how the Dirac procedure works is the quantization of the electromagnetic field, which, as remarked in § \ref{EGT} is a gauge theory of the compact $U(1)$ group. 

Canonical quantum electrodynamics is usually constructed by imposing a gauge condition on the electromagnetic potential, as initially suggested by Fermi. Specifically, the quantization usually chosen is the so-called Lorentz gauge $\partial_{\mu}A^{\mu}=0$. By imposing the gauge in the action, the definition of the momenta conjugated to the electromagnetic potential do not generate any primary constraint. So, the theory can be quantized and the gauge condition weakly imposed on the Fock quantum states, reducing the physical degrees of freedom to the two polarizations of the photon. Another possibility is to consider the full canonical classical theory. As we have calculated in § \ref{GSC} a first class Gauss constraint appears and can be treated by fixing the gauge. A possible choice is the Coloumb gauge $\partial_{\alpha}A^{\alpha}=0$. The dynamical variables $(A_{\alpha},E^{\beta})$ have to satisfy both the gauge fixing condition and the Gauss constraint. As we said in Sec. \ref{CCD}, the general conditions that a consistent gauge fixing has to satisfy imply that the Gauss constraint and the gauge condition form a set of second class constraints. Thus, the degrees of freedom in configuration space are exactly reduced to those corresponding to the two polarizations of the photon field. Finally, once the non-physical degrees of freedom have been eliminated, the system can be quantized by promoting the canonical variables to operators, satisfying the relations derived from promoting the Dirac brackets to quantum commutators.

This procedure works well if applied to linear physical systems, but it presents some complicated issues when applied to non-linear systems as, for example, gravity or Yang-Mills $SU(N)$ gauge theories for $N\geq 2$. The reason is connected with the existence of the so called Gribov ambiguities, which are produced by the complicated geometrical aspects of the complete phase space associated with the dynamics of gauge theories. In fact, in general, the geometry of the constraints surface and gauge orbits could be such that the gauge fixing surface cuts some of the orbits more than once and it does not intersect at all some of the others \cite{HenTei1992}. So the gauge fixing surface works properly only locally, in general, it is impossible to find a global suitable gauge condition. This fact is generally referred as the Gribov obstruction and represents a shared characteristic of all the non-abelian gauge theories. Also gravity is affected by this problem, indeed, studying the classical canonical aspects of the theory, we pointed out that the attempts to solve the Cauchy problem in GR reveals the non-existence of global gauge conditions. In fact, it is possible to find a solution of the Cauchy problem at most locally.

For these reasons it is important to develop a theory of first class constraints without being obliged to fix the gauge. The way, suggested by Dirac, is to impose the first class constraints after the quantization, namely directly on the quantum states. In other words, the idea is to set up a Schr\"odinger-like equation by promoting the first class Hamiltonian to a quantum operator acting on the states of the theory. The classical first class constraints, in stead, are imposed on the state functional as supplementary conditions, i.e.
\begin{equation}\label{dirac equation}
\widehat{C}_I\left|\psi\right\rangle=0\,.
\end{equation}
The action of the first class Hamiltonian and the first class constraints on the state functional is dictated by the upgrade of the classical Dirac brackets to quantum commutators. 

In this way, every quantum state remains unchanged under a transformation generated by the constraints, namely we are reintroducing the gauge invariance at a quantum level. In fact the condition above implies, as a consequence, that the quantum states are invariant under finite gauge transformations in the sector connected with the identity,\footnote{The properties of the physical states under ``large gauge transformation'', that is those not connected with the identity, are not contained in the action principle, indeed no constraints are generated by them. So, requiring the invariance of the physical states under this larger class of transformation would be an extra assumption. See Appendix \ref{App B} for more details.} i.e.
\begin{equation}
\exp\left\{i\alpha_I\widehat{C}_I\right\}\left|\psi\right\rangle=\left|\psi\right\rangle\,.
\end{equation}
We are assuming that the set of classical constraints $C_I$ is first class, i.e.
\begin{equation}
\left\{C_I,C_J\right\}=f_{IJ}^{\ \ K}C_K\,,
\end{equation}
if this relation is preserved by the quantization we have
\begin{equation}\label{anomalies}
\left[\widehat{C}_I,\widehat{C}_J\right]=i\hbar\widehat{f}_{IJ}^{\ \ K}\widehat{C}_K\,,
\end{equation}
but in general it is possible that the first class conditions above show the presence of additional terms of quantum mechanical origin, i.e. we could have
\begin{equation}
\left[\widehat{C}_I,\widehat{C}_J\right]=i\hbar\widehat{f}_{IJ}^{\ \ K}\widehat{C}_K+\hbar^2\widehat{A}_{I J}\,.
\end{equation}
If this is the case the physical states, namely the states invariant under finite gauge transformation connected with the identity, must satisfy the additional condition
\begin{equation}
\widehat{A}_{I J}\left|\psi\right\rangle=0\,,
\end{equation}
which has not a classical analogue and in general restrict the phase space too much. In particular if the operator $\widehat{A}_{I J}$ is invertible, it would imply that the space of the physical states is empty. So, on the one hand, we cannot impose such a condition without drastically affecting the content of the theory. But, on the other hand, if we do not pose that condition the operators $\widehat{C}_I$ are not first class any longer; so, they no longer generate gauge transformations. In other words, the gauge invariance is broken at a quantum level, i.e. the quantization of the system has produced a gauge anomaly. Summarizing, if quantum effects break down gauge invariance, then it is meaningless to search for gauge invariant physical states, i.e. we cannot impose equation (\ref{dirac equation}). We can finally say that if a gauge anomaly is present the Dirac quantization method cannot be applied and a different quantization procedure, e.g. BRST, must be considered, with the hope that it could improve the situation in view of a consistent quantum theory.

\subsection{Wheeler-DeWitt equation}\label{WDWE}

The Wheeler-DeWitt equation is essentially the result of the Dirac quantization procedure as applied to gravity. There is one peculiar aspect that makes this argument interesting, namely the fact that the first class Hamiltonian of GR is a combination of constraints; so that the equivalent of the Schr\"odinger equation does not exist in quantum gravity. This aspect is well known and, as remarked previously, it is often referred as the problem of time.\footnote{It is worth stressing that the observables of the theory have to be gauge invariant, namely they must commute with all the constraints. This implies that the observables of the gravitational field commute with the Hamiltonian, which leaves them invariant rather than generating their time evolution as in ordinary quantum mechanics (frozen formalism).} Nevertheless, the quantization can be formally performed, by following the standard procedure.

Firstly, let us define the smeared ADM variables, 
\begin{subequations}
\begin{align}
Q(h)&=\int\limits_{\Sigma}d^3\!x\, h_{\alpha\beta}Q^{\alpha\beta}\,,
\\
P(f)&=\int\limits_{\Sigma}d^3\!x\, p^{\alpha\beta}f_{\alpha\beta}\,,
\end{align}
\end{subequations}
where $Q^{\alpha\beta}$ and $f_{\gamma\delta}$ are smooth tensor valued function, while $h_{\alpha\beta}$ and $p^{\gamma\delta}$ are the canonical variables defined in § \ref{3+1SOST}. As is well known the wave function depends only on half of the elementary variables, since in this case a natural separation between configuration variables and momenta exists, then the ``polarization of the symplectic manifold'' is pretty natural, i.e. the wave function will depend on $Q(h)$. Now, once a suitable quantum configuration space $\mathcal{C}$ has been introduced, it has to be equipped with the structure of a Hilbert space. This consists in choosing a suitable measure $d\mu_0$, in such a way that $\mathcal{C}$ becomes naturally an $L_2$ space. Obviously, the present Hilbert space does not know about the dynamics, so it will be referred as kinematic Hilbert space. Now the quantization proceeds in the usual way, namely requiring that the operator representation of the elementary variables, i.e. $\widehat{Q}(h)$ and $\widehat{P}(f)$, acting as linear operator on a common dense domain of the kinematic Hilbert space $\mathcal{H}(L_2,d\mu_0)$ generate an irreducible representation of the canonical commutation relation. In other words, we require that
\begin{equation}
\left[\widehat{Q}(h),\widehat{P}(f)\right]=i\hbar\widehat{\left\{Q(h),P(f)\right\}}\,.
\end{equation}
As usual, $\widehat{Q}(h)$ operates by multiplication when valuated on the quantum configuration space, i.e.
\begin{equation}
\left\langle h\right|\widehat{Q}(h)\left|\psi\right\rangle=Q(h)\left\langle h\right|\left.\!\!\psi\right\rangle=Q(h)\Psi(h)\,,
\end{equation}
while $\widehat{P}(f)$ acts as a functional derivative operator, i.e.
\begin{equation}
\left\langle h\right|\widehat{P}(f)\left|\psi\right\rangle=\frac{\hbar}{i}\int d^3\!x\,f_{\alpha\beta}\frac{\delta}{\delta h_{\alpha\beta}}\left\langle h\right|\left.\!\!\psi\right\rangle=\frac{\hbar}{i}\int d^3\!x\,f_{\alpha\beta}\frac{\delta}{\delta h_{\alpha\beta}}\Psi(h)\,.
\end{equation}
Now, the naive quantization of the system follows from the translation of the classical constraints to quantum operators, in accordance with the prescription above, i.e.
\begin{align} 
\vec{H}(\vec{N})\approx 0&\qquad\Longrightarrow\qquad \widehat{H}(\vec{N})\Psi\left[Q(h)\right]=0\,,
\\
H(N)\approx 0&\qquad\Longrightarrow\qquad \widehat{H}(N)\Psi\left[Q(h)\right]=0\,.
\end{align}
But this procedure presents a lot of shortcomings, some of them are of a general nature, while others are specific for the gravitational case \cite{Thi2001}; we summarize in what follows the main ones.
\begin{itemize}
	\item We know that in the construction of the quantum phase space functions we can arbitrarily add to the elementary variables terms proportional to the constant $\hbar$ without affecting the classical limit of the theory. This ambiguity in the choice of the phase space function is known as \emph{factor ordering ambiguity}. Divergences can arise in gauge theories where a bad factor ordering is fixed, a simple example is provided by QED. In fact, only after the choice of a suitable factor ordering, the Hamiltonian operator results well defined, being otherwise divergent and nowhere defined.
	\item In general the divergences of an operator are of a worse kind and can be reabsorbed only after a regularization and renormalization procedure. It is worth noting that the renormalization is connected with the free possibility of adding localized terms to the quantum operators.
	\item It is important choose the factor ordering in such a way that the quantum operators be self-adjoint. This is a crucial step in theories with true Hamiltonian, aiming to guarantee that the eigenvalues of the operator be real. It could be neglected in theories with a constrained Hamiltonian like gravity, on condition that the eigenvalue zero is contained in the spectrum. Working with self-adjoint operators is however advantageous.
	\item The Hamiltonian quantum operator of GR depends neither polynomially nor analytically on the metric field. This fact poses a serious problem, because in general operator valued distributions multiplied at the same point gives divergent results; so, a regularization procedure is required. Even worse the presence of distribution in the denominator poses a more difficult problem of formal definition. Anyway, we can try to seek formal solutions of the Wheeler-DeWitt operator, being aware of the fact that a regularization procedure is however required.
	\item Let us suppose to neglect initially the amount of technical issues and bad definitions of the Wheeler-DeWitt approach, hopefully solvable after a suitable regularization and renormalization procedure. Let us suppose that we succeed in finding a solution of this equation, then a conceptual and interpretative issue arises. Indeed, we do not know, in general, how to interpret the result. This fact is strictly connected with the problem of time and, even though a quantum mechanics without time can be constructed, then a further amount of work is required to correctly interpret the result of quantization. 
A possible solution to this interpretative problem is the relational evolution. Namely, a Schr\"odinger-like equation can be constructed in place of the Wheeler-DeWitt one, showing that the presence of time in the quantum equations reflect on the classical theory generating matter fields \cite{MerMon2003-2}. The other way around is to couple matter to gravity, e.g. a free scalar field or a dust of particles and extract a relational time variables related to the momentum of the scalar field, by using, e.g., the Brown-Kucha\v{r} procedure \cite{BroKuc1995,Thi2006}. It is worth noting that in this kind of approaches the evolution parameter is not external with respect to the physical system, as, e.g., in background dependent theories; here the evolution is referred to an ``internal time'', as one would expect in a background independent framework.
	\item Finally we stress that a central issue should be faced and regards the presence of gauge anomalies. In GR, indeed, the problem is particularly complicate, because the group structure constants are replaced by structure function depending on the metric field. So, in computing equation (\ref{anomalies}) it is possible that an anomalous factor comes out.
\end{itemize}

All the above described issues have led to seek for a better formalization of the problem. In particular, since the choice of the elementary variables is dictated only by the convenience and simplicity of the resulting constraints, the use of Ashtekar--Barbero variables turned out to be extremely useful in facing some of these problems. In fact, a consistent anomaly free quantization is possible and has given interesting results. For example, the spectra of regularized self-adjoint operators related to geometrical quantities are exactly in line with the results one would expect from a quantum General Relativity theory.

\subsection{The program of Loop Quantum Gravity}

As we showed in Section \ref{sec5}, canonical GR in the Ashtekar--Barbero formulation is characterized by the following set of first class constraints:
\begin{align}
G_{i}&=\partial_{\alpha}P^{\alpha}_{\phantom1k}+\epsilon_{k i}^{\phantom1\phantom2j}\mathcal{A}^{i}_{\alpha}P^{\alpha}_{\phantom1j}\approx 0\,,
\\
C_{\alpha}&=P_{i}^{\gamma}\mathcal{F}^{i}_{\alpha\gamma}\approx 0\,,
\\
C&=\frac{1}{\left(\det[P_{\alpha}^{\phantom1 i}]\right)^{1/2}}P^{\alpha}_{i}P^{\gamma}_{j}\,\left[\epsilon^{i j}_{\phantom1\phantom1k}\mathcal{F}^{k}_{\alpha\gamma}-2\left(\beta^{2}+1\right)K_{[\alpha}^{i}K_{\gamma]}^{j}\right]\approx 0\,,
\end{align}
where, to simplify the notation, we defined $P_{i}^{\gamma}=\,^{(\beta)}\!\!E_{i}^{\gamma}$ and we dropped the upper left $\beta$ in the connection, i.e. $\mathcal{A}_{\alpha}^i=\beta K_{\alpha}^i+\Gamma_{\alpha}^i$. For convenience we rewrite here the symplectic structure as well
\begin{equation}
\left\{\mathcal{A}_{\alpha}^i(t,x),P_{k}^{\gamma}(t,x^{\prime})\right\}=\delta^{i}_k\delta_{\alpha}^{\gamma}\delta(x,x^{\prime})\,,\quad\left\{\mathcal{A}_{\alpha}^i(t,x),\mathcal{A}^{k}_{\gamma}(t,x^{\prime})\right\}=0\,,\quad\left\{P^{\alpha}_i(t,x),P_{k}^{\gamma}(t,x^{\prime})\right\}=0\,.
\end{equation}
The above formulation of the classical canonical GR is the starting point of LQG. Remarkably, the AB formulation, allowing to rewrite GR as a theory of connections, provides a sort of \emph{kinematical unification} with the other forces. Other interactions are, in fact, successfully described by Yang--Mills gauge theories, namely as theories of connections valued on compact groups of the $SU(N)$ family. Nevertheless, it has to be emphasized that, from a dynamical perspective, a profound different can be immediately recognized: In Yang--Mills theories the metric of space-time plays a central role, e.g. in the n-point functions, while in QG no background metric is assumed \emph{a priori}; better to say that in QG there is no space-time at all. 

According to the Dirac prescriptions previously described, the theory can be quantized by suitably defining a quantum representation of the canonical algebra and then imposing the operator translation of the canonical constraints on the state functional, $\Psi(\mathcal{A})$, representing the states of the theory. A possible choice for the representation is the one suggested by the old Wheeler--DeWitt approach described before, which, even though formally correct, cannot be made rigorous. 

The program of Loop Quantum Gravity goes, in fact, in a different direction. The idea is to
use a different set of fundamental variables,which are more suitable for quantization. In this respect, let us introduce the holonomies, $h_{\gamma}\left[\mathcal{A}\right]$, of the connection $\mathcal{A}^i_{\alpha}$ and the fluxes, $P\left[\Sigma,f\right]$, of the momentum $P_k^{\gamma}$ respectively as
\begin{subequations}
\begin{equation}\label{hol}
h_{\gamma}\left[\mathcal{A}\right]=\mathcal{P}\exp\left\{-\int\limits_{\gamma} \mathcal{A}^i_{\alpha}\tau_i\frac{dx^{\alpha}}{ds}\,ds\right\}\,,
\end{equation}
and 
\begin{equation}
P\left[S,f\right]=\int\limits_{S}P^{\alpha}_if^i\epsilon_{\alpha\beta\gamma}\frac{dx^{\beta}}{ds_1}\frac{dx^{\gamma}}{ds_2}\,ds_1 ds_2\,.
\end{equation}
\end{subequations}
Above, the $2\times 2$ matrices $\tau_i=\frac{1}{2}\sigma_i$ are the generators of the $SU(2)$ group, $\sigma_i$ being the Pauli matrices; while $f^i$ is an $SU(2)$ valued smearing function. The symbol $\gamma$ denotes the parametric oriented curve on which the holonomy is valued; while $S$ represent a 2-dimensional surface in $\Sigma^3$. It is easy to demonstrate that the holonomy has the following properties:
\begin{equation}
h_{\gamma_1\circ \gamma_2}\left[\mathcal{A}\right]=h_{\gamma_1}\left[\mathcal{A}\right]h_{\gamma_2}\left[\mathcal{A}\right]\,\quad\text{and}\quad h_{\gamma^{-1}}\left[\mathcal{A}\right]=h^{-1}_{\gamma}\left[\mathcal{A}\right]\,,
\end{equation}
where $\gamma_1\circ \gamma_2$ corresponds to join together the end point of $\gamma_1$ and the initial point of $\gamma_2$, while $\gamma^{-1}$ denotes a change in the orientation of the curve. Notice that the holonomy (\ref{hol}) is an element of the group $SU(2)$.

Our purpose is to describe the canonical dynamics of the gravitational system by using the new variables defined above. The first step in this program is the evaluation of the Poisson brackets between the new configuration observables $h_{\gamma}\left[\mathcal{A}\right]$ and momenta $P[S,f]$. Specifically, the Poisson brackets between two configuration variables vanish, while, considering that any
edge $\gamma$ with $\gamma\cap S\neq\emptyset$ can be trivially written as the union of elementary edges which either lie in S, or intersect S in exactly one of their end-points, then, for each of these elementary edges $\gamma$ which
intersect S at a point p, we have:
\begin{equation}\left\{h_e\left[\mathcal{A}\right],P(\Sigma,f)\right\}=-\kappa(S,\gamma)\times
\left\{\begin{array}{cc}
h_{\gamma}\left[\mathcal{A}\right]\tau_i f^i(p)&\text{if $p$ is the source of $\gamma$}
\\
-f^i(p)\tau_i h_{\gamma}\left[\mathcal{A}\right]&\text{if $p$ is the target of $\gamma$}
\end{array}\right.\,,
\end{equation}
where
\begin{equation}
\kappa(S,\gamma)=\left\{\begin{array}{cc}
0&\text{if $\gamma\cap\Sigma=\emptyset$\  \text{or}\  $\gamma\cap\Sigma=\gamma$}
\\
\pm 1&\text{if $\gamma\cap\Sigma\neq\emptyset$}
\end{array}\right.\,.
\end{equation}
Notice that the Poisson brackets between two momenta is non-trivial; the reason being related to the fact that new variables are still distributional quantities even though they are smeared, since the smearing is made respectively on one and two dimensional functions for holonomies and fluxes, so that a particular care has to be taken in handling with Poisson brackets involving two momenta. 

According to the usual procedure, the quantum fundamental operators in the auxiliary Hilbert space will be required to satisfy the algebra originating from the commutation relations.
So, one of the main issue one has to face is to find a consistent representation for the quantum algebra. It is worth saying that the properties of the resulting quantum geometry can be extracted through the momentum operators $\widehat{P}\left[\Sigma,f\right]$ in that representation. The momentum operator, in fact, is related to the classical orthonormal basis via the definition $P^{\alpha}_i=\beta^{-1}E^{\alpha}_i$. Surprisingly enough, the diffeomorphisms invariance requirement sort out a unique representation of the quantum algebra. This result is often referred as LOST theorem by the acronym of Lewandowski, Okolow, Sahlmann, and Thiemann \cite{LOST05} (an independent result with the same physical content has been given by Fleischhack \cite{Fle04})

After this brief introduction, a schematic description of how the Dirac procedure applies to gravity in the LQG program is in order.
\begin{itemize}
	\item Holonomies of the connection are chosen as configuration variables. In particular, the auxiliary Hilbert space can be constructed and consists of a set of functionals of the holonomies, square integrable in the Ashtekar--Lewandowski measure.
	\item The Gauss and vectorial constraints have a natural action on the states of the theory. In particular, the space of solutions of the Gauss and vectorial constraints is well understood. 
	\item The situation becomes much more involved as far as the scalar constraint is considered. The main problem is that the scalar constraint is highly non-linear. Some strategies have been developed to deal with the scalar constraint, particularly through the Thiemann's ``master constraint program'' \cite{Thi03}, but many unsolved issues are still present in the theory. Nevertheless, well defined version of the scalar constraint (in symmetric systems) have been constructed, leading to striking results, which answer some long-standing question about (quantum) gravity. 
	\item At the present stage of the development of the theory, physical observables are known only in some special cases.
\end{itemize}
In order to be more specific, let us give a brief account of these steps by introducing the so-called spin-networks representation.

In order to be as clear as possible, let us start by defining an abstract graph, $\Gamma$, which is intended as a collection of paths $\gamma\,\in\,\Sigma$ meeting at most at their end-points. Given a graph $\Gamma$, we denote by $1,2,\cdots,N$ its edges $\gamma$, i.e. $\Gamma=\bigcup_{k=1}^N\gamma_k$. We call cylindrical functions of generalized connections a functional of the holonomies valued on the edges of the graph to complex numbers, \begin{equation}
F:SU(2)^N\to\mathbb{C}\,,
\end{equation}
defined as 
\begin{equation}
\psi_{\Gamma,F}\left[\mathcal{A}\right]=F\left(h_{\gamma_1}\left[\mathcal{A}\right],\cdots,h_{\gamma_N}\left[\mathcal{A}\right]\right)\,.
\end{equation}
As a simple example, consider a closed loop $\gamma$ and the functional
\begin{equation}
W_{\gamma}\left[\mathcal{A}\right]=\psi_{\gamma,{\rm tr}}\left[\mathcal{A}\right]={\rm tr}\left\{h_{\gamma}\left[\mathcal{A}\right]\right\}\,,
\end{equation}
this is often referred as Wilson loop and belongs to the space of cylindrical functions, i.e. $W_{\gamma}\left[\mathcal{A}\right]\,\in\,{\rm Cyl}_{\Gamma}$.

Let us now denote as $\mathcal{F}$ the linear space of all functionals $\psi_{\Gamma,f}[A]$ for all $\Gamma$ and $f$. The space $\mathcal{F}$ can be equipped with a scalar product through the following procedure. Define a new state $\mu_{AL}$ as 
\begin{equation}
\mu_{AL}\left(\psi_{\Gamma,F}\right)=\int\prod_{\gamma\subset\Gamma}d h_{\gamma} F\left(h_{\gamma_1},\dots,h_{\gamma_N}\right)\,,
\end{equation}
where $d h_e$ is the normalized Haar measure of $SU(2)$. The state $\mu_{AL}\left(\psi_{\Gamma,F}\right)$ is normalized, i.e., $\mu_{AL}\left(1\right)=1$, because the Haar measure is normalized, and positive, i.e. 
\begin{equation}
\mu_{AL}\left(\overline{\psi}_{\Gamma,F}\psi_{\Gamma,F}\right)=\int\prod_{\gamma\subset\Gamma}d h_{\gamma} F^{*}\left(h_{\gamma_1},\dots,h_{\gamma_N}\right)F\left(h_{\gamma_1},\dots,h_{\gamma_N}\right)\geq 0\,.
\end{equation}
As a consequence, a scalar product on $\mathcal{F}$ can be defined as
\begin{equation}
\left\langle\psi_{\Gamma,F}|\psi_{\Gamma^{\prime},F^{\prime}}\right\rangle=\mu_{AL}\left(\overline{\psi_{\Gamma,F}}\psi_{\Gamma^{\prime},F^{\prime}}\right)=\int\prod_{e\subset\Gamma\cup\Gamma^{\prime}}d h_e F^{*}\left(h_{e_1},\dots,h_{e_N}\right)F^{\prime}\left(h_{e_1},\dots,h_{e_N}\right)\,.
\end{equation}
Usually $\mu_{AL}$ is called Ashtekar-Lewandowski measure. The above scalar product gives to the kinematical state space the structure of an auxiliary Hilbert space. Furthermore, the kinematical scalar product is invariant under the automorphisms of the local bundle and 3-diffeomorphisms, so that the kinematical state space carries a \emph{unitary} representation of local $SU(2)$ and 3-diffeomorphisms.

At this point, the states of the theory, represented by the functionals $\psi_{\Gamma,F}\,\in\,\mathcal{F}$, have to be restricted by imposing the constraints. In particular, by imposing the Gauss and vectorial constraints, the state of the theory will be invariant under the local $SU(2)$ symmetry, correlated with the (double cover of the) group of spatial rotations, and under the 3-diffs. But, in order to rigorously implement in the quantum theory the following formal equations 
\begin{subequations}
\begin{align}
\widehat{G}_i\psi_{\Gamma,F}\left[\mathcal{A}\right]&=0\,,
\\
\widehat{C}_{\alpha}\psi_{\Gamma,F}\left[\mathcal{A}\right]&=0\,,
\end{align}
\end{subequations}
it is necessary to find a quantum operator representation of the classical elementary variables. In this respect, the introduction of a suitable basis for the kinematical states space is particularly useful. Without entering in the details, we can just use the result of the Peter--Weyl theorem, stating that a basis on the Hilbert space of $L_2$ functions on $SU(2)$ is given by the matrix elements of the irreducible representations of the group.
We indicate the matrix elements in the $j$-representation as $R^{(j)m}_{\ \ \ \ \ n}$, where $m,n,\dots$ denote the matrix elements of the specific representation. Therefore, a basis for each graph $\Gamma$ is simply obtained by ``tensoring'' the basis above, i.e.
\begin{equation}
\psi_{\Gamma,F}[\mathcal{A}]=\sum_{j_1\cdots j_N}f_{j_1\cdots j_N}^{m_1\cdots m_N,n_1\cdots n_N}R^{(j_1)}_{m_1n_1}\left(h_{\gamma_1}\left[\mathcal{A}\right]\right)\cdots R^{(j_N)}_{m_N n_N}\left(h_{\gamma_N}\left[\mathcal{A}\right]\right)\,.
\end{equation}
The symbol $j$ labels the irreducible representation of $SU(2)$, which can be characterized by half-integer spins, pictorially associated to each edge of the graph $\Gamma$. The coefficients of the expansion $f_{j_1\cdots j_N}$ are
restricted by the gauge invariance. We can easily construct a simple example for one of the elements of the sum above by considering a graph $\Gamma$ made up of three edges $\gamma_1,\gamma_2,\gamma_3$, to which we respectively associate the representations $1,\frac{1}{2},\frac{1}{2}$ of the group $SU(2)$, then we have:
\begin{equation}
\psi_{(\gamma_1\cup \gamma_2\cup \gamma_3)}[\mathcal{A}]=R^{(1)}\left(h_{\gamma_1}\left[\mathcal{A}\right]\right)^{ij}R^{(1/2)}\left(h_{\gamma_2}\left[\mathcal{A}\right]\right)_{AB}R^{(1/2)}\left(h_{\gamma_3}\left[\mathcal{A}\right]\right)_{CD}\sigma_i^{AC}\sigma_j^{BD}\,,
\end{equation}
where $i,j=1,2,3$ are vector indexes, while $A,B,C,D=1,2$ are spinor indexes, and $\sigma_i^{AC}$ are the $2\times 2$ Pauli matrices. It is easy to check that the expression above is gauge invariant, in fact, the Pauli matrices are invariant tensors in the tensor product representation $1\otimes 1/2\otimes 1/2$ acting on the nodes of the graph $\Gamma=\gamma_1\cup\gamma_2\cup\gamma_3$.
Generally, we can write a gauge invariant state function as
\begin{equation}\nonumber
\psi_{\Gamma}[\mathcal{A}]=\left(\bigotimes_l R^{(j_l)}\left(h_{e_l}[\mathcal{A}]\right)\right)\cdot\left(\bigotimes_n i_n\right)\,,
\end{equation}
where the invariant tensors $i_n$ assigned on the nodes of the graph are called intertwiners 
between the representations $j_1,\cdots,j_N$ associated to the edges joining in a node.
The graph $\Gamma$, the labels $j_k$ ``coloring'' the links, and the intertwiners $i_n$ ``coloring'' the nodes completely define a state; in particular, a state defined by the triplet $(\Gamma,j_k,i_n)$ is called \emph{spin-network}.

Now, in order to physically characterize the states of the theory, represented by spin-networks, we construct some geometrical operators acting on them, defining their action on the single holonomy (the action on the complete state can be extracted by composition). The first operator we wish to define is the momentum operator. It is easily correlated to the triad, which has a precise geometrical interpretation as stressed above. The momentum $P^{\alpha}_i$ naturally acts on the holonomies as a functional derivative, i.e.
\begin{equation}
\widehat{P}\left[S,f\right]=\frac{\hbar}{i}\int\limits_{S}ds^{\beta}ds^{\gamma}\epsilon_{\alpha\beta\gamma}f^k\frac{\delta}{\delta\mathcal{A}_{\alpha}^k}\,,
\end{equation}
where we introduced the Planck constant $\hbar$ (it is worth recalling that we set $8\pi G=1$ and $c=1$ from the very beginning). 
To compute the result of the action of the momentum operator on the holonomy, let us firstly note that
\begin{equation}
\frac{\delta}{\delta\mathcal{A}^i_{\alpha}(y)}h_{\gamma}\left[\mathcal{A}\right]=\int d s \delta\left(x(s),y\right)\frac{d x^{\alpha}}{d s} h_{\gamma_1}\left[\mathcal{A}\right]\tau_i h_{\gamma_2}\left[\mathcal{A}\right]\,,
\end{equation}
namely, the action of the functional derivative ``cuts'' the link $\gamma$ at the point $y$ where the derivative operator acts, inserting an $SU(2)$ generators in the middle of the holonomies valued on the two resulting pieces of the original holonomy. 
Given that, we easily get
\begin{align}
\widehat{P}\left[S,f\right]h_{\gamma}\left[\mathcal{A}\right]=\frac{\hbar}{i}\int ds^{1}ds^{2}ds^3&\epsilon_{\alpha\beta\gamma}\frac{dx^{\alpha}}{ds^1}\frac{dx^{\beta}}{ds^2}\frac{dx^{\gamma}}{ds^3}
\delta\left(\underline{x}(s^3),\underline{y}(s^1,s^2)\right)f^i h_{\gamma_1}\left[\mathcal{A}\right]\tau_i h_{\gamma_2}\left[\mathcal{A}\right]\,.
\end{align}
Notice that the integration is made in 3-dimensions so that the $\delta$ distribution can be safely integrated, moreover its presence ensures that the above integral vanishes if the link $\gamma$ does not intersect the surface $S$. The points of intersection between links and surfaces are usually called \emph{punctures}. By using the scalar product defined before, one can show that, in fact, the triad operator is self-adjoint.

After having established the action of the triad operator on the holonomy, we can now define an interesting quantum operator, namely the area operator $\widehat{\mathbf{A}}$.
Let me initially refer to a single holonomy in the fundamental representation. Neglecting some (very important) subtleties, we have that:
\begin{align}\nonumber
\widehat{P}_i\left[S\right]\widehat{P}^i\left[S\right]h_{\gamma}\left[\mathcal{A}\right]=-{\hbar}^2h_{\gamma_1}\left[\mathcal{A}\right]\tau^i\tau_i h_{\gamma_2}\left[\mathcal{A}\right]=\frac{3}{4}\hbar^2 h_{\gamma}\left[\mathcal{A}\right]\,.
\end{align}
Considering a general irreducible representation of $SU(2)$, the action of the square of the momentum operator on the holonomy turns out to be
\begin{align}\nonumber
\widehat{P}_i\left[S\right]\widehat{P}^i\left[S\right]R^{(j)}\left(h_{\gamma}\left[\mathcal{A}\right]\right)=-\hbar^2 j\left(j+1\right)R^{(j)}\left(h_{\gamma}\left[\mathcal{A}\right]\right)\,,
\end{align}
where we assumed that the surface $S$ is punctured only once. For a generic surface $S$ in space the situation is slightly more complicated and can be reduced to the previous simple case by the following procedure. Divide the surface $S$ in $N$ cells and consider the full area as a limit, $A_S=\lim_{N\to\infty}A_S^N$, where
\begin{equation}
A_S^N=\beta^2\sum_{I=1}^N\sqrt{|P_{i}(S_I)P^i(S_I)|}\,,
\end{equation}
$P_i(S_I)$ being the flux through the $I$-th cell. The factor $\beta$ stems from the definition of the classical momentum, $P^{\alpha}_i=\,^{(\beta)}\!\!E^{\alpha}_i=E^{\alpha}_i/\beta$, remembering that the area operator has to be defined with the geometrical triad $E^{\alpha}_i$. So that, the quantum area operator then simply becomes $\widehat{A}_S=\lim_{N\to\infty}\widehat{A}_S^N$.

Considering the result obtained above, the area operator turns out to be diagonal in the basis of spin-networks and reintroducing the physical constant, its spectrum is given by
\begin{equation}
\widehat{A}_S\left|\psi\right>=\beta\ell^2_{\rm Pl}\sum_{p}\sqrt{j_p\left(j_p+1\right)}\left|\psi\right>\,,
\end{equation}
where $\ell_{\rm Pl}$ is the Planck length. Notice that the cellular decomposition is made in such a way that, in the limit $N\to\infty$, each cell is punctured at most in a single point.

An analog procedure allows to define the kinematic volume operator
$\widehat{V}$. The classical volume of a region, $\mathcal{R}$, of space can be written as
\begin{equation}
V(\mathcal{R})=\int\limits_{\mathcal{R}}d^3 x\sqrt{\frac{1}{3!}\left|\epsilon_{\alpha\beta\gamma}\epsilon^{i j k}E^{\alpha}_i E^{\beta}_j E^{\gamma}_k\right|}\,,
\end{equation}
which corresponds to a complicated quantum operator. In particular, to calculate the spectrum of the volume operator, a cellular decomposition analog to that performed in the case of the area operator results to be very useful. Specifically, the volume operator 
\begin{equation}
\widehat{\mathbf{V}}=\lim_{N\to\infty}\beta^{3/2}\ell_{\rm Pl}^3\int\limits_{\mathcal{R}}d^3 x\sqrt{\frac{1}{3!}\left|\epsilon_{\alpha\beta\gamma}\epsilon^{i j k}\widehat{P}^{\alpha}_i \widehat{P}^{\beta}_j \widehat{P}^{\gamma}_k\right|}\,,
\end{equation}
can be evaluated and its spectrum results to be discrete depending on the quantum numbers coloring the nodes of the graph. It is worth noting that the Gauss constraint obliges the flux operator at a node to vanish, so that the volume of a three-valent node vanishes as well.

The fact that the quanta of area depend on the quantum numbers associated with the links or edges of the graph, while the quanta of volume depend on the quantum numbers of the nodes, suggests a natural physical interpretation of a graph. Specifically, any node of a graph represent a chunk of volume of the quantum space-time, while links describe the quantum properties of the surfaces between two volumes. This means that quantum space-time at a kinematical level is made up of quanta of volume separated by quanta of area.

It remains to describe how the canonical constraints can be implemented on the quantum states of the theory. This argument deserves to be carefully analyzed and is far from the scopes of this paper. Nevertheless, it is important to say that the Gauss and vectorial constraints can be implemented and solved at the present stage of the development of the theory. They, in fact, have a pretty natural action on spin-networks, but serious difficulties appear as soon as the scalar constraint is regarded. Here I want to digress on the general procedure used to deal with such a problem. 

As we said before, the space $\mathcal{F}$ can be equipped with the structure of an auxiliary Hilbert space by defining a normalized positive defined kinematical scalar product. Our final purpose is to define a physical scalar product, namely between states which satisfy the constraints. In order to give a general brief description of the problem, let us refer to a general constraint $\widehat{\mathcal{C}}$ and define the following projection operator 
\begin{equation}
S_{\widehat{\mathcal{C}}}=\int \delta N e^{i\int d^3x N\widehat{\mathcal{C}}}\,.
\end{equation}
This operator allows to formally define a physical inner product (for details and rigorous procedures see \cite{AshLew2004}). The idea is that $S_{\widehat{\mathcal{C}}}$ is formally equivalent to a delta function of the constraint operator $\widehat{\mathcal{C}}$, so it can select the states of the theory that satisfy the constraint, i.e. 
\begin{equation}
\left(\left\langle\psi_{\Gamma,F}|\psi_{\Gamma^{\prime},F^{\prime}}\right\rangle\right)_{\rm Phys}=\left\langle\psi_{\Gamma,F}|\delta\left({\widehat{\mathcal{C}}}\right)|\psi_{\Gamma^{\prime},F^{\prime}}\right\rangle\,.
\end{equation} 
So, formally, the physical inner product corresponds to the following expression
\begin{equation}
\left\langle\psi_{\Gamma,F}|S_{\widehat{\mathcal{C}}}|\psi_{\Gamma^{\prime},F^{\prime}}\right\rangle=\left\langle\psi_{\Gamma,F}\left|\int \delta N e^{i\int d^3x N\widehat{\mathcal{C}}}\right|\psi_{\Gamma^{\prime},F^{\prime}}\right\rangle=\left(\left\langle\psi_{\Gamma,F}|\psi_{\Gamma^{\prime},F^{\prime}}\right\rangle\right)_{\rm Phys}\,.
\end{equation} 
which can be made rigorous through the group averaging procedure. 

Concluding, we stress that an important result obtained in the framework of LQG is the discreteness of the eigenvalues of geometric operators. But, simultaneously, this fact introduces an interesting issue. In fact, in the classical theory, to associate a precise physical meaning to geometrical quantities as the area and volume of a region of space-time, one has to define the surfaces and regions operationally, e.g., by using matter fields. Once this is done, one can simply calculate values of these observables using the geometrical formulas. An analogous situation characterizes the quantum theory. For instance, the area of the isolated horizon is a Dirac observable in the classical theory and the application of the quantum geometry area formula to this surface leads to physical results \cite{AshLew2004}. In this situation, the operators and their eigenvalues correspond to the proper lengths, areas an
volumes of physical relevant objects. 

Finally, answering to a question asked more than once during the School, it is important to emphasize that no tension exists between the discreteness of the eigenvalues of geometrical operators and Lorentz invariance. A simple example from quantum mechanics should clarify this point. Consider, e.g., the angular momentum operator in ordinary quantum mechanics, its eigenvalues are discrete and this is perfectly compatible with the rotational invariance of the theory.

\newpage

\acknowledgments

The author wants to express his gratitude to the organizers of the 5th International School on Field Theory and Gravitation. Their exquisite hospitality, friendliness, and the stimulating atmosphere they have been able to create, have made this school extremely useful for both students and speakers. Particularly, for me and the other people who experienced the unexpected bath in the alligators' lake, this school has been and will literally be unforgettable. In this respect, I want to thank especially Carlos Pinheiro and Gentil Pires for their priceless help in solving all the bureaucratic troubles and for their friendly support.

I want also to thank Abhay Ashtekar for having suggested my name to the Organizers of the school to substitute him.
\\[20pt]
This research was supported in part by NSF grant PHY0854743, The George A. and Margaret M. Downsbrough Endowment and the Eberly research funds of Penn State.

\newpage

\appendix

\section{Differential forms}\label{App A}
In this appendix, we have collected the main definitions and formulas useful to deal with differential forms. This language has become pretty common in the recent Literature and sometimes confuses students used to the index notation. We should say that in a theory like GR, where the coordinates do not have any physical meaning, differential forms represent the most natural mathematical formalism, even though, in some problems, the index notation is preferable. So, often one has the necessity to switch from one formalism to the other, going from forms to indexes and vice versa. This has induced me to collect some formulas in few pages, with the hope they will be useful to the readers as they has been for me. 

As a disclaimer, I stress that many different notations are used in the Literature, anyone valid and motivated by precise choices. The definitions and formulas below are in accordance with a notation commonly used in Physics and refers to an arbitrary number of dimensions (unless differently specified) and to any signature of the $n$-dimensional manifold. 

Let $M$ be an $n$-dimensional manifold with signature $s$.\footnote{Namely, $s$ corresponds to the number of minus signs in the metric.} Denote the $\left(\begin{array}{c}n\\p\end{array}\right)$-dimensional space of $p$-forms on the cotangent bundle as $\Lambda^{p}(T^{*}M^n)$. Let $e^a=e^a_{\mu}dx^{\mu}$ be a 1-form transforming under the vectorial representation of the local symmetry group $SO(n-s,s)$. The canonical basis for $\Lambda^{p}(T^{*}M^n)$ is naturally induced by the local basis $e^a$, through the wedge product. Specifically, a basis for $p$-forms in $n$ dimensions is given by the collection of all the possible linearly independent $p$-forms which can be formed by wedging the $n$ vectors $e^a$. For example, the natural basis for $3$-forms in four dimensions is made up of the $\left(\begin{array}{c}4\\3\end{array}\right)$ $3$-forms given below:
\begin{equation}
e^{012}=e^{0}\wedge e^{1}\wedge e^{2}\,,\quad e^{013}=e^{0}\wedge e^{1}\wedge e^{3}\,,\quad e^{023}=e^{0}\wedge e^{2}\wedge e^{3}\,,\quad e^{123}=e^{1}\wedge e^{2}\wedge e^{3}\,.
\end{equation}
Any p-form $\eta\,\in\,\Lambda^{p}_{*}(TM^n)$ can be expanded on the canonical basis according to the following definition
\begin{equation}\label{definition forms}
\eta=\frac{1}{p!}\,\eta_{[a_1\cdots a_p]}\,e^{a_1}\wedge\cdots\wedge e^{a_p}\,,
\end{equation}
where the square brackets denote anti-symmetrization. An $n$-form can be naturally integrated on the $n$-dimensional manifold $M$, by considering that it contains the natural volume element according to the following definition
\begin{equation}
e^{a_1}\wedge\cdots\wedge e^{a_n}=e^{\ a_1}_{[\mu_1}\cdots e^{\ a_n}_{\mu_n]}dx^{\mu_1}\wedge\cdots\wedge dx^{\mu_n}=(-1)^{s}\epsilon^{a_1\cdots a_n}dV\,,
\end{equation}
where $dV=\sqrt{|g|}\,dx^1\cdots dx^n$ denotes the volume element and $\epsilon_{a_1\dots a_n}$ the totally antisymmetric symbol, with the condition that $\epsilon_{a_1\dots a_n}=1$ for $a_i<a_{i+1}$.

We introduce now the internal or scalar product between differential $p$-forms and vectors $v$ defined on the tangent bundle $TM$. Let $\omega\,\in\,\Lambda^{p}(T^*M^n)$ and $v=v^a\tilde{e}_a\,\in\,TM$, where the vector fields $\tilde{e}_a=e^{\mu}_{\ a}\partial_{\mu}$ are a local basis on $TM$. By definition we have $e^a\lrcorner\,\tilde{e}_b=\delta^{a}_{b}$. The following prescription allows to evaluate any differential form in particular directions represented by vector fields, obtaining a $(p-1)$-form, according to the following prescription:
\begin{align}\nonumber
\omega(v)&=\frac{1}{p!}\,\omega_{a_1\cdots a_p}v^b\left[e^{a_1}\wedge\cdots\wedge e^{a_p}\right]\lrcorner\, e_b
\\\nonumber
&=\frac{1}{p!}\sum_{i=1}^p(-1)^{p-i}\delta^{a_i}_{b}\omega_{a_1\cdots a_i\cdots a_p}v^b e^{a_1}\wedge\cdots\wedge e^{a_{i-1}}\wedge e^{a_{i+1}}\wedge\cdots\wedge e^{a_p}
\\
&=\frac{1}{(p-1)!}\,(-1)^{p-1}\omega_{b a_1\cdots a_{p-1}}v^b e^{a_1}\wedge\cdots\wedge e^{a_{p-1}}\,,
\end{align}
where in the last line we moved the index saturated with the components of the vector $v$ on the left by using the antisymmetry of the indexes of $\omega$ and renamed the others. By using the above formula we can extract the components of a $p$-form by evaluating it on $p$ vectors of the local basis, i.e. 
\begin{equation}
\eta(e_{a_1},\cdots, e_{a_p})=(-1)^{\frac{p}{2}(p+1)}\eta_{[a_1\cdots a_p]}\,,\qquad\forall\,\eta\,\in\,\Lambda^{p}(T^{*}M^n)\,,
\end{equation}
namely, the p-form $\eta$ is a smooth map that at any point $x\in M$ associates an antisymmetric tensor of type $(0,p)$.

Let us now introduce the exterior or wedge product ``$\,\wedge\,$'' between two generic differential forms. The wedge product is a map $\wedge:\Lambda^{p}(T^*M^n)\times \Lambda^{q}(T^*M^n)\rightarrow \Lambda^{p+q}(T^*M^n)$ ($p+q\leq n$) defined as
\begin{align}\nonumber
\omega\wedge\eta&=\frac{1}{p!q!}\omega_{[a_1\cdots a_p]}\eta_{[b_1\cdots b_q]}e^{a_1}\wedge e^{a_p}\wedge e^{b_1}\wedge e^{b_q}
\\
&=\frac{1}{(p+q)!}\left(\frac{(p+q)!}{p!q!}\,\omega_{[a_1\cdots a_p}\eta_{a_{p+1}\cdots a_{p+q}]}\right)e^{a_1}\wedge e^{a_p}\wedge e^{a_{p+1}}\wedge e^{a_{p+q}}\,,
\end{align}
so that the components of the resulting $(p+q)$-form are
\begin{equation}
\omega\wedge\eta(e_{a_1},\cdots,e_{a_{p+q}})=\frac{(p+q)!}{p!q!}\,\omega_{[a_1\cdots a_p}\eta_{a_{p+1}\cdots a_{p+q}]}\,.
\end{equation}
Another useful operator we want to introduce is the so-called \emph{Hodge dual}, usually denoted by the symbol ``$\,\star\,$''. The Hodge dual is a map $\star:\Lambda^{p}(T^*M^n)\rightarrow \Lambda^{n-p}(T^*M^n)$, acting on the canonical basis according to the following prescription
\begin{align}\label{dual basis}\nonumber
\star\,\left(e^{a_1}\wedge\cdots\wedge e^{a_p}\right)=\frac{1}{(n-p)!}\epsilon^{\ \ \ \ \ \ \ \ \ \ a_1\cdots a_p}_{a_{p+1}\cdots a_{n}}e^{a_{p+1}}\wedge\cdots\wedge e^{a_n}
\\
=\frac{1}{(n-p)!}(-1)^{p(n-p)}\epsilon^{a_1\cdots a_p}_{\ \ \ \ \ \ \ \ a_{p+1}\cdots a_{n}}e^{a_{p+1}}\wedge\cdots\wedge e^{a_n}\,.
\end{align}
Notice that the above definition slightly differs from the standard one, but it results particularly convenient for a reason that will be clear soon. In fact, an interesting consequence of the definition above is that the wedge product of a $p$-form with its Hodge dual generates the volume element according to the following formula:
\begin{equation}\label{star-wedge product}
\star\,\left(e^{a_1}\wedge\cdots\wedge e^{a_p}\right)\wedge\left(e_{b_1}\wedge\cdots\wedge e_{b_p}\right)=p!\delta^{a_1\cdots a_p}_{[b_1\cdots b_p]}d V
\end{equation}
where $dV$ is the natural volume element on the $n$-dimensional manifold defined above. It is worth noting that no dependence on the signature or dimensions appear in the formula above, so that it will be particularly convenient to rewrite actions in terms of differential forms. Notice, also, that operating twice with the Hodge dual one obtains the initial form apart for a possible sign factor, i.e.
\begin{equation}
\star\,\star\,\left(e^{a_1}\wedge\cdots\wedge e^{a_p}\right)=(-1)^{s+p(n-p)}e^{a_1}\wedge\cdots\wedge e^{a_p}\,.
\end{equation}
By using the definitions (\ref{definition forms}) and (\ref{dual basis}), we can easily extract the expression of the dual of a generic $p$-form. Specifically,
\begin{equation}\label{dual}
\star\,\omega=\frac{1}{p!}\,\omega_{a_1\cdots a_p}\star\left(e^{a_1}\wedge\cdots\wedge e^{a_p}\right)=\frac{1}{(n-p)!}\left(\frac{1}{p!}(-1)^{p(n-p)}\omega_{a_1\cdots a_p}\epsilon^{a_1\cdots a_p}_{\ \ \ \ \ \ \ \ \ a_{p+1}\cdots a_{n}}\right)e^{a_{p+1}}\wedge\cdots\wedge e^{a_n}\,.
\end{equation}
In other words, the dual of a generic $p$-form is the $(n-p)$-form of components
\begin{equation}
\star\,\omega(e_{a_1},\cdots e_{a_{n-p}})=\frac{1}{p!}(-1)^{p(n-p)}\omega_{a_1\cdots a_p}\epsilon^{a_1\cdots a_p}_{\ \ \ \ \ \ \ \ \ a_{p+1}\cdots a_{n}}
\end{equation}
So let $\omega$ and $\eta$ $\in\,\Lambda^{p}_{*}(TM^n)$ be two $p$-form, we have by the formula above that:
\begin{equation}\label{Ste1}
\star\,\omega\wedge\eta=\frac{1}{p!}\omega_{a_1\cdots a_p}\eta^{a_1\cdots a_p}d V\,.
\end{equation} 
Hence, apart for the factor $1/p!$, the wedge product in (\ref{Ste1}) corresponds to the scalar product between the components of the two $p$-forms multiplied by the natural volume element. This can be rewritten as:
\begin{equation}
\star\,\omega\wedge\eta=\frac{1}{p!}\left(\omega(e_{a_1},\cdots,e_{a_p}),\eta(e_{a_1},\cdots,e_{a_p})\right)dV\,,
\end{equation}
where the symbol $\left(\dots,\dots\right)$ denotes the internal product. We remark that wedging the $p$-form with the canonical basis the factorial of $p$ disappears.

The exterior derivative operator $d$ is a map from $\Lambda^{p}(T^*M^n)$ to $\Lambda^{p+1}(T^*M^n)$ defined as
\begin{equation}
\Lambda^{p+1}(T^*M^n)\,\ni\,\eta=d\omega=\frac{1}{(p+1)!}\left((p+1)\partial_{[b}\omega_{a_1\cdots a_p]}\right)e^b\wedge e^{a_1}\wedge\cdots\wedge e^{a_p}\,,\quad\omega\,\in\,\Lambda^{p}(T^*M^n)\,,
\end{equation}
where, as usual, we contained in parentheses the components of the resulting $(p+1)$-form. 
By the definition given above we can immediately extract an important property of the exterior derivatives, i.e. $d\circ d=0$, namely the composition of two derivative operators is the vanishing operator. Moreover, assuming that $\omega\,\in\,\Lambda^{p}_{*}(TM^n)$ and $\eta\,\in\,\Lambda^{q}_{*}(TM^n)$, it is very easy to show the following formula
\begin{equation}
d(\omega\wedge\eta)=(d\omega)\wedge\eta+(-1)^p\omega\wedge d\eta\,.
\end{equation}
In general, the presence of a local symmetry requires the definition of a covariant derivative. In this framework a local $SO(s,n-s)$ symmetry is present, therefore we have to define a new exterior derivative operator acting on $SO(s,n-s)$ valued $p$-forms which generates $SO(s,n-s)$ valued $(p+1)$-forms. Namely, by using the language introduced in § \ref{EGT}, the derivative operator has to transform in the adjoint representation of the local symmetry group. In this respect, let us introduce a $SO(s,n-s)$ valued connection $1$-form $\omega^{a b}$ and define the new derivative operator $d^{(\omega)}$ as
\begin{equation}
d^{(\omega)}\cdots=d\cdots+\omega\wedge\cdots\,.
\end{equation}
We claim that the above derivative operator has exactly the property required, as can be easily demonstrated. In order to operatively define the covariant derivative operator, we firstly specify its action on the basis $1$-form $e^a$, we have
\begin{equation}
d^{(\omega)}e^a=d e^a+\omega^{a}_{\ b}\wedge e^b\,,
\end{equation}
which, as can be easily recognized, is the definition of the torsion $2$-form $T^a$. Specifically,
\begin{equation}
T^a:=d^{(\omega)}e^a=d e^a+\omega^{a}_{\ b}\wedge e^b\,.
\end{equation}
So, in the presence of torsion the covariant exterior derivative operator fails in annihilating the basis element. Sometimes, the equation $d^{(\omega)}e^a=T^a$ is referred as second Cartan structure equation. It is worth noting that the composition of two covariant exterior derivative does not trivially vanish, rather we have
\begin{equation}
d^{(\omega)}\circ d^{(\omega)}e^a=R^{a}_{\ b}\wedge e^{b}\,,
\end{equation}
which allows to extract the following expression for the curvature $2$-form
\begin{equation}
R^{a b}=d\omega^{a b}+\omega^{a}_{\ c}\wedge\omega^{c b}\,,
\end{equation}
known as first Cartan structure equation. It is worth remarking that if
\begin{equation}
R^{ab}=0\,\Longrightarrow\, \omega^{a}_{\ b}=\left(\Lambda^{-1}\right)^{a}_{\ c}d\Lambda^{c}_{\ b}\,,
\end{equation}
namely the connection is a pure gauge, $\Lambda^{ab}=-\Lambda^{ba}$ being a representation of the local symmetry group. Then, one can demonstrate that by assuming  
\begin{equation}
R^{ab}=0\,\Longrightarrow\,de^a=0\quad\text{iff}\ T^a=0\,,
\end{equation}
which implies that $e^a=dx^a$, where $x^a$ are functions of the original set of coordinates. Moreover, we have $e^{a}_{\mu}=\partial_{\mu}x^a$, so that the components of the local basis simply represent the soldering forms in flat space between two local arbitrary accelerated reference frames, with the origin placed at the same point of the tangent bundle. 

Two useful identities can be easily derived from the above definitions, i.e.
\begin{subequations}
\begin{align}
d^{(\omega)}R^{a}_{\ b}&=0\,,
\\
d^{(\omega)}T^{a}_{\ b}&=R^{a}_{\ b}\wedge e^{b}\,,
\end{align}
\end{subequations}
respectively known as first and second Bianchi identity.

We refer now to a specific case: we assume that $n=4$ and $s=1$, which means that we are referring to a 4-dimensional pseudo-Riemannian manifold $M^4$, which is locally isomorphic to Minkowski space-time with signature $(-,+,+,+)$, so the local symmetry group is $SO(3,1)$. In this framework the Hilbert-Palatini action for General Relativity can be rewritten as 
\begin{equation}
S(e,\omega)=\frac{1}{2}\int e_a\wedge e_b\wedge\star R^{a b}\,.
\end{equation}
Remembering definition (\ref{definition forms}) and formula (\ref{star-wedge product}) we can easily write:
\begin{equation}\label{demonstration Einstein-Hilbert}
S(e,\omega)=\frac{1}{2}\int e_a\wedge e_b\wedge\star R^{a b}=\frac{1}{2}\int\frac{1}{2}R_{\ \ \ c d}^{a b} e_a\wedge e_b\wedge\star\left(e^c\wedge e^d\right)=\frac{1}{2}\int d^4x\det(e) R(\omega)\,,
\end{equation}
where we used $dV=\det(e)d^4x$.

An analog procedure allows to rewrite also the Dirac action in the formalism of differential forms. But before doing that, we need to define the action of the exterior covariant derivative on a spinor field, which is a $0$-form transforming under the spinor representation of the $SO(3,1)$ local group. We do not enter in the details about the construction of the spinor bundle, we only say that the exterior covariant derivative operator acts on the spinor fields $\psi$ and $\overline{\psi}$ according to the following rules
\begin{subequations}
\begin{align}\label{spinor cov der}
D\psi=d\psi-\frac{i}{4}\,\omega^{a b}\Sigma_{a b}\psi\,,
\\
\overline{D\psi}=d\overline{\psi}+\frac{i}{4}\,\overline{\psi}\Sigma_{a
b}\omega^{a b}\,,
\end{align}
\end{subequations}
where
\begin{equation}
\Sigma^{a b}=\frac{i}{2}\left[\gamma^a,\gamma^b\right]\,
\end{equation}
are the generators of the Lorentz group. Now we claim that the Dirac action can be written as
\begin{equation}
S\left(\psi,\overline{\psi}\right)=\frac{i}{2}\int\star e_{a}\left[\overline{\psi}\gamma^a D\psi-\overline{D\psi}\gamma^a\psi+\frac{i}{2}\,me^{a}\overline{\psi}\psi\right]\,.
\end{equation}
Remembering that, according to our notation, $\star e_{a}\wedge e^b=\delta_a^b dV$, the demonstration follows immediately.

As a final remark, we recall that particular care has to be used in rewriting the physical actions in other possible signatures. For example, many books on quantum field theory use the signature $(+,-,-,-)$, which is preferred by particle physicists. The change of signature can change the sign in front the action according to the change occurring in the equations of motion. As an example try to write the action of a scalar field in both signatures and note a difference in the sign in front of the kinetic term.

\section{Large gauge transformations in Yang-Mills gauge theories}\label{App B} 
Let the $SU(N)$ valued connection $A_{\alpha}=\sum_I A_{\alpha}^I\lambda^I$ and its associated electric field $E^{\gamma}=\sum_{K}E^{\gamma}_K\lambda^K$ (where $I,J,K,\cdots$ are internal indexes running on $1,2,\cdots,N^2-1$) be a couple of conjugate variables in the framework of a canonical formulation of Yang-Mills gauge theories (see § \ref{GSC}). The evolution of the system is limited to a restricted region of the phase space by the first class Gauss constraint, expressed by the following weak equation:
\begin{equation}\label{Gauss Y-M}
G_I:=D_{\alpha}E^{\alpha}_I=\partial_{\alpha}E^{\alpha}_I+f_{I J}^{\ \ K}A^{J}_{\alpha}E^{\alpha}_K\approx 0\,.
\end{equation}
According to the Dirac quantization procedure \cite{Dir1964,HenTei1992}, the state functional describing the quantum physical system must satisfy the Gauss constraint (\ref{Gauss Y-M}), namely we have to require that
\begin{equation}
\widehat{G}_I\Phi(A)=-i D_{\alpha}\frac{\delta}{\delta A_{\alpha}^I}\Phi(A)=0\,,
\end{equation}
where the usual quantum representation of the operators has been assumed. 

The Gauss constraint in Eq.(\ref{Gauss Y-M}) formalizes the request of gauge invariance of the quantum state describing the physical system, namely it is equivalent to requiring that the state functional be invariant under the small component of the gauge group $G=SU(N)$, as can be easily realized. Since the global structure of the gauge group is non-trivial, in view of quantization, it is particularly interesting to study the behavior of the state functional under the large gauge transformations. A non-trivial global structure of the gauge group, in fact, can produce striking effects in the non-perturbative theory, as, e.g., $P$ and $CP$ violations, physically motivating this extension of the theory.

In this respect, let $\widehat{\mathcal{G}}$ be the generator of the large gauge transformations, acting on the state functional $\Phi(A)$. Considering that the Hamiltonian operator, $\widehat{\mathcal{H}}$, is invariant under the full gauge group (or, more formally, it commutes with the operator $\widehat{\mathcal{G}}$), we can construct a set of eigenstates for the quantum theory by diagonalizing simultaneously $\widehat{\mathcal{H}}$ and $\widehat{\mathcal{G}}$. In other words, the following equation
\begin{equation}\label{large operator}
\widehat{\mathcal{G}}\Phi_w(A)=\Phi_w(A^{g})=e^{i\theta w}\Phi_w(A)\,,
\quad\text{where}\quad
A^{g}=g Ag^{-1}+g dg^{-1}\,,
\end{equation}
is a super-selection rule for the states of the theory, which are now labeled by the \emph{winding number} $w=w(g)$, according to their behavior under the action of the large gauge transformation operator. The constant $\theta$ introduced in Eq. (\ref{large operator}) is an angular parameter, which indicates how much the state functional ``rotates'' under the action of the large gauge transformations operator. Specifically, it represents a quantization ambiguity connected with the non-trivial global structure of the gauge group.

Eq.(\ref{large operator}) implies that the wave functionals either have to satisfy suitable $\theta$-dependent boundary conditions passing from one ``slab'' to the next in the configuration space; or, a fully gauge invariant state functional can be constructed, transferring the $\theta$ dependence in the momentum operator. In this respect, we recall that the so-called \emph{Chern-Simons functional},
\begin{equation}
\mathcal{Y}(A)=\frac{1}{8\pi^2}\int{\rm tr}\left(F\wedge A-\frac{1}{3}A\wedge A\wedge A\right)\,,
\end{equation}
is characterized by the following remarkable property: 
\begin{equation}
\mathcal{Y}\left(A^{g}\right)=\mathcal{Y}\left(A\right)+w(g)\,.
\end{equation}
In other words, the Chern--Simons functional under a large gauge transformations turns out to be modified by a quantity exactly corresponding to the winding number, expressed by the Maurer--Cartan integral
\begin{equation}
w(g)=\frac{1}{24\pi^2}\int{\rm tr}(g^{-1}d g)\wedge (g^{-1}d g)\wedge (g^{-1}d g)\,.
\end{equation}
This directly implies that the new state functional, 
\begin{equation}\label{rescaling Y-M}
\Phi^{\prime}(A)=e^{-i\theta\mathcal{Y}(A)}\Phi_w(A)\,,
\end{equation}
will be invariant under the full gauge group, as can be easily demonstrated. In other words we have 
\begin{equation}
\widehat{\mathcal{G}}\Phi^{\prime}(A)=\Phi^{\prime}(A)\,.
\end{equation}

So, by using the rescaling (\ref{rescaling Y-M}), we have obtained a new fully gauge invariant quantum state functional, at the price of modifying the momentum operator, namely, the $\theta$-dependence has been transferred from the boundary conditions to the momentum operator, which becomes:
\begin{align}
E^{\prime\alpha}\Phi^{\prime}(A)=e^{-i\theta\mathcal{Y}(A)}E^{\alpha}e^{i\theta\mathcal{Y}(A)}\Phi^{\prime}(A)=-i\left[\frac{\delta}{\delta A_{\alpha}}-\frac{i\theta}{8\pi^2}\,\epsilon^{\alpha\beta\gamma}F_{\beta\gamma}\right]\Phi^{\prime}(A)\,.
\end{align} 
The above modification in the conjugate momentum reflects on the Hamiltonian operator, i.e.
\begin{align}
H^{\prime}=\int d^3x\,{\rm tr}\left[ \frac{1}{2}\,\left(E^{\alpha}-\frac{\theta}{8\pi^2}\,\epsilon^{\alpha\beta\gamma}F_{\beta\gamma}\right)\left(E_{\alpha}-\frac{\theta}{8\pi^2}\,\epsilon_{\alpha}^{\ \beta\gamma}F_{\beta\gamma}\right)+\frac{1}{4}\,F_{\alpha\beta}F^{\alpha\beta}\right]\,,
\end{align}
generating a pseudo-vectorial term which prevents the new Hamiltonian $H^{\prime}$ from being invariant under the CP discrete symmetry. 

The new Hamiltonian corresponds to a topological modification of the classical action, consisting in the presence of an additional term belonging to the Pontryagin class, i.e.
\begin{equation}
S_{\rm new}(A)=-\frac{1}{4}\int{\rm tr} \star F\wedge F+\frac{\theta}{8\pi^2}\int{\rm tr} F\wedge F\,.
\end{equation}
The $\theta$ parameter appears as a multiplicative constant in front of the modification. It is worth mentioning that the new term does not affect the classical equations of motion, as we have already noticed in § \ref{EGT}, but modifies the vacuum to vacuum amplitude in the path-integral formulation of the quantum theory. In other words, it allows to take into account possible tunneling phenomena between distinct vacua characterized by different winding numbers, violating the CP discrete symmetry.

\newpage



\end{document}